\documentclass[reprint,superscriptaddress]{revtex4-1}
\usepackage{graphicx}  
\usepackage{dcolumn}   
\usepackage{bm}        
\usepackage{amssymb}   
\usepackage{amsmath}
\usepackage{xspace}
\usepackage[dvipsnames]{xcolor}
\usepackage{float}
\usepackage [T1]{fontenc}
\usepackage [utf8]{inputenc}
\usepackage[colorlinks]{hyperref}
\usepackage{braket}
\usepackage[paperwidth=210mm,paperheight=297mm,centering,hmargin=2cm,vmargin=2.5cm]{geometry}
\usepackage{multirow}
\newcommand{\tw}{\ensuremath{t_\mathrm{w}}\xspace}

\newcommand{\teff}{\ensuremath{t^\mathrm{eff}}\xspace}

\newcommand{\Tm}{\ensuremath{T_\mathrm{m}}\xspace}

\newcommand{\Tc}{\ensuremath{T_\mathrm{c}}\xspace}
\newcommand{\NR}{\ensuremath{N_{\text{R}}}\xspace}

\newcommand{\NTherm}{\ensuremath{N_{\text{thermal}}}\xspace}

\newcommand{\Tg}{\ensuremath{T_\mathrm{g}}\xspace}

\begin{document}

\title{Memory and rejuvenation in spin glasses: aging systems are ruled by more than one length scale}

\author{M.~Baity-Jesi}\affiliation{Eawag, Überlandstrasse 133, CH-8600 Dübendorf, Switzerland}

\author{E.~Calore}\affiliation{Dipartimento di Fisica e Scienze della
  Terra, Universit\`a di Ferrara and INFN, Sezione di Ferrara, I-44122
  Ferrara, Italy}

\author{A.~Cruz}\affiliation{Departamento de F\'\i{}sica Te\'orica,
  Universidad de Zaragoza, 50009 Zaragoza,
  Spain}\affiliation{Instituto de Biocomputaci\'on y F\'{\i}sica de
  Sistemas Complejos (BIFI), 50018 Zaragoza, Spain}

\author{L.A.~Fernandez}\affiliation{Departamento de F\'\i{}sica
  Te\'orica, Universidad Complutense, 28040 Madrid,
  Spain}\affiliation{Instituto de Biocomputaci\'on y F\'{\i}sica de
  Sistemas Complejos (BIFI), 50018 Zaragoza, Spain}

\author{J.M.~Gil-Narvion}\affiliation{Instituto de Biocomputaci\'on y
  F\'{\i}sica de Sistemas Complejos (BIFI), 50018 Zaragoza, Spain}

\author{I.~Gonzalez-Adalid Pemartin}\affiliation{Departamento  de F\'\i{}sica Te\'orica, Universidad Complutense, 28040 Madrid, Spain}

\author{A.~Gordillo-Guerrero}\affiliation{Departamento de
  Ingenier\'{\i}a El\'ectrica, Electr\'onica y Autom\'atica, U. de
  Extremadura, 10003, C\'aceres, Spain}\affiliation{Instituto de
  Computaci\'on Cient\'{\i}fica Avanzada (ICCAEx), Universidad de
  Extremadura, 06006 Badajoz, Spain}\affiliation{Instituto de
  Biocomputaci\'on y F\'{\i}sica de Sistemas Complejos (BIFI), 50018
  Zaragoza, Spain}

\author{D.~I\~niguez}\affiliation{Instituto de Biocomputaci\'on y
  F\'{\i}sica de Sistemas Complejos (BIFI), 50018 Zaragoza,
  Spain}\affiliation{Fundaci\'on ARAID, Diputaci\'on General de
  Arag\'on, 50018 Zaragoza, Spain}

\author{A.~Maiorano}\affiliation{Dipartimento di Biotecnologie, Chimica e
  Farmacia, Universit\`a degli studi di Siena, 53100 Siena,
  Italy and INFN, Sezione di Roma 1, 00185 Rome,
  Italy}\affiliation{Instituto de Biocomputaci\'on y F\'{\i}sica de Sistemas
Complejos (BIFI), 50018 Zaragoza, Spain}

\author{E.~Marinari}\affiliation{Dipartimento di Fisica, Sapienza
  Universit\`a di Roma, and CNR-Nanotec, Rome unit and
  INFN, Sezione di Roma 1, 00185 Rome, Italy}

\author{V.~Martin-Mayor}\affiliation{Departamento de F\'\i{}sica
  Te\'orica, Universidad Complutense, 28040 Madrid,
  Spain}\affiliation{Instituto de Biocomputaci\'on y F\'{\i}sica de
  Sistemas Complejos (BIFI), 50018 Zaragoza, Spain}

\author{J.~Moreno-Gordo}\affiliation{Instituto de Biocomputaci\'on y
  F\'{\i}sica de Sistemas Complejos (BIFI), 50018 Zaragoza,
  Spain}\affiliation{Departamento de F\'\i{}sica Te\'orica,
  Universidad de Zaragoza, 50009 Zaragoza, Spain}\affiliation{Departamento de F\'{\i}sica,
  Universidad de Extremadura, 06006 Badajoz,
  Spain}\affiliation{Instituto de Computaci\'on Cient\'{\i}fica
  Avanzada (ICCAEx), Universidad de Extremadura, 06006 Badajoz,
  Spain}

\author{A.~Mu\~noz~Sudupe}\affiliation{Departamento de F\'\i{}sica
  Te\'orica, Universidad Complutense, 28040 Madrid,
  Spain}\affiliation{Instituto de Biocomputaci\'on y F\'{\i}sica de
  Sistemas Complejos (BIFI), 50018 Zaragoza, Spain}

\author{D.~Navarro}\affiliation{Departamento de Ingenier\'{\i}a,
  Electr\'onica y Comunicaciones and I3A, U. de Zaragoza, 50018
  Zaragoza, Spain}

\author{I.~Paga}\email{ilaria.paga@gmail.com}\affiliation{Dipartimento di Fisica, Sapienza
  Universit\`a di Roma, and CNR-Nanotec, Rome unit, Sezione di Roma 1, 00185 Rome, Italy}

\author{G.~Parisi}\affiliation{Dipartimento di Fisica, Sapienza
  Universit\`a di Roma, and CNR-Nanotec, Rome unit and
  INFN, Sezione di Roma 1, 00185 Rome, Italy}

\author{S.~Perez-Gaviro}\affiliation{Departamento de
  F\'\i{}sica Te\'orica, Universidad de Zaragoza, 50009 Zaragoza, Spain}\affiliation{Instituto de Biocomputaci\'on y F\'{\i}sica de Sistemas
  Complejos (BIFI), 50018 Zaragoza, Spain}

\author{F.~Ricci-Tersenghi}\affiliation{Dipartimento di Fisica, Sapienza
  Universit\`a di Roma, and CNR-Nanotec, Rome unit and
  INFN, Sezione di Roma 1, 00185 Rome, Italy}

\author{J.J.~Ruiz-Lorenzo}\affiliation{Departamento de F\'{\i}sica,
  Universidad de Extremadura, 06006 Badajoz,
  Spain}\affiliation{Instituto de Computaci\'on Cient\'{\i}fica
  Avanzada (ICCAEx), Universidad de Extremadura, 06006 Badajoz,
  Spain}\affiliation{Instituto de Biocomputaci\'on y F\'{\i}sica de
  Sistemas Complejos (BIFI), 50018 Zaragoza, Spain}

\author{S.F.~Schifano}\affiliation{Dipartimento di Scienze Chimiche e Farmaceutiche, Universit\`a di Ferrara e INFN  Sezione di Ferrara, I-44122 Ferrara, Italy}

\author{B.~Seoane}\affiliation{Departamento de F\'\i{}sica
  Te\'orica, Universidad Complutense, 28040 Madrid,
  Spain}\affiliation{Instituto de Biocomputaci\'on y F\'{\i}sica de
  Sistemas Complejos (BIFI), 50018 Zaragoza, Spain}

\author{A.~Tarancon}\affiliation{Departamento de F\'\i{}sica
  Te\'orica, Universidad de Zaragoza, 50009 Zaragoza,
  Spain}\affiliation{Instituto de Biocomputaci\'on y F\'{\i}sica de
  Sistemas Complejos (BIFI), 50018 Zaragoza, Spain}

\author{D.~Yllanes}\affiliation{Chan Zuckerberg Biohub, San Francisco, CA, 94158}
\affiliation{Instituto de Biocomputaci\'on y F\'{\i}sica de
  Sistemas Complejos (BIFI), 50018 Zaragoza, Spain}

\collaboration{Janus Collaboration}

\date{\today}   

\begin{abstract}
Memory and rejuvenation effects in the magnetic response of
off-equilibrium spin glasses have been widely regarded as the 
doorway into the experimental exploration of ultrametricity and
temperature chaos (maybe the most exotic features in glassy
free-energy landscapes). Unfortunately, despite more than twenty
years of theoretical efforts following the experimental discovery of
memory and rejuvenation, these effects have thus far been impossible to 
simulate reliably. Yet, three recent developments convinced us to
accept this challenge: first, the custom-built Janus II supercomputer
makes it possible to carry out ``numerical experiments'' in which the
very same quantities that can be measured in
single crystals of CuMn are computed from the simulation,
allowing for parallel analysis of the
simulation/experiment data. Second, Janus II simulations have taught
us how numerical and experimental length scales should be
compared. Third, we have recently understood how 
temperature chaos materializes in aging dynamics. All three aspects have proved crucial for reliably reproducing rejuvenation and memory effects on the computer. Our analysis shows that (at least) three different length scales play a key role in aging dynamics, while essentially all theoretical analyses of the aging dynamics emphasize the presence and the crucial role of a single glassy correlation length.
\end{abstract}


\maketitle

The remarkable off-equilibrium behavior of glass formers at low
temperatures has been described with terms such as
\emph{aging}~\cite{struik:80} or \emph{memory}
and~\emph{rejuvenation}~\cite{jonason:98,lundgren:83,jonsson:00,hammann:00},
which seem more suitable for living beings than for inert chunks
of matter. In this context, spin glasses (which are disordered
magnetic alloys, see, \textit{e.g.},~\cite{mydosh:93}) enjoy a privileged
status. On the experimental side, their magnetic response can be
studied with great accuracy using a superconducting quantum
interference device (SQUID).  Rejuvenation and memory (see the description below) are,
furthermore, remarkably strong in spin glasses, probably because of the large correlation length $\xi$  of
the coherent spin domains. The values of $\xi$ reached in single-crystal samples~\cite{zhai:19,zhai:22,zhai-janus:20a,zhai-janus:21} is
much larger than in other glass-forming materials
(for instance, the $\xi$ measured in supercooled glycerol or propylene
carbonate~\cite{albert:16} is smaller by a factor $\sim 100$). On the
other hand, spin-glass theory~\cite{mezard:87} has proved applicable 
to distant fields that also feature rugged free-energy landscapes, 
such as combinatorial optimization, machine
learning, biology, financial markets or social dynamics.

\begin{figure*}[t]
 \centering
 \includegraphics[width=0.86\textwidth]{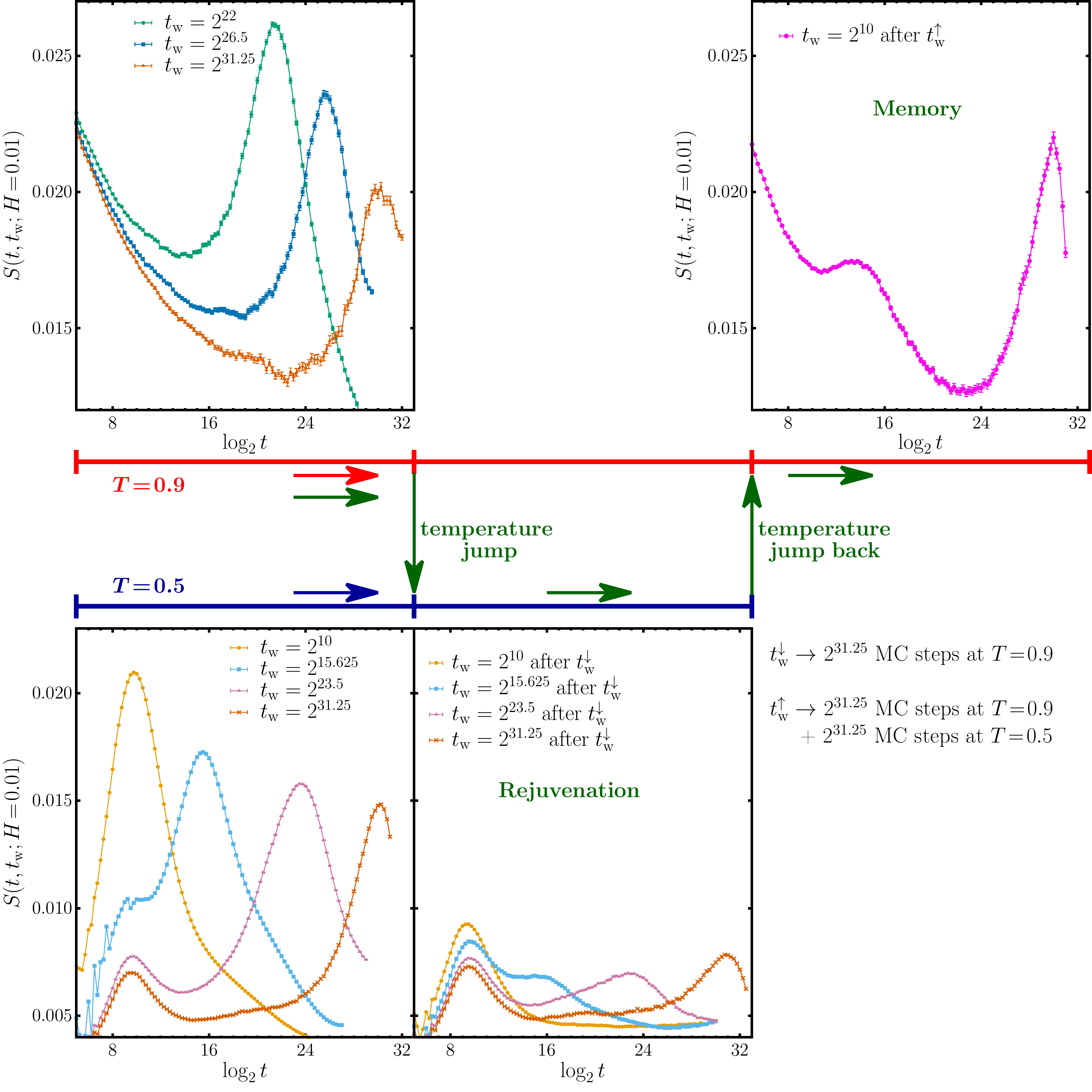}
  \caption{\textbf{The zero-field-cooling (ZFC) numerical experiment measuring rejuvenation and memory.}  The starting random spin configuration is placed instantaneously at the working temperature and it relaxes for a time $\tw$ without a field.  At time $\tw$, a magnetic field $H=0.01$ is applied and the magnetic density, $M_\mathrm{ZFC}(t,\tw;H)$, is recorded.  \textbf{Left panels} show the relaxation function $S_\mathrm{ZFC}(t,\tw;H)$, Eq.~\eqref{eq:S-def}, for the \emph{native} runs at the warmer, $T_1=0.9$, and colder, $T_2=0.5$, temperatures (both below the glass temperature $\Tg=1.102(3)$~\citep{janus:13}). The physically interesting peak of $ S_\mathrm{ZFC}(t,\tw;H) $ defines $\teff_H \simeq \tw$ (in some cases there is also a peak at shorter times, see {\bf Methods}). In our protocol (schematized by the green arrows), after a waiting time $\tw^\downarrow=2^{31.25}$, the temperature is abruptly  dropped from the initial temperature $T_1=0.9$ to the colder temperature $T_2=0.5$. Then, the system relaxes at $T_2$ for an additional time, after which the magnetic field is switched on and the function $S_\mathrm{ZFC}(t,\tw;H)$ shown in the \textbf{bottom-center panel} is measured. Waiting times for these \emph{jump} runs are reported in the legend; the \emph{rejuvenation} effect is clearly visible, since $\teff_H \ll \tw^\downarrow$ and similar to the time $\tw$ spent at $T_2$ (we use \tw for the time spent at the last
  temperature in a given protocol).   Finally, after the waiting time $\tw^\uparrow= 2\tw^\downarrow=2^{32.25}$ [\emph{i.e.}, the system has spent half of its life at the initial temperature $T_1$ and half at the colder temperature $T_2$ without a field], the spin glass is suddenly heated back to $T_1$.  We let the system relax for a short time, $\tw=2^{10}\ll\tw^\downarrow$, after which the magnetic field is switched on. The $S_\mathrm{ZFC}(t,\tw;H)$ measured after the \emph{jump back} and shown in the \textbf{top-right panel} has a peak very similar to the one before the first jump (see \textbf{top-left panel}), evincing the \emph{memory} of the aging at the initial temperature $T_1$, notwithstanding the rejuvenation observed when staying at the lower temperature $T_2$. In all cases, error bars are one standard deviation.}
  \label{fig:thermal_protocol}
\end{figure*}

It is worth stressing that the main part of spin-glass
experimental studies is carried out under off-equilibrium conditions~\cite{vincent:97}.
In the simplest setting, the so-called zero-field-cooling (ZFC) protocol, the system is initially at equilibrium at some very high temperature.
Eventually the spin glass is abruptly cooled to the working
temperature $T<\Tg$ and relaxes for a waiting time $\tw$
($\Tg$ is the glass temperature, while $\tw$ ranges from minutes to several hours). At time $\tw$ a magnetic field $H$ is
switched on and the growing magnetization $M_\text{ZFC}(t,\tw)$ is
recorded at later times $t+\tw$. $M_\text{ZFC}(t,\tw)$
has turned out to have a significant dependence on $\tw$ 
for as long as researchers have had the patience to wait. The relaxation rate
\begin{equation}\label{eq:S-def}
  S_\text{ZFC}(t,\tw;H)=\frac{1}{H}\frac{\mathrm{d}  M_\text{ZFC}(t,\tw;H)}{\mathrm{d}\log t}\,
\end{equation}
peaks at a time $\teff_H$ roughly equal to $\tw$ (see, \textit{e.g.},
Refs.~\cite{zhai-janus:20a,zhai-janus:21} for experimental
results). The only relevant time scale that can be identified
is the glass's age, namely $\tw$ (hence the term \emph{aging}). The left
panels in Fig.~\ref{fig:thermal_protocol} show our results for this
comparatively simple fixed-temperature protocol, which will be named
\emph{native} hereafter. The native setup is used as a
standard for comparison.

\subsection*{The quest for rejuvenation and memory.}
An even more interesting behavior appears when temperature is made
to vary with time. In fact, we shall consider here only the
simplest protocol for which rejuvenation and memory have
been experimentally found~\cite{djurberg:99} (see our temperature-time
scheme in the central part of Fig.~\ref{fig:thermal_protocol}). After
a relaxation of duration $\tw^\downarrow$, the temperature is
lowered abruptly from the initial temperature $T_1<\Tg$
to a lower temperature $T_2$ (the choice of $T_2$ turns out to be
critical, see below). The system is again let to relax at temperature
$T_2$ for an additional time $\tw$, after which a magnetic field is
switched on and the relaxation function $S_\text{ZFC}$ is measured at
times $\tw^\downarrow+\tw+t$. Surprisingly enough, one finds that the
initial relaxation at $T_1$ has been essentially
forgotten: the long-time peak of $S_\text{ZFC}$ is found at times
$\teff_H \sim\tw$, which can be substantially shorter than
$\tw^\downarrow$. This is the rejuvenation effect, which was
experimentally found more than 20 years ago and which we are reporting
in the bottom-central panel of Fig.~\ref{fig:thermal_protocol} for the
first time in a simulation.

Yet, rejuvenation is not the end of the story. After a
total time of $\tw^{\uparrow}=2\tw^{\downarrow}$,
half of it spent at $T_1$ and half at $T_2$, the system is suddenly
heated back to the original temperature $T_1$, where it is left to relax for a 
time $\tw\ll\tw^{\downarrow}$, after which the magnetic field is
switched on and the relaxation function measured. The $S_\text{ZFC}$
is found to peak again at time $\sim \tw^\downarrow$, as if the excursion to
temperature $T_2$ never happened (Fig.~\ref{fig:thermal_protocol}, top-right
panel). This is the memory effect, which at first sight seems quite contradictory with the rejuvenation effect.

The physical origin of memory and rejuvenation in spin glasses has not been identified yet. Then, it is perhaps unsurprising that all past attempts to reproduce these effects in computer simulations have failed~\cite{komori:00,picco:01,berthier:02,takayama:02,maiorano:05,jimenez:05}, which has even raised questions about the validity of the standard model of finite-dimensional spin glasses, the Edwards-Anderson model~\cite{edwards:75,edwards:76}. Fortunately, the Janus~II dedicated supercomputer~\cite{janus:14} has changed this situation, attaining realistic time and length scales and allowing for the first time a thorough examination of spin-glass dynamics both in the vicinity of the critical temperature $\Tg$ and in the low-temperature regime.

The spin-glass dynamics at $T<\Tg$ consists in the growth of (glassy)
magnetic domains of linear size $\xi(\tw)$~\cite{marinari:96,joh:99,janus:08b}
(we shall later refer to this length as $\xi_\text{micro}$). The
non-equilibrium nature of the process is evident in the growth of $\xi(\tw)$
as $\tw$ varies, which is never-ending and extremely slow. In fact, the lower
the temperature, the more sluggish the growth of $\xi(\tw)$ is, see,
\emph{e.g.},~\cite{janus:18,zhai:19}. Janus~II has reached unprecedentedly large values
of $\xi(\tw)$, enabling safe extrapolations from the numerical
time scale of tenths of a second (when $\xi\sim 20\, a_0$, where $a_0$ is the
typical spin-spin distance) to the experimental scale of
hours~\cite{janus:18,zhai:19} (when $\xi\sim 200\, a_0$). This special-purpose 
computer has also made it possible to simulate~\cite{janus:17b} the
experimental protocol for extracting the spin-glass coherence length from
the Zeeman effect~\cite{joh:99}, thus showing the consistency between the Zeeman method and the microscopic approach.
Janus~II allowed us to perform \emph{computer experiments} with a native (\emph{i.e.}, fixed-temperature) protocol and
make a direct comparison of the $S_\text{ZFC}$~\eqref{eq:S-def} obtained in the simulation with that from real experiments on a single crystal of
CuMn~\cite{zhai-janus:20a,zhai-janus:21,zhai-janus:22}. The Edwards-Anderson model and CuMn
turned out to be governed by the same scaling laws, where $\xi$ is the all-important scaling variable. This agreement between simulations and
experiment, however, was established only for native protocols. We need to understand what happens when temperature is varied.

Experimentalists are prone to attribute the rejuvenation effect to
temperature chaos (see, \emph{e.g.}, Ref.~\cite{djurberg:99}; explanations not invoking temperature chaos have
been also proposed~\cite{cugliandolo:98,berthier:03}). Temperature
chaos~\cite{mckay:82,bray:87b,kondor:89} is an equilibrium notion stating that
spin configurations typical from the Boltzmann distribution at temperature $T_1$
would be very atypical for temperature $T_2$, no matter how close $T_1$ and
$T_2$ are (provided that $T_1,T_2<\Tg$). Temperature chaos could explain
why the relaxation at temperature $T_1$ seems useless at $T_2$
(\textit{i.e.}, rejuvenation). Yet, even in the mean-field approximation,
showing that temperature chaos is really present in equilibrium has been a
real \emph{tour-de-force}~\cite{rizzo:03,parisi:10}.  Furthermore, extending
the equilibrium concept of temperature chaos to the experimentally relevant
context of off-equilibrium dynamics is a very recent
achievement~\cite{janus:21}.

\begin{figure*}[t]
  \centering
  \includegraphics[width=0.86\textwidth]{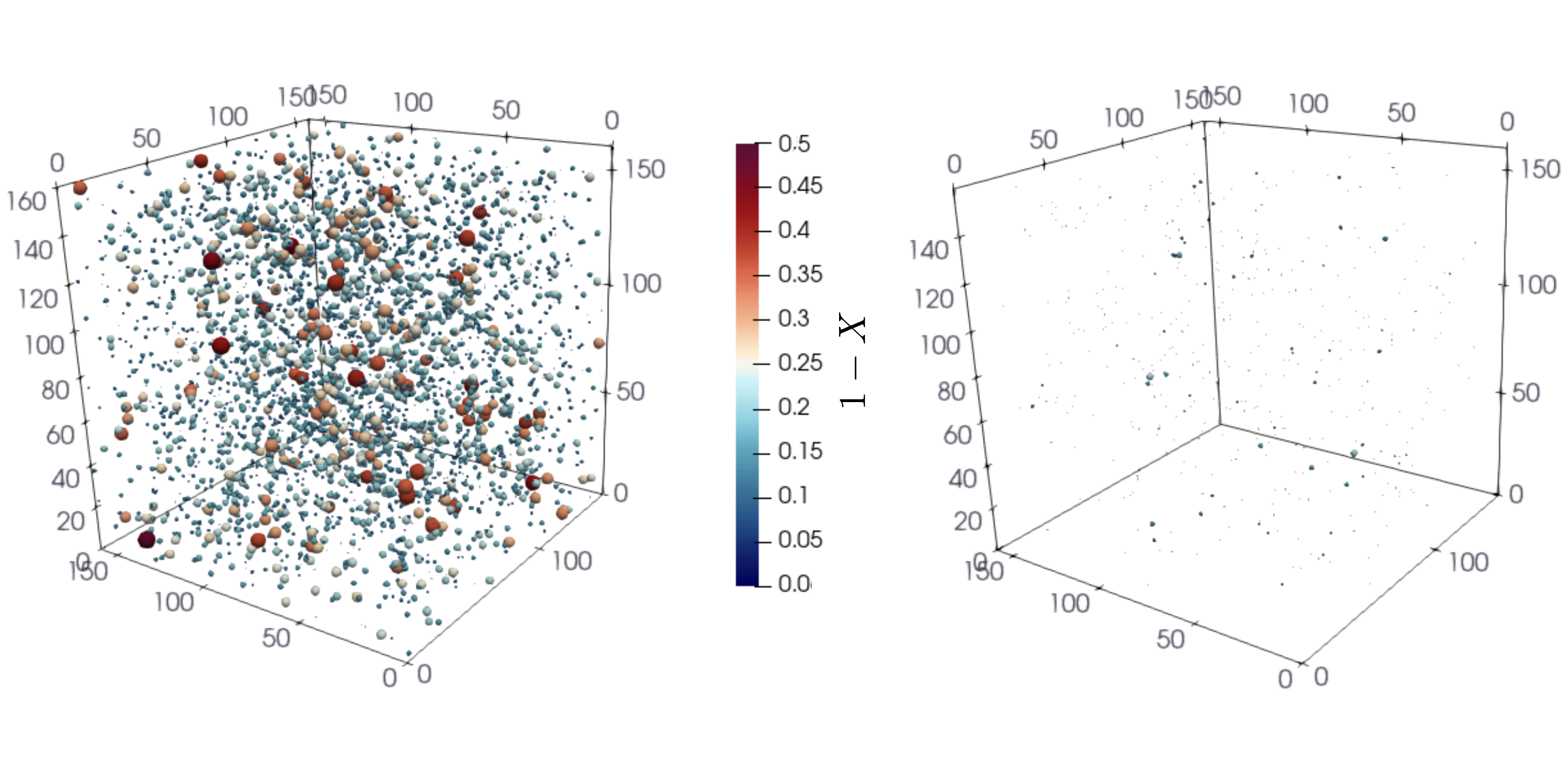}
  \caption{\textbf{Temperature chaos is spatially heterogeneous when it is clearly present}.
  The 8000 randomly chosen spheres in a sample of size $L=160$ are depicted with a color code depending on $1-X$ ($X$ is the chaotic correlation parameter as computed for spheres of radius $R=5 \, a_0$, see \textbf{Methods}). For visualization purposes, spheres are represented with a radius $12(1-X)$, so that only fully chaotic spheres (\textit{i.e.}, $X=0$) would have the largest size. In order to avoid cluttering, we draw only spheres with $X<0.97$. \textbf{On the left}, we calculate the chaotic correlation parameter $X$ between the native system at $T=0.5$ (\emph{i.e.}, a fixed-temperature protocol: a completely disordered system is put at temperature $T=0.5$ and let to evolve at this temperature for a time $\tw=2^{31.25}$) and the jump system at the same temperature $T=0.5$ (recall the central part of Fig.~\ref{fig:thermal_protocol}: the jump system has spent the first  half of its life, $\tw^\downarrow=2^{31.25}$, at the hot temperature $T_1=0.9$ and the second half, $\tw=2^{31.25}$, at the cold temperature $T=0.5$). Very strong chaotic heterogeneity is found. \textbf{On the right}, we calculate the chaotic correlation parameter $X$ between the native system at the hot temperature $T=0.9$ and the jump-back system at the same temperature $T=0.9$ (the jump-back system has spent  a time $\tw^\downarrow=2^{31.25}$ at the hot temperature $T_1=0.9$, a time $\tw^\uparrow-\tw^\downarrow=2^{31.25}$ at the cold temperature $T_2=0.5$, and then $\tw=2^{10}$ again at $T_1=0.9$ --- see the temperature protocol of Fig.~\ref{fig:thermal_protocol}). After the cycle the system does not display chaotic heterogeneity since almost every sphere has a large correlation parameter $X$, i.e., a strong \textit{memory} (more examples can be found in Supplementary Note V).
 }
\label{fig:chaotic_heterogeneity}
\end{figure*}

Dynamic temperature chaos is spatially extremely heterogeneous (see Fig.~\ref{fig:chaotic_heterogeneity}).
To measure it we choose many spheres of linear size $R$ in random positions within the sample.  We compare
\emph{within each sphere} spin configurations obtained at temperature $T_1$
and time $\tw^{T_1}$ with configurations from temperature $T_2$ and time
$\tw^{T_2}$ (the simplifying choice $\xi(\tw^{T_1},T_1)=\xi(\tw^{T_2},T_2)= \xi(\tw)$
was made in \cite{janus:21}).  The comparison is quantitative, through the
computation of a correlation coefficient $X_{T_1,T_2}$, see {\bf Methods}.
Many of those spheres turn out to have very weak temperature chaos [$X_{T_1,T_2}\approx 1$]. Yet,
with low probability, one picks a chaotic sphere with a significantly smaller
$X_{T_1,T_2}$. In fact, the analysis in \cite{janus:21} identifies a
crossover length scale $\xi^*(T_1,T_2)$: for $\xi(\tw) \ll \xi^*(T_1,T_2)$
chaotic spheres are \emph{very rare} but for $\xi(\tw) \gg \xi^*(T_1,T_2)$
chaotic spheres become \emph{fairly typical}. 
A scaling law was also found: $\xi^*(T_1,T_2)\propto (T_1-T_2)^{-1/\zeta_{\text{NE}}}$ with
$\zeta_{\text{NE}}=1.19(2)$.

Our last building block comes from the experiment of Ref.~\cite{zhai:22}, which
identifies a minimal temperature jump $\Delta T_{\text{min}}$ in a CuMn
sample. Temperature chaos in that sample 
turned out to be exceedingly weak whenever 
$T_1-T_2 < \Delta T_{\text{min}}$.
It follows that, in a simulation, chaotic
spheres will be just too rare to significantly affect the overall sample
relaxation unless~\footnote{We are indebted to Prof. Orbach for this
  observation.}
  \begin{equation}\label{eq:minimal-jump}
  \left.\frac{T_1-T_2}{\Tg}\right|_{\text{sim}} \approx \left.\frac{\Delta
      T_{\text{min}}}{T_\mathrm{g}}\right|_{\text{CuMn}} \Bigg[\frac{\xi_{\text{CuMn}}(\tw)}{\xi_{\text{micro}}(\tw)}\Bigg]^{\zeta_{\text{NE}}}\,,
\end{equation} 
where the subindex \emph{micro} stands for the $\xi$ computed in the numerical simulation (see {\bf Methods}) while $T_{\mathrm{g}}$ is the glass temperature, which is different for the CuMn sample and for simulations. Plugging in typical numbers ($\Delta T_{\text{min}}=450$ mK, $T_{\mathrm{g}}=31.5$ K,
$\xi_{\text{CuMn}}(\tw)\approx 220\, a_0$ and
$\xi_{\text{micro}}(\tw)\approx 16.6\, a_0$), we conclude from
Eq.~\eqref{eq:minimal-jump} that, given the correlation length reached in our
simulations, a successful simulation of the rejuvenation effect should have
$T_1-T_2 >0.32 \Tg$.

In this work we have considered two temperature jumps, see
Table~\ref{tab:details_NUM}. The first jump, namely
$T_1=0.9 \rightarrow T_2=0.5$, meets the requirement for temperature chaos expressed in
Eq.~\eqref{eq:minimal-jump}, while the second jump,
$T_1=0.9 \rightarrow T_2=0.7$ is too small. Hence we expect to find qualitative differences
between the two.

\subsection*{Becoming quantitative: how many controlling length scales?}

Our discussion shall emphasize three different
length scales, focusing on their physical interpretation and their utility to
rationalize the rejuvenation and memory effects (many more details are provided in {\bf Methods}~\footnote{Our simulations
  are also described in the {\bf Methods}
  section, see also Table~\ref{tab:details_NUM} for crucial simulation
  details and Ref.~\cite{paga:21} for useful computational tricks.}). Only one of these scales, named $\xi_\text{Zeeman}$, can be experimentally accessed nowadays (the other two lengths, however, provide invaluable microscopic information):
  \begin{itemize}
      \item  $\xi_\text{micro}$ is the size of the (glassy)
domains within the sample (is the largest
length scale at which we can regard the system as ordered at time $\tw$).
\item $\xi_{\text{Zeeman}}$ is obtained by counting the number of spins that react coherently to an externally applied field~\cite{joh:99}. It provides a very direct quantification
of memory and rejuvenation.
\item $\zeta(t_1,t_2)$~\cite{janus:09b,castillo:02,jaubert:07} is obtained from the comparison of the same system at the two times $t_1<t_2$ [$\zeta(t_1,t_2)$ is the typical size of the regions where coherent rearrangements have occurred between $t_1$ and $t_2$].  
\end{itemize}

\begin{figure}[t]
  \centering
    \includegraphics[width=0.86\columnwidth]{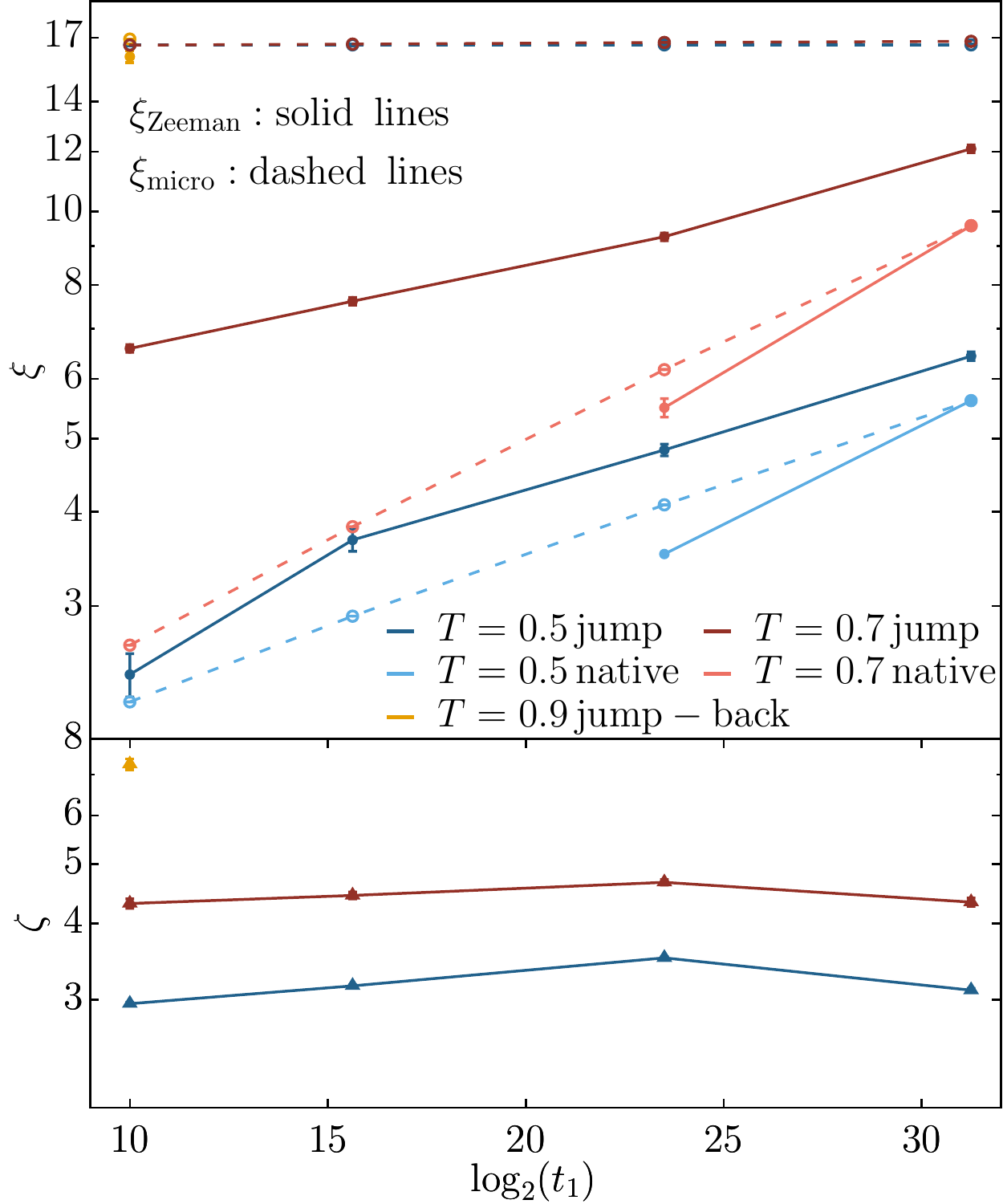}
  \caption{\textbf{\emph{Aging} dynamics is controlled by three 
  length scales (at least).}  (See text and {\bf Methods} for
extended discussion of these three length scales). The dashed lines and empty dots are for $\xi_{\mathrm{micro}}(\tw,T)$, the
continuous lines and filled dots are for $\xi_{\mathrm{Zeeman}}(\tw,T)$, and the continuous lines and filled triangles are for $\zeta(t_1,t_2)$.  \textbf{On the top}, for the
     \emph{native} protocols, $\xi_\mathrm{Zeeman}(\tw)$ follows quite closely
     the behavior of $\xi_\mathrm{micro}(\tw)$. For the \emph{jump} protocol
     with $T_2=0.5$, $\xi_\mathrm{Zeeman}(\tw)$ is extremely similar to the
     corresponding curve for the native run [this $T_2$ meets the \emph{chaos requirement}
     expressed in Eq.~\eqref{eq:minimal-jump}], which means that
     the system responds to an external magnetic field as if
     \textit{rejuvenated}, $\xi_\mathrm{Zeeman}^\mathrm{jump}(\tw) \ll
     \xi_\mathrm{micro}(\tw)$. When the system jumps back to $T_1=0.9$
     (\textit{i.e.} $T_1=0.9 \to T_2=0.5 \to T_1=0.9$),
     $\xi_\mathrm{Zeeman}^\mathrm{jump}(\tw) = \xi_\mathrm{micro}(\tw)$ after
     a extremely short time (\textit{memory}). Instead,
     $\xi_\mathrm{Zeeman}(\tw)$ never becomes small for the jump protocols
     with $T_2=0.7$.  \textbf{On the bottom}, the size of the regions
     undergoing coherent rearrangements when evolving from the initial to the final time,
     $\zeta(t_1,t_2)$, is much smaller than $\xi_\mathrm{micro}(\tw)$ for all
     our jump protocols. The earliest of the two
     times is the very last instant that the system spent at temperature $T_1=0.9$
     (\textit{i.e.}, just before the temperature jump; in the case of the
     jump-back protocol just before the first jump $T_1=0.9 \rightarrow
     T_2=0.5$). In all cases, $\zeta$ is represented as a function of $t_2-t_1$.  Error bars are one standard deviation. See Supplementary Note III for more results on $\zeta(t_1,t_2)$.}  \label{fig:three_scales}
\end{figure}

Previous analysis for native (\emph{i.e.}, fixed-temperature)  protocols tell us that $\xi_{\text{Zeeman}}$ follows quite closely the behavior of the microscopic
length $\xi_{\text{micro}}$~\cite{joh:99,janus:17b,zhai-janus:20a,zhai-janus:21}. This is what we find 
in the top panel of
Fig.~\ref{fig:three_scales}.  
There are two salient features in the time growth of either $\xi_{\text{Zeeman}}$ or $\xi_{\text{micro}}(\tw)$ at fixed temperature~\cite{janus:18,zhai:19}: the
growth slows down as $\xi_{\text{micro}}$ increases~\footnote{In fact, $\mathrm{d}\log\tw/\mathrm{d}\log\xi_{\text{micro}}$ is approximately
constant when $\tw$ varies in logarithmic scale} and the dynamics at lower
temperatures is enormously slower~\footnote{$T \mathrm{d}\log\tw/\mathrm{d}\log\xi_{\text{micro}}$ is roughly constant
when different temperatures are compared}. In fact, see
Table~\ref{tab:details_NUM} and Ref.~\cite{janus:18}, at the largest
temperature $T=0.9$ it is comparatively easy to reach a large
$\xi_{\text{micro}}\approx 16.6\,a_0$ in a native protocol. Instead, for a
similar simulation time, the native protocol at $T=0.5$ is limited to
$\xi_{\text{micro}}\approx 5.6\,a_0$.  It is then unsurprising that, when the
temperature jumps from $T_1=0.9$ to $T_2=0.5$ or $ 0.7$, see
Fig.~\ref{fig:three_scales}--top, the size of the glassy domains is locked to
their value at jump time, namely $\xi_{\text{micro}}\approx 16.6\,a_0$: the
time needed for such a large domain to grow at the lower temperature $T_2$ far
exceeds the scale of our simulations. The importance of this locking was also emphasized in Ref.~\cite{berthier:02}.

While $\xi_{\text{micro}}$ is locked at the value it has at the jump time, the behavior of $\xi_{\text{Zeeman}}$ is different in the jump protocols. In the jump complying with Eq.~\eqref{eq:minimal-jump},
$T_1=0.9 \rightarrow T_2=0.5$, $\xi_{\text{Zeeman}}(\tw)$ is 
quite similar to the corresponding curve for the
native run at $T=0.5$. From the point of view of the response to the magnetic
field, rejuvenation is almost complete for this temperature jump, because the
initial relaxation at $T_1=0.9$ (almost) does not leave a measurable trace.
Instead, for the more modest jump $T_1=0.9 \rightarrow T_2=0.7$, rejuvenation
is weaker and $\xi_{\text{Zeeman}}$ is sensibly larger than in the native runs
(see Supplementary Note II for more details).

Furthermore, it is also shown in
Fig.~\ref{fig:three_scales}--top that, when the system jumps back to $T_1=0.9$
(\textit{i.e.} $T_1=0.9 \rightarrow T_2=0.5\rightarrow T_1=0.9$, recall the top-right
panel in Fig.~\ref{fig:thermal_protocol}), the response to the magnetic-field
goes back to normal: $\xi_{\text{Zeeman}}$ catches up with $\xi_{\text{micro}}$
after a extremely short transient. This is another manifestation of the memory
effect.

As for the third length scale, see
Fig.~\ref{fig:three_scales}--bottom, for all our jump protocols
we find $\zeta\ll\xi_{\text{micro}}$, which means that
the configuration right before the jump has not been substantially modified by the
excursion to the low temperature $T_2=0.5$. Interestingly enough, for the jump-back
protocol $\zeta$ gets a lot bigger than for the single jump protocols. In other words, the dynamics is unlocked when the system
comes back to its original temperature $T_1$. Nevertheless, $\zeta$ is still
substantially smaller than $\xi_{\text{micro}}$. The original
spin configuration has, therefore, suffered only local distortions. 

\subsection*{Dynamic temperature chaos  and rejuvenation}

At this point, the elephant in the room is clear: what is the physical origin for rejuvenation and memory?

In order to answer the question, we need to compare pairs of spin
configurations. One of the configurations will be taken from the jump
protocols. The other configuration will come from the native runs at
temperatures $T_2=0.5$ or $T_2=0.7$. In an attempt  to make a fair comparison,
we shall choose the native configurations at $T_2$ at their largest possible
waiting time. In fact, the magnetic domains will be substantially smaller in
the native protocol than they are in the jump protocol (at $T_2=0.5$, for
instance, one has to compare $\xi^{\text{native}}_{\text{micro}}\approx 5.8\,a_0$ with
$\xi^{\text{jump}}_{\text{micro}}\approx 16.6\,a_0$). 

The main steps in the comparison were outlined above (for a more paused exposition see~{\bf Methods} and
Ref.~\cite{janus:21}). We pick  at random in the sample spheres of radius
$R$. The results presented in this paper
were obtained with $R=5\,a_0$ to make sure that the spheres will have a chance to fit within
the glassy domains of the native runs (we have tried
other values of $R$, finding qualitatively similar results, see Supplementary Note VI). The configurations from the two protocols are compared by
computing a correlation coefficient $X$ that takes into account only the spins
contained in the sphere. If $X$ is significantly smaller than unity we regard
that particular sphere as~\emph{chaotic}, because typical configurations from
the two protocols differ within the sphere. To be precise, we compute the
probability distribution function $F(\tilde X)$, namely the fraction of the
spheres with a correlation coefficient $X< \tilde X$.

\begin{table}[t] \centering
\begin{tabular}{c c c l l D{.}{.}{-1} l l}
\hline\hline
             $\mathrm{Type}$ && $T$ && \multicolumn{1}{c}{Waiting time} &  \multicolumn{2}{c}{$\quad\xi_\mathrm{micro}\quad$}  &  $t_\mathrm{max}$  \\ [0.5ex]
                        \hline  \\ [-1.5ex]
                        \textsc{native} &&  0.9 && $2^{31.25} (\,=\,\tw^{\downarrow})$            &  16.63(5)   && $2^{32}$  \\
                        \textsc{native} &&  0.5 && $2^{10}$     &  2.23926(2) && $2^{28}$  \\ 
                        \textsc{native} &&  0.5 && $2^{15.625}$ &  2.9090(4)  && $2^{28}$   \\ 
                        \textsc{native} &&  0.5 && $2^{23.5}$   &  4.0865(15) && $2^{30}$   \\ 
                        \textsc{native} &&  0.5 && $2^{31.25}$  &  5.6167(4) && $2^{32}$   \\ 
                        \textsc{jump}   &&  0.5 && $\tw^{\downarrow}+2^{10}$     &  16.62(12)   && $2^{28}$   \\ 
                        \textsc{jump}   &&  0.5 && $\tw^{\downarrow}+2^{15.625}$ &  16.68(12)   && $2^{28}$  \\ 
                        \textsc{jump}   &&  0.5 && $\tw^{\downarrow}+2^{23.5}$   &  16.75(13)   && $2^{31}$  \\ 
                        \textsc{jump}   &&  0.5 && $\tw^{\downarrow}+2^{31.25}$  &  16.81(13)   && $2^{33.5}$   \\ 
                        \textsc{native} &&  0.7 && $2^{10}$     &  2.6629(4)  && $2^{28}$   \\ 
                        \textsc{native} &&  0.7 && $2^{15.625}$ &  3.8230(10) && $2^{28}$  \\ 
                        \textsc{native} &&  0.7 && $2^{23.5}$   &  6.1742(4) && $2^{28}$   \\ 
                            \textsc{native} &&  0.7 && $2^{31.25}$  &  9.578(11)   && $2^{33}$    \\ 
                        \textsc{jump}   &&  0.7 && $\tw^{\downarrow}+2^{10}$     &  16.62(12)   && $2^{28}$    \\ 
                        \textsc{jump}   &&  0.7 && $\tw^{\downarrow}+2^{15.625}$ &  16.67(12)   && $2^{28}$    \\ 
                        \textsc{jump}   &&  0.7 && $\tw^{\downarrow}+2^{23.5}$   &  16.76(12)     && $2^{28}$    \\ 
                        \textsc{jump}   &&  0.7 && $\tw^{\downarrow}+2^{31.25}$  &  16.81(13)     && $2^{32}$   \\ [1ex] 
                        \hline\hline
\end{tabular}
\caption{ {\bf Basic features of our simulations}. We have simulated on the
  Janus II supercomputer the Edwards-Anderson model with nearest-neighbor
  couplings ($J=\pm1$ with 50\% probability), on simple-cubic lattices
  containing $160^3$ Ising spins $s=\pm 1$ (the lattice size is $L=160\,a_0$)
  and endowed with periodic boundary conditions.  A particular set of couplings is
  termed \emph{sample}. For every sample and every set of parameters, we have
  simulated 512 independent trajectories (\emph{i.e.}, 512 \emph{replicas}, see {\bf
    Methods}). This table lists the main parameters for
  each of our numerical simulations. Temperature-varying protocols, see the
  central part of Fig.~\ref{fig:thermal_protocol}, are named \emph{jump}
  protocols. In all cases, temperature $T$ refers to the temperature at which the relaxation
  function in Eq.~\eqref{eq:S-def} is computed. All temperatures considered
  are in the spin-glass phase: $T<\Tg=1.102(3)$ \cite{janus:13}. The waiting time
  is the period before the magnetic field
  $H$ is switched on (for jump protocols, this consists of a time $\tw^\downarrow$ at the starting temperature $T_1=0.9$, plus
  a time \tw at $T_2$).  The microscopic
  correlation length $\xi_\mathrm{micro}$ is given in $a_0=1$ units and
  computed just before the magnetic field is switched on [error bars for
  $\xi_\mathrm{micro}(\tw)$ are one standard deviation]. Finally, the longest
  simulation time in the presence of a field is $t_\mathrm{max}$.}
   \label{tab:details_NUM}
\end{table}

\begin{figure}[t] \centering 
  \centering
  \includegraphics[width=0.86\columnwidth]{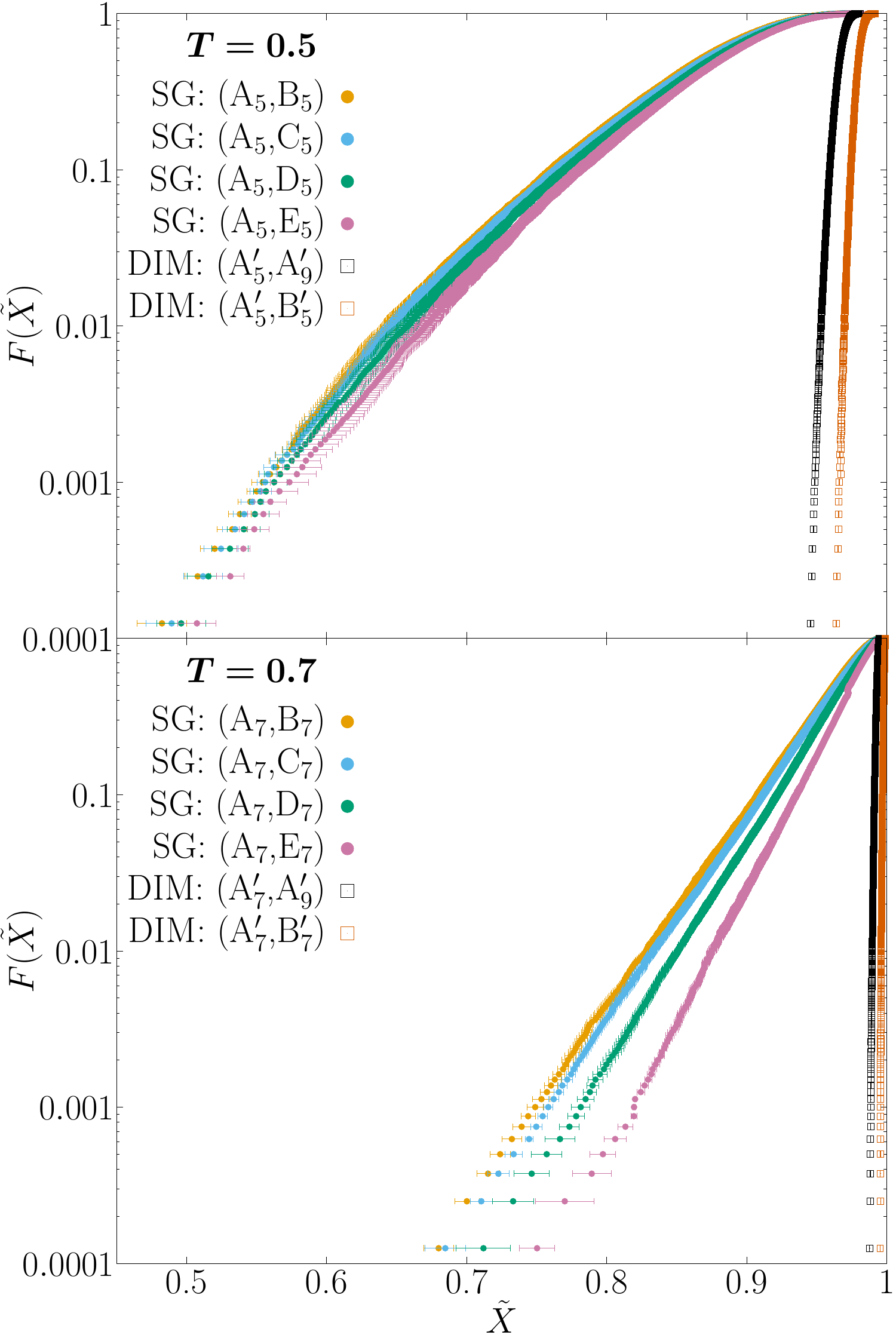} 
  \caption{\textbf{Strong temperature chaos correlates with full rejuvenation.} The figure shows the fraction of the spheres with radius $R=5\, a_0$ that have a correlation parameter $X$ smaller than $\tilde X$, $F(\tilde X)$ (see {\bf Methods}; the pairs of systems for which the correlation parameter $X$ is computed are listed in Table~\ref{tab:configurations_chaotic_comparison}).
    In the {\bf top} panel, one of the systems in the pair used to compute $X$ is always taken from the native protocol at $T=0.5$ ($T=0.7$ for the {\bf bottom} panel). In the cases reported in the top panel,
    the partner in the pair that undergoes the temperature-jump protocol experiences strong rejuvenation. Instead, recall Fig.~\ref{fig:three_scales}--top, rejuvenation is only partial for the cases reported in the bottom panel. Interestingly enough, small correlation parameters appear with high probability in the top panel while they are very rare events in the bottom panel. We also show
     a comparison with the diluted Ising Model (our null experiment, see {\bf Methods}), where temperature chaos is not expected. Indeed, in the absence of temperature chaos, the probability concentrates at $X\approx 1$. In all cases, error bars are one standard deviation.}
         \label{fig:null_exp}
\end{figure}

Our results shown in Fig.~\ref{fig:null_exp}--bottom for the jump protocol $T_1=0.9 \rightarrow T_2=0.7$  remind us of previous studies~\cite{janus:21}. The vast majority of the spheres have a very large correlation coefficient, and truly chaotic spheres are found only in the tail of the distribution (probability $0.1\%$ or smaller). 

Interestingly enough, see the left panel in
Fig.~\ref{fig:chaotic_heterogeneity} and
Fig.~\ref{fig:null_exp}--top, the situation
is radically different for the jump protocol $T_1=0.9 \rightarrow T_2=0.5$,
where the spheres in percentile 10 of the distribution are as chaotic as
the most chaotic spheres we could find for the jump
$T_1=0.9 \rightarrow T_2=0.7$. In fact, to our knowledge,
Fig.~\ref{fig:null_exp}--top reports the strongest temperature-chaos
signal ever observed in a simulation of glassy dynamics.

In order to convince ourselves that the extreme chaos is not an
artifact of the disparity in domain sizes, we have tried a
null experiment by simulating a model where no temperature chaos is
expected, namely the link-diluted ferromagnetic Ising model (we have used the
results in Ref.~\cite{berche:04} to match as closely as possible in the
diluted ferromagnet the conditions in our spin-glass simulations, with special
care in matching the size of the domains, see {\bf Methods}). As expected, see
Fig.~\ref{fig:null_exp}, the sphere distribution for the ferromagnet is
concentrated at correlation coefficient $X\approx 1$. We conclude that the
spin-glass results in Fig.~\ref{fig:null_exp}--top are genuine evidence for
dynamic temperature chaos.

It is also interesting that the distribution function in
Fig.~\ref{fig:null_exp}--top barely depends on $\tw$. This is another
manifestation of the dynamic lock-down when the temperature jumps to the lower value.

The overall conclusions of this analysis are twofold. First, the requirement
expressed by Eq.~\eqref{eq:minimal-jump}, which is based on CuMn experimental
results~\cite{zhai:22}, is sensible: strong temperature chaos is found only
when $T_1-T_2$ is as large as Eq.~\eqref{eq:minimal-jump} demands.
Second, only when temperature chaos is strong do our simulations find 
strong rejuvenation (recall Fig.~\ref{fig:three_scales}--top).

\begin{table}[t] \centering
\begin{tabular}{c c c c c c c c l}
\hline\hline
 & $\,\,\,$ & System &  $\,\,\,$ & $T$ &  $\,\,\,$ & Type &  $\,\,\,$ & Waiting time \\
\hline \\ [-1.5ex]
A$_9$ & & SG & & $0.9$ & & \textsc{native} & & $\tw^{\downarrow}=2^{31.25}$ \\
B$_9$ & & SG & & $0.9$ & & \textsc{jump-back} & & $\tw^{\downarrow}+\tw^{\uparrow}+2^{10}$ \\
A$_5$ & & SG & & $0.5$ & & \textsc{native} & & $\tw^{\downarrow}+2^{31.25}$ \\
B$_5$ & & SG & & $0.5$ & & \textsc{jump} & & $\tw^{\downarrow}+2^{10}$ \\
C$_5$ & & SG & & $0.5$ & & \textsc{jump} & & $\tw^{\downarrow}+2^{15.625}$ \\
D$_5$ & & SG & & $0.5$ & & \textsc{jump} & & $\tw^{\downarrow}+2^{23.5}$ \\
E$_5$ & & SG & & $0.5$ & & \textsc{jump} & & $\tw^{\downarrow}+2^{31.25}$ \\
\hline\\ [-1.5ex]
A$_7$ & & SG & & $0.7$ & & \textsc{native} & & $\tw^{\downarrow}+2^{31.25}$ \\
B$_7$ & & SG & & $0.7$ & & \textsc{jump} & &  $\tw^{\downarrow}+2^{10}$ \\
C$_7$ & & SG & & $0.7$ & & \textsc{jump} & & $\tw^{\downarrow}+2^{15.625}$ \\
D$_7$ & & SG & & $0.7$ & & \textsc{jump} & & $\tw^{\downarrow}+2^{23.5}$ \\
E$_7$ & & SG & & $0.7$ & & \textsc{jump} & & $\tw^{\downarrow}+2^{31.25}$ \\
\hline\\ [-1.5ex]
$\text{A}'_5$ & & DIM & & $0.5$ & & \textsc{native} & & $76$ \\
$\text{B}'_5$ & & DIM & & $0.5$ & & \textsc{jump} & & $430+69$ \\
$\text{A}'_7$ & & DIM & & $0.7$ & & \textsc{native} & & $197$ \\
$\text{B}'_7$ & & DIM & & $0.7$ & & \textsc{jump} & & $430+165$ \\
$\text{A}'_9$ & & DIM & & $0.9$ & & \textsc{native} & & $430$ \\
\hline\hline
\end{tabular}
\caption{{\bf  Identifying parameters for each of the numerical simulations appearing in
  Figure~\ref{fig:null_exp}}.  Spin glass (SG) protocols follow the notation in Figure~\ref{fig:thermal_protocol}. $T$ is the final temperature in the protocol. For the diluted Ising Model (DIM; see {\bf Methods} for the DIM temperature-naming convention) we write explicitly $\tw^{(1)} + \tw^{(2)}$ for jump protocols
  to stress that the time before the jump (at $T\!=\! 0.9$), $\tw^{(1)}$, differs from the time $\tw^{(2)}$ at 
  the
  final temperature. We choose $\tw^{(1)}$ such  that $\xi_{\text{micro}}$ coincides for both protocols in the pairs  (A$_5$,A'$_5$), (A$_7$,A'$_7$) and (A$_9$,A'$_9$).
  }
\label{tab:configurations_chaotic_comparison}
\end{table}

\subsection*{Where do we stand?}

Our simulations depict a clear picture of the rejuvenation and memory
effects. Provided that the temperature jump is large enough, see
Eq.~\eqref{eq:minimal-jump}, the spin-glass state that was forming at
temperature $T_1$ is completely alien at temperature $T_2$ (at least it looks
like an alien when compared with the native state that grows directly at $T_2$,
see Fig.~\ref{fig:chaotic_heterogeneity}).
In fact, the response to the magnetic field (which is the quantity measured in experiments~\cite{jonason:98,lundgren:83,jonsson:00,hammann:00,djurberg:99,zhai:19,zhai:22,zhai-janus:20a,zhai-janus:21})
is not qualitatively different in the alien state and in the native
state that grows from a fully disordered high-temperature state. The
system just dismisses the relaxation it achieved at the higher temperature $T_1$.

Paradoxically enough, the alien state is locked at temperature $T_2$:
the microscopic rearrangement at $T_2$, see Fig.~\ref{fig:three_scales}--bottom,
takes place on too small length scales to dissolve such foreign glassy domains.
As a consequence, when the temperature is taken back to $T_1$, the glassy domains characteristic of
$T_1$ are still there.  This seems to be the physical origin of the memory effect.
This reasoning is also consistent with recent experiments that find that the memory effect strongly depends on $\tw^\downarrow$ (\textit{i.e.}, the time spent in the first stay at $T_1$)~\footnote{Jennifer Freedberg, private communication (2022).}. Indeed, if $\tw^\downarrow$ is too small, the memory effect almost disappears. Our interpretation of this experimental finding is that the glassy domains at $T_1$ need to grow large enough as to
remain mostly unaltered at the lower temperature $T_2$. 

Looking back, we understand as well why rejuvenation has been so
difficult to find in simulations: the correlation lengths that could be
reached prior to the Janus family of supercomputers were rather limited (we
are referring here to the $\xi_{\text{micro}}$ length scale). Therefore, Eq.~\eqref{eq:minimal-jump} would demand an exceedingly large temperature jump $T_1 - T_2$ if one wants to have  a large fraction of chaotic spheres of the relevant size.

An open question is whether or not the only experimentally accessible coherence length, namely  $\xi_{\text{Zeeman}}(\tw)$, relates
to some correlation function under \emph{all} circumstances.   Indeed, in the case of native protocols, $\xi_{\text{Zeeman}}(\tw)$ behaves analogously to
$\xi_{\text{micro}}(\tw)$, which we know how  to obtain from a correlation function. However, $\xi_{\text{micro}}(\tw)$ is not a valid proxy for $\xi_{\text{Zeeman}}(\tw)$ in temperature-jump protocols.

Finally, we should also stress that the analysis of the rejuvenation and
memory effects requires considering no less than three different
length scales, which can be quite different from each other. Of course, one of
the three, the domain size $\xi_{\text{micro}}(\tw)$, acts as a cut-off for
the other lengths. Yet we have seen that $\xi_{\text{micro}}(\tw)$ is not
nearly enough to describe the variety of behaviors that an aging system may
present. In fact, $\xi_{\text{micro}}(\tw)$ has stayed essentially constant
for all the jump simulations that we have considered here! Therefore, a useful
theory of aging dynamics cannot feature just a single length scale. In this
sense, we think that our work poses a new and significant question for the
different theories of aging dynamics.

\bibliographystyle{apsrev4-1}
\bibliography{rejuvenation_and_memory_janus_v2.bbl}

\begin{thebibliography}{48}%
\makeatletter
\providecommand \@ifxundefined [1]{%
 \@ifx{#1\undefined}
}%
\providecommand \@ifnum [1]{%
 \ifnum #1\expandafter \@firstoftwo
 \else \expandafter \@secondoftwo
 \fi
}%
\providecommand \@ifx [1]{%
 \ifx #1\expandafter \@firstoftwo
 \else \expandafter \@secondoftwo
 \fi
}%
\providecommand \natexlab [1]{#1}%
\providecommand \enquote  [1]{``#1''}%
\providecommand \bibnamefont  [1]{#1}%
\providecommand \bibfnamefont [1]{#1}%
\providecommand \citenamefont [1]{#1}%
\providecommand \href@noop [0]{\@secondoftwo}%
\providecommand \href [0]{\begingroup \@sanitize@url \@href}%
\providecommand \@href[1]{\@@startlink{#1}\@@href}%
\providecommand \@@href[1]{\endgroup#1\@@endlink}%
\providecommand \@sanitize@url [0]{\catcode `\\12\catcode `\$12\catcode
  `\&12\catcode `\#12\catcode `\^12\catcode `\_12\catcode `\%12\relax}%
\providecommand \@@startlink[1]{}%
\providecommand \@@endlink[0]{}%
\providecommand \url  [0]{\begingroup\@sanitize@url \@url }%
\providecommand \@url [1]{\endgroup\@href {#1}{\urlprefix }}%
\providecommand \urlprefix  [0]{URL }%
\providecommand \Eprint [0]{\href }%
\providecommand \doibase [0]{http://dx.doi.org/}%
\providecommand \selectlanguage [0]{\@gobble}%
\providecommand \bibinfo  [0]{\@secondoftwo}%
\providecommand \bibfield  [0]{\@secondoftwo}%
\providecommand \translation [1]{[#1]}%
\providecommand \BibitemOpen [0]{}%
\providecommand \bibitemStop [0]{}%
\providecommand \bibitemNoStop [0]{.\EOS\space}%
\providecommand \EOS [0]{\spacefactor3000\relax}%
\providecommand \BibitemShut  [1]{\csname bibitem#1\endcsname}%
\let\auto@bib@innerbib\@empty
\bibitem [{\citenamefont {Struik}(1980)}]{struik:80}%
  \BibitemOpen
  \bibfield  {author} {\bibinfo {author} {\bibfnamefont {C.~L.~E.}\
  \bibnamefont {Struik}},\ }\href {\doibase 10.1002/bbpc.19780820975} {\emph
  {\bibinfo {title} {Physical Aging in Amorphous Polymers and Other
  Materials}}}\ (\bibinfo  {publisher} {Elsevier},\ \bibinfo {address}
  {Amsterdam},\ \bibinfo {year} {1980})\BibitemShut {NoStop}%
\bibitem [{\citenamefont {Jonason}\ \emph {et~al.}(1998)\citenamefont
  {Jonason}, \citenamefont {Vincent}, \citenamefont {Hammann}, \citenamefont
  {Bouchaud},\ and\ \citenamefont {Nordblad}}]{jonason:98}%
  \BibitemOpen
  \bibfield  {author} {\bibinfo {author} {\bibfnamefont {K.}~\bibnamefont
  {Jonason}}, \bibinfo {author} {\bibfnamefont {E.}~\bibnamefont {Vincent}},
  \bibinfo {author} {\bibfnamefont {J.}~\bibnamefont {Hammann}}, \bibinfo
  {author} {\bibfnamefont {J.~P.}\ \bibnamefont {Bouchaud}}, \ and\ \bibinfo
  {author} {\bibfnamefont {P.}~\bibnamefont {Nordblad}},\ }\href {\doibase
  10.1103/PhysRevLett.81.3243} {\bibfield  {journal} {\bibinfo  {journal}
  {Phys. Rev. Lett.}\ }\textbf {\bibinfo {volume} {81}},\ \bibinfo {pages}
  {3243} (\bibinfo {year} {1998})}\BibitemShut {NoStop}%
\bibitem [{\citenamefont {Lundgren}\ \emph {et~al.}(1983)\citenamefont
  {Lundgren}, \citenamefont {Svedlindh},\ and\ \citenamefont
  {Beckman}}]{lundgren:83}%
  \BibitemOpen
  \bibfield  {author} {\bibinfo {author} {\bibfnamefont {L.}~\bibnamefont
  {Lundgren}}, \bibinfo {author} {\bibfnamefont {P.}~\bibnamefont {Svedlindh}},
  \ and\ \bibinfo {author} {\bibfnamefont {O.}~\bibnamefont {Beckman}},\ }\href
  {\doibase 10.1016/0304-8853(83)90922-8} {\bibfield  {journal} {\bibinfo
  {journal} {J. Magn. Magn. Mater.}\ }\textbf {\bibinfo {volume} {31--34}},\
  \bibinfo {pages} {1349} (\bibinfo {year} {1983})}\BibitemShut {NoStop}%
\bibitem [{\citenamefont {Jonsson}\ \emph {et~al.}(1999)\citenamefont
  {Jonsson}, \citenamefont {Jonason}, \citenamefont {J{\"o}nsson},\ and\
  \citenamefont {Nordblad}}]{jonsson:00}%
  \BibitemOpen
  \bibfield  {author} {\bibinfo {author} {\bibfnamefont {T.}~\bibnamefont
  {Jonsson}}, \bibinfo {author} {\bibfnamefont {K.}~\bibnamefont {Jonason}},
  \bibinfo {author} {\bibfnamefont {P.~E.}\ \bibnamefont {J{\"o}nsson}}, \ and\
  \bibinfo {author} {\bibfnamefont {P.}~\bibnamefont {Nordblad}},\ }\href
  {\doibase 10.1103/PhysRevB.59.8770} {\bibfield  {journal} {\bibinfo
  {journal} {Phys. Rev. B}\ }\textbf {\bibinfo {volume} {59}},\ \bibinfo
  {pages} {8770} (\bibinfo {year} {1999})}\BibitemShut {NoStop}%
\bibitem [{\citenamefont {Hammann}\ \emph {et~al.}(2000)\citenamefont
  {Hammann}, \citenamefont {Vincent}, \citenamefont {Dupuis}, \citenamefont
  {Alba}, \citenamefont {Ocio},\ and\ \citenamefont {Bouchaud}}]{hammann:00}%
  \BibitemOpen
  \bibfield  {author} {\bibinfo {author} {\bibfnamefont {J.}~\bibnamefont
  {Hammann}}, \bibinfo {author} {\bibfnamefont {E.}~\bibnamefont {Vincent}},
  \bibinfo {author} {\bibfnamefont {V.}~\bibnamefont {Dupuis}}, \bibinfo
  {author} {\bibfnamefont {M.}~\bibnamefont {Alba}}, \bibinfo {author}
  {\bibfnamefont {M.}~\bibnamefont {Ocio}}, \ and\ \bibinfo {author}
  {\bibfnamefont {J.-P.}\ \bibnamefont {Bouchaud}},\ }\href@noop {} {\bibfield
  {journal} {\bibinfo  {journal} {J. Phys. Soc. Jpn.}\ ,\ \bibinfo {pages}
  {Suppl A. 206}} (\bibinfo {year} {2000})}\BibitemShut {NoStop}%
\bibitem [{\citenamefont {Mydosh}(1993)}]{mydosh:93}%
  \BibitemOpen
  \bibfield  {author} {\bibinfo {author} {\bibfnamefont {J.~A.}\ \bibnamefont
  {Mydosh}},\ }\href@noop {} {\emph {\bibinfo {title} {Spin Glasses: an
  Experimental Introduction}}}\ (\bibinfo  {publisher} {Taylor and Francis},\
  \bibinfo {address} {London},\ \bibinfo {year} {1993})\BibitemShut {NoStop}%
\bibitem [{\citenamefont {Zhai}\ \emph {et~al.}(2019)\citenamefont {Zhai},
  \citenamefont {Martin-Mayor}, \citenamefont {Schlagel}, \citenamefont
  {Kenning},\ and\ \citenamefont {Orbach}}]{zhai:19}%
  \BibitemOpen
  \bibfield  {author} {\bibinfo {author} {\bibfnamefont {Q.}~\bibnamefont
  {Zhai}}, \bibinfo {author} {\bibfnamefont {V.}~\bibnamefont {Martin-Mayor}},
  \bibinfo {author} {\bibfnamefont {D.~L.}\ \bibnamefont {Schlagel}}, \bibinfo
  {author} {\bibfnamefont {G.~G.}\ \bibnamefont {Kenning}}, \ and\ \bibinfo
  {author} {\bibfnamefont {R.~L.}\ \bibnamefont {Orbach}},\ }\href {\doibase
  10.1103/PhysRevB.100.094202} {\bibfield  {journal} {\bibinfo  {journal}
  {Phys. Rev. B}\ }\textbf {\bibinfo {volume} {100}},\ \bibinfo {pages}
  {094202} (\bibinfo {year} {2019})}\BibitemShut {NoStop}%
\bibitem [{\citenamefont {Zhai}\ \emph {et~al.}(2022)\citenamefont {Zhai},
  \citenamefont {Orbach},\ and\ \citenamefont {Schlagel}}]{zhai:22}%
  \BibitemOpen
  \bibfield  {author} {\bibinfo {author} {\bibfnamefont {Q.}~\bibnamefont
  {Zhai}}, \bibinfo {author} {\bibfnamefont {R.~L.}\ \bibnamefont {Orbach}}, \
  and\ \bibinfo {author} {\bibfnamefont {D.~L.}\ \bibnamefont {Schlagel}},\
  }\href {\doibase 10.1103/PhysRevB.105.014434} {\bibfield  {journal} {\bibinfo
   {journal} {Phys. Rev. B}\ }\textbf {\bibinfo {volume} {105}},\ \bibinfo
  {pages} {014434} (\bibinfo {year} {2022})}\BibitemShut {NoStop}%
\bibitem [{\citenamefont {Zhai}\ \emph {et~al.}(2020)\citenamefont {Zhai},
  \citenamefont {Paga}, \citenamefont {Baity-Jesi}, \citenamefont {Calore},
  \citenamefont {Cruz}, \citenamefont {Fernandez}, \citenamefont {Gil-Narvion},
  \citenamefont {Gonzalez-Adalid~Pemartin}, \citenamefont {Gordillo-Guerrero},
  \citenamefont {I\~niguez}, \citenamefont {Maiorano}, \citenamefont
  {Marinari}, \citenamefont {Martin-Mayor}, \citenamefont {Moreno-Gordo},
  \citenamefont {Mu\~noz Sudupe}, \citenamefont {Navarro}, \citenamefont
  {Orbach}, \citenamefont {Parisi}, \citenamefont {Perez-Gaviro}, \citenamefont
  {Ricci-Tersenghi}, \citenamefont {Ruiz-Lorenzo}, \citenamefont {Schifano},
  \citenamefont {Schlagel}, \citenamefont {Seoane}, \citenamefont {Tarancon},
  \citenamefont {Tripiccione},\ and\ \citenamefont {Yllanes}}]{zhai-janus:20a}%
  \BibitemOpen
  \bibfield  {author} {\bibinfo {author} {\bibfnamefont {Q.}~\bibnamefont
  {Zhai}}, \bibinfo {author} {\bibfnamefont {I.}~\bibnamefont {Paga}}, \bibinfo
  {author} {\bibfnamefont {M.}~\bibnamefont {Baity-Jesi}}, \bibinfo {author}
  {\bibfnamefont {E.}~\bibnamefont {Calore}}, \bibinfo {author} {\bibfnamefont
  {A.}~\bibnamefont {Cruz}}, \bibinfo {author} {\bibfnamefont {L.~A.}\
  \bibnamefont {Fernandez}}, \bibinfo {author} {\bibfnamefont {J.~M.}\
  \bibnamefont {Gil-Narvion}}, \bibinfo {author} {\bibfnamefont
  {I.}~\bibnamefont {Gonzalez-Adalid~Pemartin}}, \bibinfo {author}
  {\bibfnamefont {A.}~\bibnamefont {Gordillo-Guerrero}}, \bibinfo {author}
  {\bibfnamefont {D.}~\bibnamefont {I\~niguez}}, \bibinfo {author}
  {\bibfnamefont {A.}~\bibnamefont {Maiorano}}, \bibinfo {author}
  {\bibfnamefont {E.}~\bibnamefont {Marinari}}, \bibinfo {author}
  {\bibfnamefont {V.}~\bibnamefont {Martin-Mayor}}, \bibinfo {author}
  {\bibfnamefont {J.}~\bibnamefont {Moreno-Gordo}}, \bibinfo {author}
  {\bibfnamefont {A.}~\bibnamefont {Mu\~noz Sudupe}}, \bibinfo {author}
  {\bibfnamefont {D.}~\bibnamefont {Navarro}}, \bibinfo {author} {\bibfnamefont
  {R.~L.}\ \bibnamefont {Orbach}}, \bibinfo {author} {\bibfnamefont
  {G.}~\bibnamefont {Parisi}}, \bibinfo {author} {\bibfnamefont
  {S.}~\bibnamefont {Perez-Gaviro}}, \bibinfo {author} {\bibfnamefont
  {F.}~\bibnamefont {Ricci-Tersenghi}}, \bibinfo {author} {\bibfnamefont
  {J.~J.}\ \bibnamefont {Ruiz-Lorenzo}}, \bibinfo {author} {\bibfnamefont
  {S.~F.}\ \bibnamefont {Schifano}}, \bibinfo {author} {\bibfnamefont {D.~L.}\
  \bibnamefont {Schlagel}}, \bibinfo {author} {\bibfnamefont {B.}~\bibnamefont
  {Seoane}}, \bibinfo {author} {\bibfnamefont {A.}~\bibnamefont {Tarancon}},
  \bibinfo {author} {\bibfnamefont {R.}~\bibnamefont {Tripiccione}}, \ and\
  \bibinfo {author} {\bibfnamefont {D.}~\bibnamefont {Yllanes}},\ }\href
  {\doibase 10.1103/PhysRevLett.125.237202} {\bibfield  {journal} {\bibinfo
  {journal} {Phys. Rev. Lett.}\ }\textbf {\bibinfo {volume} {125}},\ \bibinfo
  {pages} {237202} (\bibinfo {year} {2020})}\BibitemShut {NoStop}%
\bibitem [{\citenamefont {Paga}\ \emph {et~al.}(2021)\citenamefont {Paga},
  \citenamefont {Zhai}, \citenamefont {Baity-Jesi}, \citenamefont {Calore},
  \citenamefont {Cruz}, \citenamefont {Fernandez}, \citenamefont {Gil-Narvion},
  \citenamefont {Gonzalez-Adalid~Pemartin}, \citenamefont {Gordillo-Guerrero},
  \citenamefont {I{\~n}iguez}, \citenamefont {Maiorano}, \citenamefont
  {Marinari}, \citenamefont {Martin-Mayor}, \citenamefont {Moreno-Gordo},
  \citenamefont {Mu{\~n}oz-Sudupe}, \citenamefont {Navarro}, \citenamefont
  {Orbach}, \citenamefont {Parisi}, \citenamefont {Perez-Gaviro}, \citenamefont
  {Ricci-Tersenghi}, \citenamefont {Ruiz-Lorenzo}, \citenamefont {Schifano},
  \citenamefont {Schlagel}, \citenamefont {Seoane}, \citenamefont {Tarancon},
  \citenamefont {Tripiccione},\ and\ \citenamefont {Yllanes}}]{zhai-janus:21}%
  \BibitemOpen
  \bibfield  {author} {\bibinfo {author} {\bibfnamefont {I.}~\bibnamefont
  {Paga}}, \bibinfo {author} {\bibfnamefont {Q.}~\bibnamefont {Zhai}}, \bibinfo
  {author} {\bibfnamefont {M.}~\bibnamefont {Baity-Jesi}}, \bibinfo {author}
  {\bibfnamefont {E.}~\bibnamefont {Calore}}, \bibinfo {author} {\bibfnamefont
  {A.}~\bibnamefont {Cruz}}, \bibinfo {author} {\bibfnamefont {L.~A.}\
  \bibnamefont {Fernandez}}, \bibinfo {author} {\bibfnamefont {J.~M.}\
  \bibnamefont {Gil-Narvion}}, \bibinfo {author} {\bibfnamefont
  {I.}~\bibnamefont {Gonzalez-Adalid~Pemartin}}, \bibinfo {author}
  {\bibfnamefont {A.}~\bibnamefont {Gordillo-Guerrero}}, \bibinfo {author}
  {\bibfnamefont {D.}~\bibnamefont {I{\~n}iguez}}, \bibinfo {author}
  {\bibfnamefont {A.}~\bibnamefont {Maiorano}}, \bibinfo {author}
  {\bibfnamefont {E.}~\bibnamefont {Marinari}}, \bibinfo {author}
  {\bibfnamefont {V.}~\bibnamefont {Martin-Mayor}}, \bibinfo {author}
  {\bibfnamefont {J.}~\bibnamefont {Moreno-Gordo}}, \bibinfo {author}
  {\bibfnamefont {A.}~\bibnamefont {Mu{\~n}oz-Sudupe}}, \bibinfo {author}
  {\bibfnamefont {D.}~\bibnamefont {Navarro}}, \bibinfo {author} {\bibfnamefont
  {R.~L.}\ \bibnamefont {Orbach}}, \bibinfo {author} {\bibfnamefont
  {G.}~\bibnamefont {Parisi}}, \bibinfo {author} {\bibfnamefont
  {S.}~\bibnamefont {Perez-Gaviro}}, \bibinfo {author} {\bibfnamefont
  {F.}~\bibnamefont {Ricci-Tersenghi}}, \bibinfo {author} {\bibfnamefont
  {J.~J.}\ \bibnamefont {Ruiz-Lorenzo}}, \bibinfo {author} {\bibfnamefont
  {S.~F.}\ \bibnamefont {Schifano}}, \bibinfo {author} {\bibfnamefont {D.~L.}\
  \bibnamefont {Schlagel}}, \bibinfo {author} {\bibfnamefont {B.}~\bibnamefont
  {Seoane}}, \bibinfo {author} {\bibfnamefont {A.}~\bibnamefont {Tarancon}},
  \bibinfo {author} {\bibfnamefont {R.}~\bibnamefont {Tripiccione}}, \ and\
  \bibinfo {author} {\bibfnamefont {D.}~\bibnamefont {Yllanes}},\ }\href
  {\doibase 10.1088/1742-5468/abdfca} {\bibfield  {journal} {\bibinfo
  {journal} {J. Stat. Mech.}\ }\textbf {\bibinfo {volume} {2021}},\ \bibinfo
  {pages} {033301} (\bibinfo {year} {2021})}\BibitemShut {NoStop}%
\bibitem [{\citenamefont {Albert}\ \emph {et~al.}(2016)\citenamefont {Albert},
  \citenamefont {Bauer}, \citenamefont {Michl}, \citenamefont {Biroli},
  \citenamefont {Bouchaud}, \citenamefont {Loidl}, \citenamefont
  {Lunkenheimer}, \citenamefont {Tourbot}, \citenamefont {Wiertel-Gasquet},\
  and\ \citenamefont {Ladieu}}]{albert:16}%
  \BibitemOpen
  \bibfield  {author} {\bibinfo {author} {\bibfnamefont {S.}~\bibnamefont
  {Albert}}, \bibinfo {author} {\bibfnamefont {T.}~\bibnamefont {Bauer}},
  \bibinfo {author} {\bibfnamefont {M.}~\bibnamefont {Michl}}, \bibinfo
  {author} {\bibfnamefont {G.}~\bibnamefont {Biroli}}, \bibinfo {author}
  {\bibfnamefont {J.-P.}\ \bibnamefont {Bouchaud}}, \bibinfo {author}
  {\bibfnamefont {A.}~\bibnamefont {Loidl}}, \bibinfo {author} {\bibfnamefont
  {P.}~\bibnamefont {Lunkenheimer}}, \bibinfo {author} {\bibfnamefont
  {R.}~\bibnamefont {Tourbot}}, \bibinfo {author} {\bibfnamefont
  {C.}~\bibnamefont {Wiertel-Gasquet}}, \ and\ \bibinfo {author} {\bibfnamefont
  {F.}~\bibnamefont {Ladieu}},\ }\href {\doibase 10.1126/science.aaf3182}
  {\bibfield  {journal} {\bibinfo  {journal} {Science}\ }\textbf {\bibinfo
  {volume} {352}},\ \bibinfo {pages} {1308} (\bibinfo {year} {2016})},\ \Eprint
  {http://arxiv.org/abs/arXiv:1606.04079} {arXiv:1606.04079} \BibitemShut
  {NoStop}%
\bibitem [{\citenamefont {M{\'e}zard}\ \emph {et~al.}(1987)\citenamefont
  {M{\'e}zard}, \citenamefont {Parisi},\ and\ \citenamefont
  {Virasoro}}]{mezard:87}%
  \BibitemOpen
  \bibfield  {author} {\bibinfo {author} {\bibfnamefont {M.}~\bibnamefont
  {M{\'e}zard}}, \bibinfo {author} {\bibfnamefont {G.}~\bibnamefont {Parisi}},
  \ and\ \bibinfo {author} {\bibfnamefont {M.}~\bibnamefont {Virasoro}},\
  }\href {\doibase 10.1142/0271} {\emph {\bibinfo {title} {Spin-Glass Theory
  and Beyond}}}\ (\bibinfo  {publisher} {World Scientific},\ \bibinfo {address}
  {Singapore},\ \bibinfo {year} {1987})\BibitemShut {NoStop}%
\bibitem [{\citenamefont {Baity-Jesi}\ \emph {et~al.}(2013)\citenamefont
  {Baity-Jesi}, \citenamefont {Ba\~{n}os}, \citenamefont {Cruz}, \citenamefont
  {Fernandez}, \citenamefont {Gil-Narvion}, \citenamefont {Gordillo-Guerrero},
  \citenamefont {Iniguez}, \citenamefont {Maiorano}, \citenamefont {Mantovani},
  \citenamefont {Marinari}, \citenamefont {Mart\'{i}n-Mayor}, \citenamefont
  {Monforte-Garcia}, \citenamefont {Mu{\~n}oz~Sudupe}, \citenamefont {Navarro},
  \citenamefont {Parisi}, \citenamefont {Perez-Gaviro}, \citenamefont
  {Pivanti}, \citenamefont {Ricci-Tersenghi}, \citenamefont {Ruiz-Lorenzo},
  \citenamefont {Schifano}, \citenamefont {Seoane}, \citenamefont {Tarancon},
  \citenamefont {Tripiccione},\ and\ \citenamefont {Yllanes}}]{janus:13}%
  \BibitemOpen
  \bibfield  {author} {\bibinfo {author} {\bibfnamefont {M.}~\bibnamefont
  {Baity-Jesi}}, \bibinfo {author} {\bibfnamefont {R.~A.}\ \bibnamefont
  {Ba\~{n}os}}, \bibinfo {author} {\bibfnamefont {A.}~\bibnamefont {Cruz}},
  \bibinfo {author} {\bibfnamefont {L.~A.}\ \bibnamefont {Fernandez}}, \bibinfo
  {author} {\bibfnamefont {J.~M.}\ \bibnamefont {Gil-Narvion}}, \bibinfo
  {author} {\bibfnamefont {A.}~\bibnamefont {Gordillo-Guerrero}}, \bibinfo
  {author} {\bibfnamefont {D.}~\bibnamefont {Iniguez}}, \bibinfo {author}
  {\bibfnamefont {A.}~\bibnamefont {Maiorano}}, \bibinfo {author}
  {\bibfnamefont {F.}~\bibnamefont {Mantovani}}, \bibinfo {author}
  {\bibfnamefont {E.}~\bibnamefont {Marinari}}, \bibinfo {author}
  {\bibfnamefont {V.}~\bibnamefont {Mart\'{i}n-Mayor}}, \bibinfo {author}
  {\bibfnamefont {J.}~\bibnamefont {Monforte-Garcia}}, \bibinfo {author}
  {\bibfnamefont {A.}~\bibnamefont {Mu{\~n}oz~Sudupe}}, \bibinfo {author}
  {\bibfnamefont {D.}~\bibnamefont {Navarro}}, \bibinfo {author} {\bibfnamefont
  {G.}~\bibnamefont {Parisi}}, \bibinfo {author} {\bibfnamefont
  {S.}~\bibnamefont {Perez-Gaviro}}, \bibinfo {author} {\bibfnamefont
  {M.}~\bibnamefont {Pivanti}}, \bibinfo {author} {\bibfnamefont
  {F.}~\bibnamefont {Ricci-Tersenghi}}, \bibinfo {author} {\bibfnamefont
  {J.~J.}\ \bibnamefont {Ruiz-Lorenzo}}, \bibinfo {author} {\bibfnamefont
  {S.~F.}\ \bibnamefont {Schifano}}, \bibinfo {author} {\bibfnamefont
  {B.}~\bibnamefont {Seoane}}, \bibinfo {author} {\bibfnamefont
  {A.}~\bibnamefont {Tarancon}}, \bibinfo {author} {\bibfnamefont
  {R.}~\bibnamefont {Tripiccione}}, \ and\ \bibinfo {author} {\bibfnamefont
  {D.}~\bibnamefont {Yllanes}} (\bibinfo {collaboration} {Janus
  Collaboration}),\ }\href {\doibase 10.1103/PhysRevB.88.224416} {\bibfield
  {journal} {\bibinfo  {journal} {Phys. Rev. B}\ }\textbf {\bibinfo {volume}
  {88}},\ \bibinfo {pages} {224416} (\bibinfo {year} {{2013}})},\ \Eprint
  {http://arxiv.org/abs/arXiv:1310.2910} {arXiv:1310.2910} \BibitemShut
  {NoStop}%
\bibitem [{\citenamefont {Vincent}\ \emph {et~al.}(1997)\citenamefont
  {Vincent}, \citenamefont {Hammann}, \citenamefont {Ocio}, \citenamefont
  {Bouchaud},\ and\ \citenamefont {Cugliandolo}}]{vincent:97}%
  \BibitemOpen
  \bibfield  {author} {\bibinfo {author} {\bibfnamefont {E.}~\bibnamefont
  {Vincent}}, \bibinfo {author} {\bibfnamefont {J.}~\bibnamefont {Hammann}},
  \bibinfo {author} {\bibfnamefont {M.}~\bibnamefont {Ocio}}, \bibinfo {author}
  {\bibfnamefont {J.-P.}\ \bibnamefont {Bouchaud}}, \ and\ \bibinfo {author}
  {\bibfnamefont {L.~F.}\ \bibnamefont {Cugliandolo}},\ }in\ \href {\doibase
  10.1007/BFb0104827} {\emph {\bibinfo {booktitle} {Complex Behavior of Glassy
  Systems}}},\ \bibinfo {series and number} {\bibinfo {series} {Lecture Notes
  in Physics}\ No.\ \bibinfo {number} {492}},\ \bibinfo {editor} {edited by\
  \bibinfo {editor} {\bibfnamefont {M.}~\bibnamefont {Rub{\'{\i}}}}\ and\
  \bibinfo {editor} {\bibfnamefont {C.}~\bibnamefont {P{\'e}rez-Vicente}}}\
  (\bibinfo  {publisher} {Springer},\ \bibinfo {year} {1997})\BibitemShut
  {NoStop}%
\bibitem [{\citenamefont {Djurberg}\ \emph {et~al.}(1999)\citenamefont
  {Djurberg}, \citenamefont {Jonason},\ and\ \citenamefont
  {Nordblad}}]{djurberg:99}%
  \BibitemOpen
  \bibfield  {author} {\bibinfo {author} {\bibfnamefont {C.}~\bibnamefont
  {Djurberg}}, \bibinfo {author} {\bibfnamefont {K.}~\bibnamefont {Jonason}}, \
  and\ \bibinfo {author} {\bibfnamefont {P.}~\bibnamefont {Nordblad}},\ }\href
  {\doibase https://doi.org/10.1007/s100510050824} {\bibfield  {journal}
  {\bibinfo  {journal} {Eur. Phys. J. B}\ }\textbf {\bibinfo {volume} {10}},\
  \bibinfo {pages} {15} (\bibinfo {year} {1999})}\BibitemShut {NoStop}%
\bibitem [{\citenamefont {Komori}\ \emph {et~al.}(2000)\citenamefont {Komori},
  \citenamefont {Yoshino},\ and\ \citenamefont {Takayama}}]{komori:00}%
  \BibitemOpen
  \bibfield  {author} {\bibinfo {author} {\bibfnamefont {T.}~\bibnamefont
  {Komori}}, \bibinfo {author} {\bibfnamefont {H.}~\bibnamefont {Yoshino}}, \
  and\ \bibinfo {author} {\bibfnamefont {H.}~\bibnamefont {Takayama}},\ }\href
  {\doibase 10.1143/JPSJ.69.1192} {\bibfield  {journal} {\bibinfo  {journal}
  {Journal of the Physical Society of Japan}\ }\textbf {\bibinfo {volume}
  {69}},\ \bibinfo {pages} {1192} (\bibinfo {year} {2000})}\BibitemShut
  {NoStop}%
\bibitem [{\citenamefont {Picco}\ \emph {et~al.}(2001)\citenamefont {Picco},
  \citenamefont {Ricci-Tersenghi},\ and\ \citenamefont {Ritort}}]{picco:01}%
  \BibitemOpen
  \bibfield  {author} {\bibinfo {author} {\bibfnamefont {M.}~\bibnamefont
  {Picco}}, \bibinfo {author} {\bibfnamefont {F.}~\bibnamefont
  {Ricci-Tersenghi}}, \ and\ \bibinfo {author} {\bibfnamefont {F.}~\bibnamefont
  {Ritort}},\ }\href {\doibase 10.1103/PhysRevB.63.174412} {\bibfield
  {journal} {\bibinfo  {journal} {Phys. Rev. B}\ }\textbf {\bibinfo {volume}
  {63}},\ \bibinfo {pages} {174412} (\bibinfo {year} {2001})}\BibitemShut
  {NoStop}%
\bibitem [{\citenamefont {Berthier}\ and\ \citenamefont
  {Bouchaud}(2002)}]{berthier:02}%
  \BibitemOpen
  \bibfield  {author} {\bibinfo {author} {\bibfnamefont {L.}~\bibnamefont
  {Berthier}}\ and\ \bibinfo {author} {\bibfnamefont {J.-P.}\ \bibnamefont
  {Bouchaud}},\ }\href {\doibase 10.1103/PhysRevB.66.054404} {\bibfield
  {journal} {\bibinfo  {journal} {Phys. Rev. B}\ }\textbf {\bibinfo {volume}
  {66}},\ \bibinfo {pages} {054404} (\bibinfo {year} {2002})}\BibitemShut
  {NoStop}%
\bibitem [{\citenamefont {Takayama}\ and\ \citenamefont
  {Hukushima}(2002)}]{takayama:02}%
  \BibitemOpen
  \bibfield  {author} {\bibinfo {author} {\bibfnamefont {H.}~\bibnamefont
  {Takayama}}\ and\ \bibinfo {author} {\bibfnamefont {K.}~\bibnamefont
  {Hukushima}},\ }\href {\doibase 10.1143/JPSJ.71.3003} {\bibfield  {journal}
  {\bibinfo  {journal} {Journal of the Physical Society of Japan}\ }\textbf
  {\bibinfo {volume} {71}},\ \bibinfo {pages} {3003} (\bibinfo {year}
  {2002})}\BibitemShut {NoStop}%
\bibitem [{\citenamefont {Maiorano}\ \emph {et~al.}(2005)\citenamefont
  {Maiorano}, \citenamefont {Marinari},\ and\ \citenamefont
  {Ricci-Tersenghi}}]{maiorano:05}%
  \BibitemOpen
  \bibfield  {author} {\bibinfo {author} {\bibfnamefont {A.}~\bibnamefont
  {Maiorano}}, \bibinfo {author} {\bibfnamefont {E.}~\bibnamefont {Marinari}},
  \ and\ \bibinfo {author} {\bibfnamefont {F.}~\bibnamefont
  {Ricci-Tersenghi}},\ }\href {\doibase 10.1103/PhysRevB.72.104411} {\bibfield
  {journal} {\bibinfo  {journal} {Phys. Rev. B}\ }\textbf {\bibinfo {volume}
  {72}},\ \bibinfo {pages} {104411} (\bibinfo {year} {2005})}\BibitemShut
  {NoStop}%
\bibitem [{\citenamefont {Jim\'enez}\ \emph {et~al.}(2005)\citenamefont
  {Jim\'enez}, \citenamefont {Mart\'{\i}n-Mayor},\ and\ \citenamefont
  {P\'erez-Gaviro}}]{jimenez:05}%
  \BibitemOpen
  \bibfield  {author} {\bibinfo {author} {\bibfnamefont {S.}~\bibnamefont
  {Jim\'enez}}, \bibinfo {author} {\bibfnamefont {V.}~\bibnamefont
  {Mart\'{\i}n-Mayor}}, \ and\ \bibinfo {author} {\bibfnamefont
  {S.}~\bibnamefont {P\'erez-Gaviro}},\ }\href {\doibase
  10.1103/PhysRevB.72.054417} {\bibfield  {journal} {\bibinfo  {journal} {Phys.
  Rev. B}\ }\textbf {\bibinfo {volume} {72}},\ \bibinfo {pages} {054417}
  (\bibinfo {year} {2005})}\BibitemShut {NoStop}%
\bibitem [{\citenamefont {Edwards}\ and\ \citenamefont
  {Anderson}(1975)}]{edwards:75}%
  \BibitemOpen
  \bibfield  {author} {\bibinfo {author} {\bibfnamefont {S.~F.}\ \bibnamefont
  {Edwards}}\ and\ \bibinfo {author} {\bibfnamefont {P.~W.}\ \bibnamefont
  {Anderson}},\ }\href {\doibase 10.1088/0305-4608/5/5/017} {\bibfield
  {journal} {\bibinfo  {journal} {Journal of Physics F: Metal Physics}\
  }\textbf {\bibinfo {volume} {5}},\ \bibinfo {pages} {965} (\bibinfo {year}
  {1975})}\BibitemShut {NoStop}%
\bibitem [{\citenamefont {Edwards}\ and\ \citenamefont
  {Anderson}(1976)}]{edwards:76}%
  \BibitemOpen
  \bibfield  {author} {\bibinfo {author} {\bibfnamefont {S.~F.}\ \bibnamefont
  {Edwards}}\ and\ \bibinfo {author} {\bibfnamefont {P.~W.}\ \bibnamefont
  {Anderson}},\ }\href {\doibase 10.1088/0305-4608/6/10/022} {\bibfield
  {journal} {\bibinfo  {journal} {J. Phys. F}\ }\textbf {\bibinfo {volume}
  {6}},\ \bibinfo {pages} {1927} (\bibinfo {year} {1976})}\BibitemShut
  {NoStop}%
\bibitem [{\citenamefont {Baity-Jesi}\ \emph {et~al.}(2014)\citenamefont
  {Baity-Jesi}, \citenamefont {Ba\~{n}os}, \citenamefont {Cruz}, \citenamefont
  {Fernandez}, \citenamefont {Gil-Narvion}, \citenamefont {Gordillo-Guerrero},
  \citenamefont {Iniguez}, \citenamefont {Maiorano}, \citenamefont {Mantovani},
  \citenamefont {Marinari}, \citenamefont {Mart\'{i}n-Mayor}, \citenamefont
  {Monforte-Garcia}, \citenamefont {Mu{\~n}oz~Sudupe}, \citenamefont {Navarro},
  \citenamefont {Parisi}, \citenamefont {Perez-Gaviro}, \citenamefont
  {Pivanti}, \citenamefont {Ricci-Tersenghi}, \citenamefont {Ruiz-Lorenzo},
  \citenamefont {Schifano}, \citenamefont {Seoane}, \citenamefont {Tarancon},
  \citenamefont {Tripiccione},\ and\ \citenamefont {Yllanes}}]{janus:14}%
  \BibitemOpen
  \bibfield  {author} {\bibinfo {author} {\bibfnamefont {M.}~\bibnamefont
  {Baity-Jesi}}, \bibinfo {author} {\bibfnamefont {R.~A.}\ \bibnamefont
  {Ba\~{n}os}}, \bibinfo {author} {\bibfnamefont {A.}~\bibnamefont {Cruz}},
  \bibinfo {author} {\bibfnamefont {L.~A.}\ \bibnamefont {Fernandez}}, \bibinfo
  {author} {\bibfnamefont {J.~M.}\ \bibnamefont {Gil-Narvion}}, \bibinfo
  {author} {\bibfnamefont {A.}~\bibnamefont {Gordillo-Guerrero}}, \bibinfo
  {author} {\bibfnamefont {D.}~\bibnamefont {Iniguez}}, \bibinfo {author}
  {\bibfnamefont {A.}~\bibnamefont {Maiorano}}, \bibinfo {author}
  {\bibfnamefont {F.}~\bibnamefont {Mantovani}}, \bibinfo {author}
  {\bibfnamefont {E.}~\bibnamefont {Marinari}}, \bibinfo {author}
  {\bibfnamefont {V.}~\bibnamefont {Mart\'{i}n-Mayor}}, \bibinfo {author}
  {\bibfnamefont {J.}~\bibnamefont {Monforte-Garcia}}, \bibinfo {author}
  {\bibfnamefont {A.}~\bibnamefont {Mu{\~n}oz~Sudupe}}, \bibinfo {author}
  {\bibfnamefont {D.}~\bibnamefont {Navarro}}, \bibinfo {author} {\bibfnamefont
  {G.}~\bibnamefont {Parisi}}, \bibinfo {author} {\bibfnamefont
  {S.}~\bibnamefont {Perez-Gaviro}}, \bibinfo {author} {\bibfnamefont
  {M.}~\bibnamefont {Pivanti}}, \bibinfo {author} {\bibfnamefont
  {F.}~\bibnamefont {Ricci-Tersenghi}}, \bibinfo {author} {\bibfnamefont
  {J.~J.}\ \bibnamefont {Ruiz-Lorenzo}}, \bibinfo {author} {\bibfnamefont
  {S.~F.}\ \bibnamefont {Schifano}}, \bibinfo {author} {\bibfnamefont
  {B.}~\bibnamefont {Seoane}}, \bibinfo {author} {\bibfnamefont
  {A.}~\bibnamefont {Tarancon}}, \bibinfo {author} {\bibfnamefont
  {R.}~\bibnamefont {Tripiccione}}, \ and\ \bibinfo {author} {\bibfnamefont
  {D.}~\bibnamefont {Yllanes}} (\bibinfo {collaboration} {Janus
  Collaboration}),\ }\href {\doibase 10.1016/j.cpc.2013.10.019} {\bibfield
  {journal} {\bibinfo  {journal} {Comp. Phys. Comm}\ }\textbf {\bibinfo
  {volume} {185}},\ \bibinfo {pages} {550} (\bibinfo {year} {2014})},\ \Eprint
  {http://arxiv.org/abs/arXiv:1310.1032} {arXiv:1310.1032} \BibitemShut
  {NoStop}%
\bibitem [{\citenamefont {Marinari}\ \emph {et~al.}(1996)\citenamefont
  {Marinari}, \citenamefont {Parisi}, \citenamefont {Ruiz-Lorenzo},\ and\
  \citenamefont {Ritort}}]{marinari:96}%
  \BibitemOpen
  \bibfield  {author} {\bibinfo {author} {\bibfnamefont {E.}~\bibnamefont
  {Marinari}}, \bibinfo {author} {\bibfnamefont {G.}~\bibnamefont {Parisi}},
  \bibinfo {author} {\bibfnamefont {J.}~\bibnamefont {Ruiz-Lorenzo}}, \ and\
  \bibinfo {author} {\bibfnamefont {F.}~\bibnamefont {Ritort}},\ }\href
  {\doibase 10.1103/PhysRevLett.76.843} {\bibfield  {journal} {\bibinfo
  {journal} {Phys. Rev. Lett.}\ }\textbf {\bibinfo {volume} {76}},\ \bibinfo
  {pages} {843} (\bibinfo {year} {1996})}\BibitemShut {NoStop}%
\bibitem [{\citenamefont {Joh}\ \emph {et~al.}(1999)\citenamefont {Joh},
  \citenamefont {Orbach}, \citenamefont {Wood}, \citenamefont {Hammann},\ and\
  \citenamefont {Vincent}}]{joh:99}%
  \BibitemOpen
  \bibfield  {author} {\bibinfo {author} {\bibfnamefont {Y.~G.}\ \bibnamefont
  {Joh}}, \bibinfo {author} {\bibfnamefont {R.}~\bibnamefont {Orbach}},
  \bibinfo {author} {\bibfnamefont {G.~G.}\ \bibnamefont {Wood}}, \bibinfo
  {author} {\bibfnamefont {J.}~\bibnamefont {Hammann}}, \ and\ \bibinfo
  {author} {\bibfnamefont {E.}~\bibnamefont {Vincent}},\ }\href {\doibase
  10.1103/PhysRevLett.82.438} {\bibfield  {journal} {\bibinfo  {journal} {Phys.
  Rev. Lett.}\ }\textbf {\bibinfo {volume} {82}},\ \bibinfo {pages} {438}
  (\bibinfo {year} {1999})}\BibitemShut {NoStop}%
\bibitem [{\citenamefont {Belletti}\ \emph {et~al.}(2008)\citenamefont
  {Belletti}, \citenamefont {Cotallo}, \citenamefont {Cruz}, \citenamefont
  {Fernandez}, \citenamefont {Gordillo-Guerrero}, \citenamefont {Guidetti},
  \citenamefont {Maiorano}, \citenamefont {Mantovani}, \citenamefont
  {Marinari}, \citenamefont {Mart\'{i}n-Mayor}, \citenamefont {Sudupe},
  \citenamefont {Navarro}, \citenamefont {Parisi}, \citenamefont
  {Perez-Gaviro}, \citenamefont {Ruiz-Lorenzo}, \citenamefont {Schifano},
  \citenamefont {Sciretti}, \citenamefont {Tarancon}, \citenamefont
  {Tripiccione}, \citenamefont {Velasco},\ and\ \citenamefont
  {Yllanes}}]{janus:08b}%
  \BibitemOpen
  \bibfield  {author} {\bibinfo {author} {\bibfnamefont {F.}~\bibnamefont
  {Belletti}}, \bibinfo {author} {\bibfnamefont {M.}~\bibnamefont {Cotallo}},
  \bibinfo {author} {\bibfnamefont {A.}~\bibnamefont {Cruz}}, \bibinfo {author}
  {\bibfnamefont {L.~A.}\ \bibnamefont {Fernandez}}, \bibinfo {author}
  {\bibfnamefont {A.}~\bibnamefont {Gordillo-Guerrero}}, \bibinfo {author}
  {\bibfnamefont {M.}~\bibnamefont {Guidetti}}, \bibinfo {author}
  {\bibfnamefont {A.}~\bibnamefont {Maiorano}}, \bibinfo {author}
  {\bibfnamefont {F.}~\bibnamefont {Mantovani}}, \bibinfo {author}
  {\bibfnamefont {E.}~\bibnamefont {Marinari}}, \bibinfo {author}
  {\bibfnamefont {V.}~\bibnamefont {Mart\'{i}n-Mayor}}, \bibinfo {author}
  {\bibfnamefont {A.~M.}\ \bibnamefont {Sudupe}}, \bibinfo {author}
  {\bibfnamefont {D.}~\bibnamefont {Navarro}}, \bibinfo {author} {\bibfnamefont
  {G.}~\bibnamefont {Parisi}}, \bibinfo {author} {\bibfnamefont
  {S.}~\bibnamefont {Perez-Gaviro}}, \bibinfo {author} {\bibfnamefont {J.~J.}\
  \bibnamefont {Ruiz-Lorenzo}}, \bibinfo {author} {\bibfnamefont {S.~F.}\
  \bibnamefont {Schifano}}, \bibinfo {author} {\bibfnamefont {D.}~\bibnamefont
  {Sciretti}}, \bibinfo {author} {\bibfnamefont {A.}~\bibnamefont {Tarancon}},
  \bibinfo {author} {\bibfnamefont {R.}~\bibnamefont {Tripiccione}}, \bibinfo
  {author} {\bibfnamefont {J.~L.}\ \bibnamefont {Velasco}}, \ and\ \bibinfo
  {author} {\bibfnamefont {D.}~\bibnamefont {Yllanes}} (\bibinfo
  {collaboration} {Janus Collaboration}),\ }\href {\doibase
  10.1103/PhysRevLett.101.157201} {\bibfield  {journal} {\bibinfo  {journal}
  {Phys. Rev. Lett.}\ }\textbf {\bibinfo {volume} {101}},\ \bibinfo {pages}
  {157201} (\bibinfo {year} {2008})},\ \Eprint
  {http://arxiv.org/abs/arXiv:0804.1471} {arXiv:0804.1471} \BibitemShut
  {NoStop}%
\bibitem [{\citenamefont {Baity-Jesi}\ \emph {et~al.}(2018)\citenamefont
  {Baity-Jesi}, \citenamefont {Calore}, \citenamefont {Cruz}, \citenamefont
  {Fernandez}, \citenamefont {Gil-Narvion}, \citenamefont {Gordillo-Guerrero},
  \citenamefont {I\~niguez}, \citenamefont {Maiorano}, \citenamefont
  {Marinari}, \citenamefont {Martin-Mayor}, \citenamefont {Moreno-Gordo},
  \citenamefont {Mu\~noz Sudupe}, \citenamefont {Navarro}, \citenamefont
  {Parisi}, \citenamefont {Perez-Gaviro}, \citenamefont {Ricci-Tersenghi},
  \citenamefont {Ruiz-Lorenzo}, \citenamefont {Schifano}, \citenamefont
  {Seoane}, \citenamefont {Tarancon}, \citenamefont {Tripiccione},\ and\
  \citenamefont {Yllanes}}]{janus:18}%
  \BibitemOpen
  \bibfield  {author} {\bibinfo {author} {\bibfnamefont {M.}~\bibnamefont
  {Baity-Jesi}}, \bibinfo {author} {\bibfnamefont {E.}~\bibnamefont {Calore}},
  \bibinfo {author} {\bibfnamefont {A.}~\bibnamefont {Cruz}}, \bibinfo {author}
  {\bibfnamefont {L.~A.}\ \bibnamefont {Fernandez}}, \bibinfo {author}
  {\bibfnamefont {J.~M.}\ \bibnamefont {Gil-Narvion}}, \bibinfo {author}
  {\bibfnamefont {A.}~\bibnamefont {Gordillo-Guerrero}}, \bibinfo {author}
  {\bibfnamefont {D.}~\bibnamefont {I\~niguez}}, \bibinfo {author}
  {\bibfnamefont {A.}~\bibnamefont {Maiorano}}, \bibinfo {author}
  {\bibfnamefont {E.}~\bibnamefont {Marinari}}, \bibinfo {author}
  {\bibfnamefont {V.}~\bibnamefont {Martin-Mayor}}, \bibinfo {author}
  {\bibfnamefont {J.}~\bibnamefont {Moreno-Gordo}}, \bibinfo {author}
  {\bibfnamefont {A.}~\bibnamefont {Mu\~noz Sudupe}}, \bibinfo {author}
  {\bibfnamefont {D.}~\bibnamefont {Navarro}}, \bibinfo {author} {\bibfnamefont
  {G.}~\bibnamefont {Parisi}}, \bibinfo {author} {\bibfnamefont
  {S.}~\bibnamefont {Perez-Gaviro}}, \bibinfo {author} {\bibfnamefont
  {F.}~\bibnamefont {Ricci-Tersenghi}}, \bibinfo {author} {\bibfnamefont
  {J.~J.}\ \bibnamefont {Ruiz-Lorenzo}}, \bibinfo {author} {\bibfnamefont
  {S.~F.}\ \bibnamefont {Schifano}}, \bibinfo {author} {\bibfnamefont
  {B.}~\bibnamefont {Seoane}}, \bibinfo {author} {\bibfnamefont
  {A.}~\bibnamefont {Tarancon}}, \bibinfo {author} {\bibfnamefont
  {R.}~\bibnamefont {Tripiccione}}, \ and\ \bibinfo {author} {\bibfnamefont
  {D.}~\bibnamefont {Yllanes}} (\bibinfo {collaboration} {Janus
  Collaboration}),\ }\href {\doibase 10.1103/PhysRevLett.120.267203} {\bibfield
   {journal} {\bibinfo  {journal} {Phys. Rev. Lett.}\ }\textbf {\bibinfo
  {volume} {120}},\ \bibinfo {pages} {267203} (\bibinfo {year}
  {2018})}\BibitemShut {NoStop}%
\bibitem [{\citenamefont {Baity-Jesi}\ \emph {et~al.}(2017)\citenamefont
  {Baity-Jesi}, \citenamefont {Calore}, \citenamefont {Cruz}, \citenamefont
  {Fernandez}, \citenamefont {Gil-Narvion}, \citenamefont {Gordillo-Guerrero},
  \citenamefont {I\~niguez}, \citenamefont {Maiorano}, \citenamefont
  {Marinari}, \citenamefont {Martin-Mayor}, \citenamefont {Monforte-Garcia},
  \citenamefont {Mu\~noz Sudupe}, \citenamefont {Navarro}, \citenamefont
  {Parisi}, \citenamefont {Perez-Gaviro}, \citenamefont {Ricci-Tersenghi},
  \citenamefont {Ruiz-Lorenzo}, \citenamefont {Schifano}, \citenamefont
  {Seoane}, \citenamefont {Tarancon}, \citenamefont {Tripiccione},\ and\
  \citenamefont {Yllanes}}]{janus:17b}%
  \BibitemOpen
  \bibfield  {author} {\bibinfo {author} {\bibfnamefont {M.}~\bibnamefont
  {Baity-Jesi}}, \bibinfo {author} {\bibfnamefont {E.}~\bibnamefont {Calore}},
  \bibinfo {author} {\bibfnamefont {A.}~\bibnamefont {Cruz}}, \bibinfo {author}
  {\bibfnamefont {L.~A.}\ \bibnamefont {Fernandez}}, \bibinfo {author}
  {\bibfnamefont {J.~M.}\ \bibnamefont {Gil-Narvion}}, \bibinfo {author}
  {\bibfnamefont {A.}~\bibnamefont {Gordillo-Guerrero}}, \bibinfo {author}
  {\bibfnamefont {D.}~\bibnamefont {I\~niguez}}, \bibinfo {author}
  {\bibfnamefont {A.}~\bibnamefont {Maiorano}}, \bibinfo {author}
  {\bibfnamefont {E.}~\bibnamefont {Marinari}}, \bibinfo {author}
  {\bibfnamefont {V.}~\bibnamefont {Martin-Mayor}}, \bibinfo {author}
  {\bibfnamefont {J.}~\bibnamefont {Monforte-Garcia}}, \bibinfo {author}
  {\bibfnamefont {A.}~\bibnamefont {Mu\~noz Sudupe}}, \bibinfo {author}
  {\bibfnamefont {D.}~\bibnamefont {Navarro}}, \bibinfo {author} {\bibfnamefont
  {G.}~\bibnamefont {Parisi}}, \bibinfo {author} {\bibfnamefont
  {S.}~\bibnamefont {Perez-Gaviro}}, \bibinfo {author} {\bibfnamefont
  {F.}~\bibnamefont {Ricci-Tersenghi}}, \bibinfo {author} {\bibfnamefont
  {J.~J.}\ \bibnamefont {Ruiz-Lorenzo}}, \bibinfo {author} {\bibfnamefont
  {S.~F.}\ \bibnamefont {Schifano}}, \bibinfo {author} {\bibfnamefont
  {B.}~\bibnamefont {Seoane}}, \bibinfo {author} {\bibfnamefont
  {A.}~\bibnamefont {Tarancon}}, \bibinfo {author} {\bibfnamefont
  {R.}~\bibnamefont {Tripiccione}}, \ and\ \bibinfo {author} {\bibfnamefont
  {D.}~\bibnamefont {Yllanes}} (\bibinfo {collaboration} {Janus
  Collaboration}),\ }\href {\doibase 10.1103/PhysRevLett.118.157202} {\bibfield
   {journal} {\bibinfo  {journal} {Phys. Rev. Lett.}\ }\textbf {\bibinfo
  {volume} {118}},\ \bibinfo {pages} {157202} (\bibinfo {year}
  {2017})}\BibitemShut {NoStop}%
\bibitem [{\citenamefont {Paga}\ \emph {et~al.}()\citenamefont {Paga},
  \citenamefont {Zhai}, \citenamefont {Baity-Jesi}, \citenamefont {Calore},
  \citenamefont {Cruz}, \citenamefont {Fernandez}, \citenamefont {Gil-Narvion},
  \citenamefont {Gonzalez-Adalid~Pemartin}, \citenamefont {Gordillo-Guerrero},
  \citenamefont {I{\~n}iguez}, \citenamefont {Maiorano}, \citenamefont
  {Marinari}, \citenamefont {Martin-Mayor}, \citenamefont {Moreno-Gordo},
  \citenamefont {Mu{\~n}oz-Sudupe}, \citenamefont {Navarro}, \citenamefont
  {Orbach}, \citenamefont {Parisi}, \citenamefont {Perez-Gaviro}, \citenamefont
  {Ricci-Tersenghi}, \citenamefont {Ruiz-Lorenzo}, \citenamefont {Schifano},
  \citenamefont {Schlagel}, \citenamefont {Seoane}, \citenamefont {Tarancon},\
  and\ \citenamefont {Yllanes}}]{zhai-janus:22}%
  \BibitemOpen
  \bibfield  {author} {\bibinfo {author} {\bibfnamefont {I.}~\bibnamefont
  {Paga}}, \bibinfo {author} {\bibfnamefont {Q.}~\bibnamefont {Zhai}}, \bibinfo
  {author} {\bibfnamefont {M.}~\bibnamefont {Baity-Jesi}}, \bibinfo {author}
  {\bibfnamefont {E.}~\bibnamefont {Calore}}, \bibinfo {author} {\bibfnamefont
  {A.}~\bibnamefont {Cruz}}, \bibinfo {author} {\bibfnamefont {L.~A.}\
  \bibnamefont {Fernandez}}, \bibinfo {author} {\bibfnamefont {J.~M.}\
  \bibnamefont {Gil-Narvion}}, \bibinfo {author} {\bibfnamefont
  {I.}~\bibnamefont {Gonzalez-Adalid~Pemartin}}, \bibinfo {author}
  {\bibfnamefont {A.}~\bibnamefont {Gordillo-Guerrero}}, \bibinfo {author}
  {\bibfnamefont {D.}~\bibnamefont {I{\~n}iguez}}, \bibinfo {author}
  {\bibfnamefont {A.}~\bibnamefont {Maiorano}}, \bibinfo {author}
  {\bibfnamefont {E.}~\bibnamefont {Marinari}}, \bibinfo {author}
  {\bibfnamefont {V.}~\bibnamefont {Martin-Mayor}}, \bibinfo {author}
  {\bibfnamefont {J.}~\bibnamefont {Moreno-Gordo}}, \bibinfo {author}
  {\bibfnamefont {A.}~\bibnamefont {Mu{\~n}oz-Sudupe}}, \bibinfo {author}
  {\bibfnamefont {D.}~\bibnamefont {Navarro}}, \bibinfo {author} {\bibfnamefont
  {R.~L.}\ \bibnamefont {Orbach}}, \bibinfo {author} {\bibfnamefont
  {G.}~\bibnamefont {Parisi}}, \bibinfo {author} {\bibfnamefont
  {S.}~\bibnamefont {Perez-Gaviro}}, \bibinfo {author} {\bibfnamefont
  {F.}~\bibnamefont {Ricci-Tersenghi}}, \bibinfo {author} {\bibfnamefont
  {J.~J.}\ \bibnamefont {Ruiz-Lorenzo}}, \bibinfo {author} {\bibfnamefont
  {S.~F.}\ \bibnamefont {Schifano}}, \bibinfo {author} {\bibfnamefont {D.~L.}\
  \bibnamefont {Schlagel}}, \bibinfo {author} {\bibfnamefont {B.}~\bibnamefont
  {Seoane}}, \bibinfo {author} {\bibfnamefont {A.}~\bibnamefont {Tarancon}}, \
  and\ \bibinfo {author} {\bibfnamefont {D.}~\bibnamefont {Yllanes}},\
  }\href@noop {} {\enquote {\bibinfo {title} {Magnetic-field symmetry breaking
  in spin glasses},}\ }\bibinfo {note} {In preparation (2022)}\BibitemShut
  {NoStop}%
\bibitem [{\citenamefont {Cugliandolo}\ and\ \citenamefont
  {Kurchan}(1999)}]{cugliandolo:98}%
  \BibitemOpen
  \bibfield  {author} {\bibinfo {author} {\bibfnamefont {L.~F.}\ \bibnamefont
  {Cugliandolo}}\ and\ \bibinfo {author} {\bibfnamefont {J.}~\bibnamefont
  {Kurchan}},\ }\href {\doibase 10.1103/PhysRevB.60.922} {\bibfield  {journal}
  {\bibinfo  {journal} {Phys. Rev. B}\ }\textbf {\bibinfo {volume} {60}},\
  \bibinfo {pages} {922} (\bibinfo {year} {1999})}\BibitemShut {NoStop}%
\bibitem [{\citenamefont {Berthier}\ and\ \citenamefont
  {Bouchaud}(2003)}]{berthier:03}%
  \BibitemOpen
  \bibfield  {author} {\bibinfo {author} {\bibfnamefont {L.}~\bibnamefont
  {Berthier}}\ and\ \bibinfo {author} {\bibfnamefont {J.-P.}\ \bibnamefont
  {Bouchaud}},\ }\href {\doibase 10.1103/PhysRevLett.90.059701} {\bibfield
  {journal} {\bibinfo  {journal} {Phys. Rev. Lett}\ }\textbf {\bibinfo {volume}
  {90}},\ \bibinfo {pages} {059701} (\bibinfo {year} {2003})}\BibitemShut
  {NoStop}%
\bibitem [{\citenamefont {McKay}\ \emph {et~al.}(1982)\citenamefont {McKay},
  \citenamefont {Berker},\ and\ \citenamefont {Kirkpatrick}}]{mckay:82}%
  \BibitemOpen
  \bibfield  {author} {\bibinfo {author} {\bibfnamefont {S.~R.}\ \bibnamefont
  {McKay}}, \bibinfo {author} {\bibfnamefont {A.~N.}\ \bibnamefont {Berker}}, \
  and\ \bibinfo {author} {\bibfnamefont {S.}~\bibnamefont {Kirkpatrick}},\
  }\href {\doibase 10.1103/PhysRevLett.48.767} {\bibfield  {journal} {\bibinfo
  {journal} {Phys. Rev. Lett.}\ }\textbf {\bibinfo {volume} {48}},\ \bibinfo
  {pages} {767} (\bibinfo {year} {1982})}\BibitemShut {NoStop}%
\bibitem [{\citenamefont {Bray}\ and\ \citenamefont {Moore}(1987)}]{bray:87b}%
  \BibitemOpen
  \bibfield  {author} {\bibinfo {author} {\bibfnamefont {A.~J.}\ \bibnamefont
  {Bray}}\ and\ \bibinfo {author} {\bibfnamefont {M.~A.}\ \bibnamefont
  {Moore}},\ }\href {\doibase 10.1103/PhysRevLett.58.57} {\bibfield  {journal}
  {\bibinfo  {journal} {Phys. Rev. Lett.}\ }\textbf {\bibinfo {volume} {58}},\
  \bibinfo {pages} {57} (\bibinfo {year} {1987})}\BibitemShut {NoStop}%
\bibitem [{\citenamefont {Kondor}(1989)}]{kondor:89}%
  \BibitemOpen
  \bibfield  {author} {\bibinfo {author} {\bibfnamefont {I.}~\bibnamefont
  {Kondor}},\ }\href {\doibase 10.1088/0305-4470/22/5/005} {\bibfield
  {journal} {\bibinfo  {journal} {J. Phys. A}\ }\textbf {\bibinfo {volume}
  {22}},\ \bibinfo {pages} {L163} (\bibinfo {year} {1989})}\BibitemShut
  {NoStop}%
\bibitem [{\citenamefont {Rizzo}\ and\ \citenamefont
  {Crisanti}(2003)}]{rizzo:03}%
  \BibitemOpen
  \bibfield  {author} {\bibinfo {author} {\bibfnamefont {T.}~\bibnamefont
  {Rizzo}}\ and\ \bibinfo {author} {\bibfnamefont {A.}~\bibnamefont
  {Crisanti}},\ }\href {\doibase 10.1103/PhysRevLett.90.137201} {\bibfield
  {journal} {\bibinfo  {journal} {Phys. Rev. Lett.}\ }\textbf {\bibinfo
  {volume} {90}},\ \bibinfo {pages} {137201} (\bibinfo {year}
  {2003})}\BibitemShut {NoStop}%
\bibitem [{\citenamefont {Parisi}\ and\ \citenamefont
  {Rizzo}(2010)}]{parisi:10}%
  \BibitemOpen
  \bibfield  {author} {\bibinfo {author} {\bibfnamefont {G.}~\bibnamefont
  {Parisi}}\ and\ \bibinfo {author} {\bibfnamefont {T.}~\bibnamefont {Rizzo}},\
  }\href {\doibase 10.1088/1751-8113/43/23/235003} {\bibfield  {journal}
  {\bibinfo  {journal} {Journal of Physics A: Mathematical and Theoretical}\
  }\textbf {\bibinfo {volume} {43}},\ \bibinfo {pages} {235003} (\bibinfo
  {year} {2010})}\BibitemShut {NoStop}%
\bibitem [{\citenamefont {Baity-Jesi}\ \emph {et~al.}(2021)\citenamefont
  {Baity-Jesi}, \citenamefont {Calore}, \citenamefont {Cruz}, \citenamefont
  {Fernandez}, \citenamefont {Gil-Narvion}, \citenamefont
  {Gonzalez-Adalid~Pemartin}, \citenamefont {Gordillo-Guerrero}, \citenamefont
  {I\~niguez}, \citenamefont {Maiorano}, \citenamefont {Marinari},
  \citenamefont {Martin-Mayor}, \citenamefont {Moreno-Gordo}, \citenamefont
  {Mu\~noz Sudupe}, \citenamefont {Navarro}, \citenamefont {Paga},
  \citenamefont {Parisi}, \citenamefont {Perez-Gaviro}, \citenamefont
  {Ricci-Tersenghi}, \citenamefont {Ruiz-Lorenzo}, \citenamefont {Schifano},
  \citenamefont {Seoane}, \citenamefont {Tarancon}, \citenamefont
  {Tripiccione},\ and\ \citenamefont {Yllanes}}]{janus:21}%
  \BibitemOpen
  \bibfield  {author} {\bibinfo {author} {\bibfnamefont {M.}~\bibnamefont
  {Baity-Jesi}}, \bibinfo {author} {\bibfnamefont {E.}~\bibnamefont {Calore}},
  \bibinfo {author} {\bibfnamefont {A.}~\bibnamefont {Cruz}}, \bibinfo {author}
  {\bibfnamefont {L.~A.}\ \bibnamefont {Fernandez}}, \bibinfo {author}
  {\bibfnamefont {J.~M.}\ \bibnamefont {Gil-Narvion}}, \bibinfo {author}
  {\bibfnamefont {I.}~\bibnamefont {Gonzalez-Adalid~Pemartin}}, \bibinfo
  {author} {\bibfnamefont {A.}~\bibnamefont {Gordillo-Guerrero}}, \bibinfo
  {author} {\bibfnamefont {D.}~\bibnamefont {I\~niguez}}, \bibinfo {author}
  {\bibfnamefont {A.}~\bibnamefont {Maiorano}}, \bibinfo {author}
  {\bibfnamefont {E.}~\bibnamefont {Marinari}}, \bibinfo {author}
  {\bibfnamefont {V.}~\bibnamefont {Martin-Mayor}}, \bibinfo {author}
  {\bibfnamefont {J.}~\bibnamefont {Moreno-Gordo}}, \bibinfo {author}
  {\bibfnamefont {A.}~\bibnamefont {Mu\~noz Sudupe}}, \bibinfo {author}
  {\bibfnamefont {D.}~\bibnamefont {Navarro}}, \bibinfo {author} {\bibfnamefont
  {I.}~\bibnamefont {Paga}}, \bibinfo {author} {\bibfnamefont {G.}~\bibnamefont
  {Parisi}}, \bibinfo {author} {\bibfnamefont {S.}~\bibnamefont
  {Perez-Gaviro}}, \bibinfo {author} {\bibfnamefont {F.}~\bibnamefont
  {Ricci-Tersenghi}}, \bibinfo {author} {\bibfnamefont {J.~J.}\ \bibnamefont
  {Ruiz-Lorenzo}}, \bibinfo {author} {\bibfnamefont {S.~F.}\ \bibnamefont
  {Schifano}}, \bibinfo {author} {\bibfnamefont {B.}~\bibnamefont {Seoane}},
  \bibinfo {author} {\bibfnamefont {A.}~\bibnamefont {Tarancon}}, \bibinfo
  {author} {\bibfnamefont {R.}~\bibnamefont {Tripiccione}}, \ and\ \bibinfo
  {author} {\bibfnamefont {D.}~\bibnamefont {Yllanes}},\ }\href {\doibase
  10.1038/s42005-021-00565-9} {\bibfield  {journal} {\bibinfo  {journal}
  {Commun. Phys.}\ }\textbf {\bibinfo {volume} {4}},\ \bibinfo {pages} {74}
  (\bibinfo {year} {2021})}\BibitemShut {NoStop}%
\bibitem [{Note1()}]{Note1}%
  \BibitemOpen
  \bibinfo {note} {We are indebted to Prof. Orbach for this
  observation.}\BibitemShut {Stop}%
\bibitem [{Note2()}]{Note2}%
  \BibitemOpen
  \bibinfo {note} {Our simulations are also described in the {\protect \bf
  Methods} section, see also Table~\ref {tab:details_NUM} for crucial
  simulation details and Ref.~\cite {paga:21} for useful computational
  tricks.}\BibitemShut {Stop}%
\bibitem [{\citenamefont {Belletti}\ \emph {et~al.}(2009)\citenamefont
  {Belletti}, \citenamefont {Cruz}, \citenamefont {Fernandez}, \citenamefont
  {Gordillo-Guerrero}, \citenamefont {Guidetti}, \citenamefont {Maiorano},
  \citenamefont {Mantovani}, \citenamefont {Marinari}, \citenamefont
  {Mart\'{i}n-Mayor}, \citenamefont {Monforte}, \citenamefont
  {Mu{\~n}oz~Sudupe}, \citenamefont {Navarro}, \citenamefont {Parisi},
  \citenamefont {Perez-Gaviro}, \citenamefont {Ruiz-Lorenzo}, \citenamefont
  {Schifano}, \citenamefont {Sciretti}, \citenamefont {Tarancon}, \citenamefont
  {Tripiccione},\ and\ \citenamefont {Yllanes}}]{janus:09b}%
  \BibitemOpen
  \bibfield  {author} {\bibinfo {author} {\bibfnamefont {F.}~\bibnamefont
  {Belletti}}, \bibinfo {author} {\bibfnamefont {A.}~\bibnamefont {Cruz}},
  \bibinfo {author} {\bibfnamefont {L.~A.}\ \bibnamefont {Fernandez}}, \bibinfo
  {author} {\bibfnamefont {A.}~\bibnamefont {Gordillo-Guerrero}}, \bibinfo
  {author} {\bibfnamefont {M.}~\bibnamefont {Guidetti}}, \bibinfo {author}
  {\bibfnamefont {A.}~\bibnamefont {Maiorano}}, \bibinfo {author}
  {\bibfnamefont {F.}~\bibnamefont {Mantovani}}, \bibinfo {author}
  {\bibfnamefont {E.}~\bibnamefont {Marinari}}, \bibinfo {author}
  {\bibfnamefont {V.}~\bibnamefont {Mart\'{i}n-Mayor}}, \bibinfo {author}
  {\bibfnamefont {J.}~\bibnamefont {Monforte}}, \bibinfo {author}
  {\bibfnamefont {A.}~\bibnamefont {Mu{\~n}oz~Sudupe}}, \bibinfo {author}
  {\bibfnamefont {D.}~\bibnamefont {Navarro}}, \bibinfo {author} {\bibfnamefont
  {G.}~\bibnamefont {Parisi}}, \bibinfo {author} {\bibfnamefont
  {S.}~\bibnamefont {Perez-Gaviro}}, \bibinfo {author} {\bibfnamefont {J.~J.}\
  \bibnamefont {Ruiz-Lorenzo}}, \bibinfo {author} {\bibfnamefont {S.~F.}\
  \bibnamefont {Schifano}}, \bibinfo {author} {\bibfnamefont {D.}~\bibnamefont
  {Sciretti}}, \bibinfo {author} {\bibfnamefont {A.}~\bibnamefont {Tarancon}},
  \bibinfo {author} {\bibfnamefont {R.}~\bibnamefont {Tripiccione}}, \ and\
  \bibinfo {author} {\bibfnamefont {D.}~\bibnamefont {Yllanes}} (\bibinfo
  {collaboration} {Janus Collaboration}),\ }\href {\doibase
  10.1007/s10955-009-9727-z} {\bibfield  {journal} {\bibinfo  {journal} {J.
  Stat. Phys.}\ }\textbf {\bibinfo {volume} {135}},\ \bibinfo {pages} {1121}
  (\bibinfo {year} {2009})}\BibitemShut {NoStop}%
\bibitem [{\citenamefont {Castillo}\ \emph {et~al.}(2002)\citenamefont
  {Castillo}, \citenamefont {Chamon}, \citenamefont {Cugliandolo},\ and\
  \citenamefont {Kennett}}]{castillo:02}%
  \BibitemOpen
  \bibfield  {author} {\bibinfo {author} {\bibfnamefont {H.~E.}\ \bibnamefont
  {Castillo}}, \bibinfo {author} {\bibfnamefont {C.}~\bibnamefont {Chamon}},
  \bibinfo {author} {\bibfnamefont {L.~F.}\ \bibnamefont {Cugliandolo}}, \ and\
  \bibinfo {author} {\bibfnamefont {M.~P.}\ \bibnamefont {Kennett}},\ }\href
  {\doibase 10.1103/PhysRevLett.88.237201} {\bibfield  {journal} {\bibinfo
  {journal} {Phys. Rev. Lett.}\ }\textbf {\bibinfo {volume} {88}},\ \bibinfo
  {pages} {237201} (\bibinfo {year} {2002})}\BibitemShut {NoStop}%
\bibitem [{\citenamefont {Jaubert}\ \emph {et~al.}(2007)\citenamefont
  {Jaubert}, \citenamefont {Chamon}, \citenamefont {Cugliandolo},\ and\
  \citenamefont {Picco}}]{jaubert:07}%
  \BibitemOpen
  \bibfield  {author} {\bibinfo {author} {\bibfnamefont {L.~C.}\ \bibnamefont
  {Jaubert}}, \bibinfo {author} {\bibfnamefont {C.}~\bibnamefont {Chamon}},
  \bibinfo {author} {\bibfnamefont {L.~F.}\ \bibnamefont {Cugliandolo}}, \ and\
  \bibinfo {author} {\bibfnamefont {M.}~\bibnamefont {Picco}},\ }\href
  {\doibase 10.1088/1742-5468/2007/05/P05001} {\bibfield  {journal} {\bibinfo
  {journal} {J. Stat. Mech.}\ }\textbf {\bibinfo {volume} {2007}},\ \bibinfo
  {pages} {P05001} (\bibinfo {year} {2007})}\BibitemShut {NoStop}%
\bibitem [{Note3()}]{Note3}%
  \BibitemOpen
  \bibinfo {note} {In fact, $\protect \mathrm {d}\protect \qopname \relax
  o{log}\protect \ensuremath {t_\protect \mathrm {w}}\protect \xspace /\protect
  \mathrm {d}\protect \qopname \relax o{log}\xi _{\protect \text {micro}}$ is
  approximately constant when $\protect \ensuremath {t_\protect \mathrm
  {w}}\protect \xspace $ varies in logarithmic scale}\BibitemShut {NoStop}%
\bibitem [{Note4()}]{Note4}%
  \BibitemOpen
  \bibinfo {note} {$T \protect \mathrm {d}\protect \qopname \relax
  o{log}\protect \ensuremath {t_\protect \mathrm {w}}\protect \xspace /\protect
  \mathrm {d}\protect \qopname \relax o{log}\xi _{\protect \text {micro}}$ is
  roughly constant when different temperatures are compared}\BibitemShut
  {NoStop}%
\bibitem [{\citenamefont {Berche}\ \emph {et~al.}(2004)\citenamefont {Berche},
  \citenamefont {Chatelain}, \citenamefont {Berche},\ and\ \citenamefont
  {Janke}}]{berche:04}%
  \BibitemOpen
  \bibfield  {author} {\bibinfo {author} {\bibfnamefont {P.-E.}\ \bibnamefont
  {Berche}}, \bibinfo {author} {\bibfnamefont {C.}~\bibnamefont {Chatelain}},
  \bibinfo {author} {\bibfnamefont {B.}~\bibnamefont {Berche}}, \ and\ \bibinfo
  {author} {\bibfnamefont {W.}~\bibnamefont {Janke}},\ }\href {\doibase
  10.1140/epjb/e2004-00141-x} {\bibfield  {journal} {\bibinfo  {journal} {Eur.
  Phys. J. B}\ }\textbf {\bibinfo {volume} {38}},\ \bibinfo {pages} {463}
  (\bibinfo {year} {2004})}\BibitemShut {NoStop}%
\bibitem [{Note5()}]{Note5}%
  \BibitemOpen
  \bibinfo {note} {Jennifer Freedberg, private communication
  (2022).}\BibitemShut {Stop}%
\bibitem [{\citenamefont {Paga}(2021)}]{paga:21}%
  \BibitemOpen
  \bibfield  {author} {\bibinfo {author} {\bibfnamefont {I.}~\bibnamefont
  {Paga}},\ }\emph {\bibinfo {title} {From glassy bulk systems to spin-glass
  films: simulations meet experiments}},\ \href@noop {} {Ph.D. thesis}
  (\bibinfo {year} {2021})\BibitemShut {NoStop}%
\end{thebibliography}%


\begin{thebibliography}{4}%
\makeatletter
\providecommand \@ifxundefined [1]{%
 \@ifx{#1\undefined}
}%
\providecommand \@ifnum [1]{%
 \ifnum #1\expandafter \@firstoftwo
 \else \expandafter \@secondoftwo
 \fi
}%
\providecommand \@ifx [1]{%
 \ifx #1\expandafter \@firstoftwo
 \else \expandafter \@secondoftwo
 \fi
}%
\providecommand \natexlab [1]{#1}%
\providecommand \enquote  [1]{``#1''}%
\providecommand \bibnamefont  [1]{#1}%
\providecommand \bibfnamefont [1]{#1}%
\providecommand \citenamefont [1]{#1}%
\providecommand \href@noop [0]{\@secondoftwo}%
\providecommand \href [0]{\begingroup \@sanitize@url \@href}%
\providecommand \@href[1]{\@@startlink{#1}\@@href}%
\providecommand \@@href[1]{\endgroup#1\@@endlink}%
\providecommand \@sanitize@url [0]{\catcode `\\12\catcode `\$12\catcode
  `\&12\catcode `\#12\catcode `\^12\catcode `\_12\catcode `\%12\relax}%
\providecommand \@@startlink[1]{}%
\providecommand \@@endlink[0]{}%
\providecommand \url  [0]{\begingroup\@sanitize@url \@url }%
\providecommand \@url [1]{\endgroup\@href {#1}{\urlprefix }}%
\providecommand \urlprefix  [0]{URL }%
\providecommand \Eprint [0]{\href }%
\providecommand \doibase [0]{http://dx.doi.org/}%
\providecommand \selectlanguage [0]{\@gobble}%
\providecommand \bibinfo  [0]{\@secondoftwo}%
\providecommand \bibfield  [0]{\@secondoftwo}%
\providecommand \translation [1]{[#1]}%
\providecommand \BibitemOpen [0]{}%
\providecommand \bibitemStop [0]{}%
\providecommand \bibitemNoStop [0]{.\EOS\space}%
\providecommand \EOS [0]{\spacefactor3000\relax}%
\providecommand \BibitemShut  [1]{\csname bibitem#1\endcsname}%
\let\auto@bib@innerbib\@empty
\bibitem [{\citenamefont {Zhai}\ \emph {et~al.}(2020)\citenamefont {Zhai},
  \citenamefont {Paga}, \citenamefont {Baity-Jesi}, \citenamefont {Calore},
  \citenamefont {Cruz}, \citenamefont {Fernandez}, \citenamefont {Gil-Narvion},
  \citenamefont {Gonzalez-Adalid~Pemartin}, \citenamefont {Gordillo-Guerrero},
  \citenamefont {I\~niguez}, \citenamefont {Maiorano}, \citenamefont
  {Marinari}, \citenamefont {Martin-Mayor}, \citenamefont {Moreno-Gordo},
  \citenamefont {Mu\~noz Sudupe}, \citenamefont {Navarro}, \citenamefont
  {Orbach}, \citenamefont {Parisi}, \citenamefont {Perez-Gaviro}, \citenamefont
  {Ricci-Tersenghi}, \citenamefont {Ruiz-Lorenzo}, \citenamefont {Schifano},
  \citenamefont {Schlagel}, \citenamefont {Seoane}, \citenamefont {Tarancon},
  \citenamefont {Tripiccione},\ and\ \citenamefont {Yllanes}}]{zhai-janus:20a}%
  \BibitemOpen
  \bibfield  {author} {\bibinfo {author} {\bibfnamefont {Q.}~\bibnamefont
  {Zhai}}, \bibinfo {author} {\bibfnamefont {I.}~\bibnamefont {Paga}}, \bibinfo
  {author} {\bibfnamefont {M.}~\bibnamefont {Baity-Jesi}}, \bibinfo {author}
  {\bibfnamefont {E.}~\bibnamefont {Calore}}, \bibinfo {author} {\bibfnamefont
  {A.}~\bibnamefont {Cruz}}, \bibinfo {author} {\bibfnamefont {L.~A.}\
  \bibnamefont {Fernandez}}, \bibinfo {author} {\bibfnamefont {J.~M.}\
  \bibnamefont {Gil-Narvion}}, \bibinfo {author} {\bibfnamefont
  {I.}~\bibnamefont {Gonzalez-Adalid~Pemartin}}, \bibinfo {author}
  {\bibfnamefont {A.}~\bibnamefont {Gordillo-Guerrero}}, \bibinfo {author}
  {\bibfnamefont {D.}~\bibnamefont {I\~niguez}}, \bibinfo {author}
  {\bibfnamefont {A.}~\bibnamefont {Maiorano}}, \bibinfo {author}
  {\bibfnamefont {E.}~\bibnamefont {Marinari}}, \bibinfo {author}
  {\bibfnamefont {V.}~\bibnamefont {Martin-Mayor}}, \bibinfo {author}
  {\bibfnamefont {J.}~\bibnamefont {Moreno-Gordo}}, \bibinfo {author}
  {\bibfnamefont {A.}~\bibnamefont {Mu\~noz Sudupe}}, \bibinfo {author}
  {\bibfnamefont {D.}~\bibnamefont {Navarro}}, \bibinfo {author} {\bibfnamefont
  {R.~L.}\ \bibnamefont {Orbach}}, \bibinfo {author} {\bibfnamefont
  {G.}~\bibnamefont {Parisi}}, \bibinfo {author} {\bibfnamefont
  {S.}~\bibnamefont {Perez-Gaviro}}, \bibinfo {author} {\bibfnamefont
  {F.}~\bibnamefont {Ricci-Tersenghi}}, \bibinfo {author} {\bibfnamefont
  {J.~J.}\ \bibnamefont {Ruiz-Lorenzo}}, \bibinfo {author} {\bibfnamefont
  {S.~F.}\ \bibnamefont {Schifano}}, \bibinfo {author} {\bibfnamefont {D.~L.}\
  \bibnamefont {Schlagel}}, \bibinfo {author} {\bibfnamefont {B.}~\bibnamefont
  {Seoane}}, \bibinfo {author} {\bibfnamefont {A.}~\bibnamefont {Tarancon}},
  \bibinfo {author} {\bibfnamefont {R.}~\bibnamefont {Tripiccione}}, \ and\
  \bibinfo {author} {\bibfnamefont {D.}~\bibnamefont {Yllanes}},\ }\href
  {\doibase 10.1103/PhysRevLett.125.237202} {\bibfield  {journal} {\bibinfo
  {journal} {Phys. Rev. Lett.}\ }\textbf {\bibinfo {volume} {125}},\ \bibinfo
  {pages} {237202} (\bibinfo {year} {2020})}\BibitemShut {NoStop}%
\bibitem [{\citenamefont {Paga}\ \emph {et~al.}(2021)\citenamefont {Paga},
  \citenamefont {Zhai}, \citenamefont {Baity-Jesi}, \citenamefont {Calore},
  \citenamefont {Cruz}, \citenamefont {Fernandez}, \citenamefont {Gil-Narvion},
  \citenamefont {Gonzalez-Adalid~Pemartin}, \citenamefont {Gordillo-Guerrero},
  \citenamefont {I{\~n}iguez}, \citenamefont {Maiorano}, \citenamefont
  {Marinari}, \citenamefont {Martin-Mayor}, \citenamefont {Moreno-Gordo},
  \citenamefont {Mu{\~n}oz-Sudupe}, \citenamefont {Navarro}, \citenamefont
  {Orbach}, \citenamefont {Parisi}, \citenamefont {Perez-Gaviro}, \citenamefont
  {Ricci-Tersenghi}, \citenamefont {Ruiz-Lorenzo}, \citenamefont {Schifano},
  \citenamefont {Schlagel}, \citenamefont {Seoane}, \citenamefont {Tarancon},
  \citenamefont {Tripiccione},\ and\ \citenamefont {Yllanes}}]{zhai-janus:21}%
  \BibitemOpen
  \bibfield  {author} {\bibinfo {author} {\bibfnamefont {I.}~\bibnamefont
  {Paga}}, \bibinfo {author} {\bibfnamefont {Q.}~\bibnamefont {Zhai}}, \bibinfo
  {author} {\bibfnamefont {M.}~\bibnamefont {Baity-Jesi}}, \bibinfo {author}
  {\bibfnamefont {E.}~\bibnamefont {Calore}}, \bibinfo {author} {\bibfnamefont
  {A.}~\bibnamefont {Cruz}}, \bibinfo {author} {\bibfnamefont {L.~A.}\
  \bibnamefont {Fernandez}}, \bibinfo {author} {\bibfnamefont {J.~M.}\
  \bibnamefont {Gil-Narvion}}, \bibinfo {author} {\bibfnamefont
  {I.}~\bibnamefont {Gonzalez-Adalid~Pemartin}}, \bibinfo {author}
  {\bibfnamefont {A.}~\bibnamefont {Gordillo-Guerrero}}, \bibinfo {author}
  {\bibfnamefont {D.}~\bibnamefont {I{\~n}iguez}}, \bibinfo {author}
  {\bibfnamefont {A.}~\bibnamefont {Maiorano}}, \bibinfo {author}
  {\bibfnamefont {E.}~\bibnamefont {Marinari}}, \bibinfo {author}
  {\bibfnamefont {V.}~\bibnamefont {Martin-Mayor}}, \bibinfo {author}
  {\bibfnamefont {J.}~\bibnamefont {Moreno-Gordo}}, \bibinfo {author}
  {\bibfnamefont {A.}~\bibnamefont {Mu{\~n}oz-Sudupe}}, \bibinfo {author}
  {\bibfnamefont {D.}~\bibnamefont {Navarro}}, \bibinfo {author} {\bibfnamefont
  {R.~L.}\ \bibnamefont {Orbach}}, \bibinfo {author} {\bibfnamefont
  {G.}~\bibnamefont {Parisi}}, \bibinfo {author} {\bibfnamefont
  {S.}~\bibnamefont {Perez-Gaviro}}, \bibinfo {author} {\bibfnamefont
  {F.}~\bibnamefont {Ricci-Tersenghi}}, \bibinfo {author} {\bibfnamefont
  {J.~J.}\ \bibnamefont {Ruiz-Lorenzo}}, \bibinfo {author} {\bibfnamefont
  {S.~F.}\ \bibnamefont {Schifano}}, \bibinfo {author} {\bibfnamefont {D.~L.}\
  \bibnamefont {Schlagel}}, \bibinfo {author} {\bibfnamefont {B.}~\bibnamefont
  {Seoane}}, \bibinfo {author} {\bibfnamefont {A.}~\bibnamefont {Tarancon}},
  \bibinfo {author} {\bibfnamefont {R.}~\bibnamefont {Tripiccione}}, \ and\
  \bibinfo {author} {\bibfnamefont {D.}~\bibnamefont {Yllanes}},\ }\href
  {\doibase 10.1088/1742-5468/abdfca} {\bibfield  {journal} {\bibinfo
  {journal} {J. Stat. Mech.}\ }\textbf {\bibinfo {volume} {2021}},\ \bibinfo
  {pages} {033301} (\bibinfo {year} {2021})}\BibitemShut {NoStop}%
\bibitem [{\citenamefont {Baity-Jesi}\ \emph {et~al.}(2013)\citenamefont
  {Baity-Jesi}, \citenamefont {Ba\~{n}os}, \citenamefont {Cruz}, \citenamefont
  {Fernandez}, \citenamefont {Gil-Narvion}, \citenamefont {Gordillo-Guerrero},
  \citenamefont {Iniguez}, \citenamefont {Maiorano}, \citenamefont {Mantovani},
  \citenamefont {Marinari}, \citenamefont {Mart\'{i}n-Mayor}, \citenamefont
  {Monforte-Garcia}, \citenamefont {Mu{\~n}oz~Sudupe}, \citenamefont {Navarro},
  \citenamefont {Parisi}, \citenamefont {Perez-Gaviro}, \citenamefont
  {Pivanti}, \citenamefont {Ricci-Tersenghi}, \citenamefont {Ruiz-Lorenzo},
  \citenamefont {Schifano}, \citenamefont {Seoane}, \citenamefont {Tarancon},
  \citenamefont {Tripiccione},\ and\ \citenamefont {Yllanes}}]{janus:13}%
  \BibitemOpen
  \bibfield  {author} {\bibinfo {author} {\bibfnamefont {M.}~\bibnamefont
  {Baity-Jesi}}, \bibinfo {author} {\bibfnamefont {R.~A.}\ \bibnamefont
  {Ba\~{n}os}}, \bibinfo {author} {\bibfnamefont {A.}~\bibnamefont {Cruz}},
  \bibinfo {author} {\bibfnamefont {L.~A.}\ \bibnamefont {Fernandez}}, \bibinfo
  {author} {\bibfnamefont {J.~M.}\ \bibnamefont {Gil-Narvion}}, \bibinfo
  {author} {\bibfnamefont {A.}~\bibnamefont {Gordillo-Guerrero}}, \bibinfo
  {author} {\bibfnamefont {D.}~\bibnamefont {Iniguez}}, \bibinfo {author}
  {\bibfnamefont {A.}~\bibnamefont {Maiorano}}, \bibinfo {author}
  {\bibfnamefont {F.}~\bibnamefont {Mantovani}}, \bibinfo {author}
  {\bibfnamefont {E.}~\bibnamefont {Marinari}}, \bibinfo {author}
  {\bibfnamefont {V.}~\bibnamefont {Mart\'{i}n-Mayor}}, \bibinfo {author}
  {\bibfnamefont {J.}~\bibnamefont {Monforte-Garcia}}, \bibinfo {author}
  {\bibfnamefont {A.}~\bibnamefont {Mu{\~n}oz~Sudupe}}, \bibinfo {author}
  {\bibfnamefont {D.}~\bibnamefont {Navarro}}, \bibinfo {author} {\bibfnamefont
  {G.}~\bibnamefont {Parisi}}, \bibinfo {author} {\bibfnamefont
  {S.}~\bibnamefont {Perez-Gaviro}}, \bibinfo {author} {\bibfnamefont
  {M.}~\bibnamefont {Pivanti}}, \bibinfo {author} {\bibfnamefont
  {F.}~\bibnamefont {Ricci-Tersenghi}}, \bibinfo {author} {\bibfnamefont
  {J.~J.}\ \bibnamefont {Ruiz-Lorenzo}}, \bibinfo {author} {\bibfnamefont
  {S.~F.}\ \bibnamefont {Schifano}}, \bibinfo {author} {\bibfnamefont
  {B.}~\bibnamefont {Seoane}}, \bibinfo {author} {\bibfnamefont
  {A.}~\bibnamefont {Tarancon}}, \bibinfo {author} {\bibfnamefont
  {R.}~\bibnamefont {Tripiccione}}, \ and\ \bibinfo {author} {\bibfnamefont
  {D.}~\bibnamefont {Yllanes}} (\bibinfo {collaboration} {Janus
  Collaboration}),\ }\href {\doibase 10.1103/PhysRevB.88.224416} {\bibfield
  {journal} {\bibinfo  {journal} {Phys. Rev. B}\ }\textbf {\bibinfo {volume}
  {88}},\ \bibinfo {pages} {224416} (\bibinfo {year} {{2013}})},\ \Eprint
  {http://arxiv.org/abs/arXiv:1310.2910} {arXiv:1310.2910} \BibitemShut
  {NoStop}%
\bibitem [{\citenamefont {Belletti}\ \emph {et~al.}(2009)\citenamefont
  {Belletti}, \citenamefont {Cruz}, \citenamefont {Fernandez}, \citenamefont
  {Gordillo-Guerrero}, \citenamefont {Guidetti}, \citenamefont {Maiorano},
  \citenamefont {Mantovani}, \citenamefont {Marinari}, \citenamefont
  {Mart\'{i}n-Mayor}, \citenamefont {Monforte}, \citenamefont
  {Mu{\~n}oz~Sudupe}, \citenamefont {Navarro}, \citenamefont {Parisi},
  \citenamefont {Perez-Gaviro}, \citenamefont {Ruiz-Lorenzo}, \citenamefont
  {Schifano}, \citenamefont {Sciretti}, \citenamefont {Tarancon}, \citenamefont
  {Tripiccione},\ and\ \citenamefont {Yllanes}}]{janus:09b}%
  \BibitemOpen
  \bibfield  {author} {\bibinfo {author} {\bibfnamefont {F.}~\bibnamefont
  {Belletti}}, \bibinfo {author} {\bibfnamefont {A.}~\bibnamefont {Cruz}},
  \bibinfo {author} {\bibfnamefont {L.~A.}\ \bibnamefont {Fernandez}}, \bibinfo
  {author} {\bibfnamefont {A.}~\bibnamefont {Gordillo-Guerrero}}, \bibinfo
  {author} {\bibfnamefont {M.}~\bibnamefont {Guidetti}}, \bibinfo {author}
  {\bibfnamefont {A.}~\bibnamefont {Maiorano}}, \bibinfo {author}
  {\bibfnamefont {F.}~\bibnamefont {Mantovani}}, \bibinfo {author}
  {\bibfnamefont {E.}~\bibnamefont {Marinari}}, \bibinfo {author}
  {\bibfnamefont {V.}~\bibnamefont {Mart\'{i}n-Mayor}}, \bibinfo {author}
  {\bibfnamefont {J.}~\bibnamefont {Monforte}}, \bibinfo {author}
  {\bibfnamefont {A.}~\bibnamefont {Mu{\~n}oz~Sudupe}}, \bibinfo {author}
  {\bibfnamefont {D.}~\bibnamefont {Navarro}}, \bibinfo {author} {\bibfnamefont
  {G.}~\bibnamefont {Parisi}}, \bibinfo {author} {\bibfnamefont
  {S.}~\bibnamefont {Perez-Gaviro}}, \bibinfo {author} {\bibfnamefont {J.~J.}\
  \bibnamefont {Ruiz-Lorenzo}}, \bibinfo {author} {\bibfnamefont {S.~F.}\
  \bibnamefont {Schifano}}, \bibinfo {author} {\bibfnamefont {D.}~\bibnamefont
  {Sciretti}}, \bibinfo {author} {\bibfnamefont {A.}~\bibnamefont {Tarancon}},
  \bibinfo {author} {\bibfnamefont {R.}~\bibnamefont {Tripiccione}}, \ and\
  \bibinfo {author} {\bibfnamefont {D.}~\bibnamefont {Yllanes}} (\bibinfo
  {collaboration} {Janus Collaboration}),\ }\href {\doibase
  10.1007/s10955-009-9727-z} {\bibfield  {journal} {\bibinfo  {journal} {J.
  Stat. Phys.}\ }\textbf {\bibinfo {volume} {135}},\ \bibinfo {pages} {1121}
  (\bibinfo {year} {2009})}\BibitemShut {NoStop}%
\end{thebibliography}%

\appendix
\section{Methods}

The layout of this note is as follows. In Sect.~\ref{Sec:Janus_details} we describe our simulations.
In Sect.~\ref{sec:zero-field-observables} we define some quantities characteristic of the ZFC protocol. In fact, the magnetic field plays a crucial role in the determination of the Zeeman length scale, as we explain in Sect.~\ref{sect:Zeeman}. The other two spin-glass coherence lengths, $\xi_{\text{micro}}$ and $\zeta$, are computed as explained in Sect.~\ref{sect:ximicro-zeta}. Finally, in Sect.~\ref{sect:chaos} we explain our computation of the chaotic correlation parameter.

\subsection{The models simulated}
\label{Sec:Janus_details}
We performed massive simulations on the Janus II supercomputer~\cite{janus:14} to study the three-dimensional Edwards-Anderson (EA) model on a cubic lattice with periodic boundary conditions and size $L=160 $ (in units of the lattice constant $a_0$). The main parameters describing our simulations are provided in Tab.~\ref{tab:details_NUM}. 

The $N=L^3$ Ising spins, $s_{\boldsymbol{x}} = \pm 1$, interact with their lattice nearest neighbors in presence of a magnetic field ($H$) through the Hamiltonian:
\begin{equation}\label{eqMethods:H}
{\cal H} = - \sum_{\langle\boldsymbol{x,y}\rangle} J_{\boldsymbol{xy}} s_{\boldsymbol{x}} s_{\boldsymbol{y}} - H \sum_{\boldsymbol{x}} s_{\boldsymbol{x}} \; ,
\end{equation}
where the couplings are independent, identically distributed random variables: $J_{\boldsymbol{xy}}= \pm 1$, with $50\%$ probability.
The couplings are chosen at simulation start, and remain fixed (\emph{quenched} disorder). A particular choice of the couplings is termed 
a \emph{sample}. In the absence of an external magnetic field $H=0$, this model undergoes a spin-glass transition at the critical temperature $\Tg=1.102(3)$ \cite{janus:13}.

The off-equilibrium dynamics was simulated with a Metropolis algorithm. The numerical time unit is the lattice sweep, which roughly corresponds to 1 ps of physical time.

In this work we have simulated $N_\mathrm{S}=4$ samples using a lattice size of  $L=160~a_0$. For each of these samples and for each protocol (Tab.~\ref{tab:details_NUM} in the main text) we have simulated $N_{\text{R}}=512$ {\emph replicas} (\emph{i.e.}, independent simulations carried out for a given sample, following an identical protocol).
We use replicas to account for the thermal noise controlling the simulation (each replica is controlled by an independent realization of the thermal noise). The average over the thermal noise will be represented as  $\langle\cdots\rangle$. Only afterwards, we shall perform the average over samples, which will be indicated as  $\overline{\langle\cdots\rangle}$. 

Some times, however, (most notably for the analysis in Sect.~\ref{sect:Zeeman})
final quantities are computed  for a single sample (this is, of course, the approach followed in the laboratory). In these cases, the different samples allow us to asses to which extent our results depend on the disorder realization, see Supplementary Note I.

Besides, as a null experiment for temperature chaos, we have
studied the link-diluted Ising model (DIM), also on cubic lattices of size $L=160\, a_0$ with periodic boundary conditions
and using Metropolis dynamics. Specifically, we used the Hamiltonian in Eq.~\eqref{eqMethods:H} but with couplings
$J_{\boldsymbol{xy}}=1$ (with $70\%$ probability) or $J_{\boldsymbol{xy}}=0$ (with $30\%$ probability) and magnetic field $H=0$. Since all couplings are positive or zero, this is a ferromagnetic system without frustration, for which no temperature chaos is expected.
The critical temperature for the DIM is $\Tc=3.0609(5)$~\cite{berche:04} (actually, this is twice the value reported in~\cite{berche:04} due to our use of an Ising, rather than Potts, formulation). 
In fact, with some abuse of language,
in the main text we refer to DIM temperatures as $T=0.9$, $T=0.7$ or $0.5$ rather than to their actual values $T=0.9\, \Tc/\Tg$, $T=0.7\, \Tc/\Tg$ or $0.5\, \Tc/\Tg$,
where $\Tg$ is the critical temperature for the EA model.
We follow the very same procedure, which is explained  in Sect.~\ref{subsect:ximicro}, to compute the coherence length $\xi_{\text{micro}}$ for both the spin glass and the DIM. We have chosen times for the DIM such that $\xi_{\text{micro}}$
coincides with the corresponding spin-glass value, namely
$\xi_{\text{micro}}=5.84$ (protocol A'$_5$ in Table~\ref{tab:configurations_chaotic_comparison} in the main text), $\xi_{\text{micro}}=10.11$ (protocol A'$_7$) and
$\xi_{\text{micro}}=16.63$ (protocol A'$_9$). Of course, the necessary times are extremely shorter for the DIM than for the spin glass. Given that DIM simulations were comparatively inexpensive, we simulated 16 samples (each with 512 replicas)  for this model.

\subsection{Some zero-field-cooled observables}
\label{sec:zero-field-observables}

As explained in the main text, our simulations are designed to mimic the experimental protocol named zero-field cooling (ZFC).
In ZFC protocols, a sample initially in equilibrium at some very high temperature is cooled below $\Tg$, always being kept at zero magnetic field.
In the native protocols, the system is abruptly taken to the measuring temperature, where it is let to relax for a time $\tw$.
The cooling process (always without a field) is more complex for our jump protocols, as depicted in Fig.~\ref{fig:thermal_protocol} in the main text.

For both protocols, native or jump, we let the system relax for
a time $\tw$ at the final, measuring temperature. Then, the external magnetic field, $H$, is switched on and we record the magnetic density 
\begin{equation}\label{eq:Mzfc_def}
M_{\mathrm{ZFC}}(t,\tw;H)={\frac {1}{N}} \sum_{\boldsymbol{x}}\, \langle s_{\boldsymbol x}(t+\tw;H)\rangle\,,
\end{equation}
which  grows with $t$ from its initial value $M=0$ at $t=0$. We also record the
two-time  autocorrelation function,
\begin{equation}\label{eq:Czfc_def}
C_{\mathrm{ZFC}}(t,\tw;H)={\frac {1}{N}}\sum_{\boldsymbol{x}}\, 
\langle s_{\boldsymbol{x}}(\tw;0)s_x(t+\tw;H)\rangle \, .
\end{equation}
Note that $C_{\text{ZFC}}$ is a monotonically decreasing function of time and 
$C_{\text{ZFC}}=1$ at $t=0$. 

\subsection{Measurement of  the Zeeman length through the scaling law of the effective times}
\label{sect:Zeeman}

The method introduced in Ref.~\cite{joh:99} to measure the spin-glass coherence length experimentally  has recently been
refined. Indeed,  the scaling law introduced in~\cite{zhai-janus:20a, zhai-janus:21} is a milestone for describing the magnetic response of a spin glass in both ``lab experiments'' and ``numerical experiments''. We shall name
$\xi_{\text{Zeeman}}$ the length scale extracted using these methods.

In experiments on a single-crystal CuMn sample, the main quantity evaluated is the relaxation function $S_\mathrm{ZFC}(t,\tw;H)$, which exhibits a local maximum at time $\teff_H \approx \tw$. Hence, one focuses on the $H$ dependence of
$\teff_H$. On the numerical side, we carry out massive numerical experiments spanning from picoseconds to tenths of a second on Janus II, from which we can also extract the $\teff_H$. The numerical method proceeds as follows (see~\cite{zhai-janus:20a, zhai-janus:21} for a full discussion). One first changes variable by considering $S_\mathrm{ZFC}$ as a function of $C(t,\tw;H)$, recall Eq.~\eqref{eq:Czfc_def}, rather than time. The peak is found at some $C_{\text{peak}}(\tw)$. Finally, $\teff_H$ is found by solving the equation $C(\teff_H,\tw;H)=C_{\text{peak}}(\tw)$. A crucial advantage is that this equation can also be solved directly at $H=0$.

The numerical $S_\mathrm{ZFC}(t,\tw;H)$, however, shows two peaks: a $\tw$-independent peak at very short times, and a second, physically interesting peak at $t \sim \tw$. Unfortunately,
in  fixed-temperature simulations (\emph{i.e.}, native protocols) with very short $\tw$, the two peaks cannot be resolved (see, for instance,
bottom-left of Fig.~\ref{fig:thermal_protocol}). We have not attempted to extract $\xi_{\text{Zeeman}}$ in native runs  where the two peaks cannot be resolved. However, for the shortest jump protocol with $T_2=0.5$, namely $\tw=2^{10},2^{15.625}$, we could borrow $C_\text{peak}$ from the jump with the largest $\tw$ (unfortunately, the same trick did not work for \emph{native} runs, because important consistency checks~\cite{zhai-janus:22} were not passed in this case).

From a phenomenological point of view, the effective time $\teff_H$ can be associated with the height of the largest free-energy barrier, $\varDelta_\mathrm{max}$, through the usual Arrhenius law \cite{joh:99}
\begin{equation}
\label{eq:Delta_max_arrhenius_law}
\varDelta_\mathrm{max}= k_\mathrm{B} T ( \log \teff_H - \log \tau_0) \, ,
\end{equation}
where $\tau_0$ is a characteristic exchange time, $\tau_0 \sim \hbar /k_B \Tg$.
In an external magnetic field, the free-energy barriers are lowered by the Zeeman energy $E_\mathrm{Z}$ \citep{joh:99}.
For small magnetic field, $E_\mathrm{Z}$ behaves as:
\begin{equation}
\label{eq:Zeema_energy_def}
E_\mathrm{Z} = \xi_\text{Zeeman}^{D-\theta/2}\chi_\mathrm{FC} H^2 \; ,
\end{equation}
which defines $\xi_\text{Zeeman}$. $\chi_\mathrm{FC}$ is the field-cooled magnetic susceptibility per spin, $\xi_\text{Zeeman}^{D-\theta/2}$ is the number of correlated spins, $D=3$ is the spatial dimension and $\theta$ is the replicon exponent \cite{janus:17b}.

We slightly depart from the previous approach by exploiting a scaling theory. We use the effective time $\teff_H$ to reflect the total free-energy change at magnetic fields $H$ and $H=0^+$ ~\citep{zhai-janus:20a,zhai-janus:21}:
\begin{align}\nonumber
\label{eq:scaling_law-Zeeman}
\log\bigg[{\frac {t_H^{\text {eff}}}{t_{H\rightarrow 0^+}^{\text {eff}}}}\bigg]=&{\frac {\hat S}{2 T}}\,\xi_\text{micro}^{D-\theta/2}H^2
+ \\
&\xi_\text{micro}^{-\theta/2}{\mathcal G}\big(T,\xi_\text{micro}^{D-\theta/2}H^2\big)\,,
\end{align}
where $\hat{S}$ is a constant coming from the fluctuation-dissipation relations and $\mathcal{G}(x)$ is a scaling function behaving as $\mathcal{G}(x)\sim x^2$ for small $x=\xi_\text{micro}^{D-\theta/2}H^2$.
For small-enough magnetic fields [$H\leq 0.017$], we can neglect the $\mathcal{O}(H^4)$ terms in Eq.~\eqref{eq:scaling_law-Zeeman}:
\begin{equation}\label{eq:fitting_f(x)}
\log\bigg[{\frac {t_H^{\text {eff}}}{t_{H\rightarrow 0^+}^{\text {eff}}}}\bigg] = c_2(\tw;T) H^2 \, ,
\end{equation}
where we have included all the constants in the  $c_2(\tw;T)$ coefficient.

Thus, fitting our data according to Eq.~\eqref{eq:fitting_f(x)}, we can define the \emph{Zeeman} coherence length $\xi_\mathrm{Zeeman}$ as
\begin{align}
\xi&_\mathrm{Zeeman}^\mathrm{jump}(\tw,T_1 \to \Tm) = \nonumber\\ &\left[ \frac{c_2(\tw,T_1 \to \Tm)}{c_2(\tw^*,\Tm)}   \right]^{1/(D-\theta/2)}
\xi_\mathrm{micro}(\tw^*;\Tm),\label{eq:xi_eff-jump_def}\\
\xi&_\mathrm{Zeeman}^\mathrm{native}(\tw,\Tm) =\nonumber\\& \left[ \frac{c_2(\tw, \Tm)}{c_2(\tw^*,\Tm)}    \right]^{1/(D-\theta/2)} \xi_\mathrm{micro}(\tw^*;\Tm) 	\, .
\label{eq:xi_eff-nat_def}
\end{align}
where $\xi_\mathrm{micro}(\tw^*;\Tm)$ plays the role of a reference length [the reference length allows us to avoid the precise determination of constants in Eq.~\eqref{eq:fitting_f(x)}]. The refence time $\tw^*$ is the longest available waiting time for our \emph{native} runs at the measuring temperature $\Tm$.
For the sake of clarity, we omit in Eqs.~\eqref{eq:xi_eff-jump_def} and~\eqref{eq:xi_eff-nat_def} the explicit dependence of $\theta$ on $\xi_\mathrm{micro}$ (which is dealt with as explained in Ref.~\cite{zhai:19}).

\subsection{\boldmath Numerical coherence lengths $\xi_\mathrm{micro}$ and $\zeta$}
\label{sect:ximicro-zeta}

In this paragraph, we shall consider two more length scales. One of them, $\xi_{\text{micro}}$, is computed from the correlation function for the spin-glass order parameter (hence, $\xi_{\text{micro}}$ tells us about the size of the glassy domains). The second length scale, $\zeta(t_1,t_2)$, tells us about how the system reorganizes itself when going from the earlier time $t_1$ to the later time $t_2$.

\subsubsection{The computation of $\xi_{\mathrm{micro}}$}\label{subsect:ximicro}

For the reader's convenience, let us recall the definition of the spatial autocorrelation function that we use for computing $\xi_\mathrm{micro}(\tw)$~\cite{janus:09b}
\begin{equation}\label{eq:Gr_def}
C_{4}({\boldsymbol r},t^\prime;T)=
\overline { \langle q^{(a,b)}(\boldsymbol{x},t^\prime) q^{(a,b)}(\boldsymbol{x+r},t^\prime) \rangle_T} \; ,
\end{equation}
\begin{equation}
q^{(a,b)}(\boldsymbol{x},t^\prime) \equiv \sigma^{(a)}(\boldsymbol{x},t^\prime) \sigma^{(b)}(\boldsymbol{x},t^\prime),
\end{equation}
where $t^\prime = \tw+t$, the indices $(a,b)$ label different real replicas and $\langle \cdots \rangle_T$ stands for the average over the thermal noise at temperature $T$.

The calculation of the correlation function is computationally costly since we have $N_{\text{R}}(N_{\text{R}}-1)/2$ possible choices of the pair of replicas. Fortunately,  it can be accelerated using the specific  multispin coding methods explained in Ref.~\cite{paga:21}. 

Once we have $C_4({\boldsymbol r},t^\prime;T)$, we compute the integrals~\cite{janus:08b,janus:09b,janus:18}:
\begin{equation}\label{eq:Ik_def}
I_k(t^\prime;T)=\int_0^\infty \mathrm{d}^3 r\,\,r^k C_4\big({\boldsymbol r}=(r,0,0), t^\prime;T\big)\,.
\end{equation}
A coherence length  can be computed as
\begin{equation}\label{eq:xi_micro_def}
\xi_{k,k+1}(t^\prime,T)={\frac {I_{k+1}(t^\prime,T)}{I_k(t^\prime,T)}}\,.
\end{equation}
We define  $\xi_\mathrm{micro}(t,\tw;H) =\xi_{12}(t,\tw;H)$. 

\subsubsection{The $\zeta$ length scale}

This length scale was studied in details in Ref.~\cite{janus:09b}
by refining earlier suggestions~\cite{castillo:02,jaubert:07}. 

Let us consider the thermal trajectory followed by a given replica at the two times $t_1 < t_2$. Our basic quantity will be the local
correlation
\begin{equation}
c_{\boldsymbol{x}}(t_1,t_2)=s_{\boldsymbol{x}}(t_2) s_{\boldsymbol{x}}(t_1)\,.
\end{equation}
Note that $c_{\boldsymbol{x}}(t_1,t_2)=-1$ if the spin at site
$\boldsymbol{x}$ has been flipped when going from time $t_1$ to
time $t_2$ [otherwise, $c_{\boldsymbol{x}}(t_1,t_2)=1$]. Then, the
two-time, two-site correlation function is
\begin{eqnarray}
\label{eq:C22_def}
C_{2+2}(\boldsymbol{r},t_1,t_2) &=&  \frac{1}{N}\sum_{\boldsymbol{x}} \overline{ [ \langle c_{\boldsymbol{x}}(t_1,t_2) c_{\boldsymbol{x}+\boldsymbol{r}}(t_1,t_2)\rangle}\nonumber\\ &-& \overline{C^2(t_1,t_2)]} \,,
\end{eqnarray}
where 
\begin{equation}
C(t_1,t_2)=\frac{1}{N}\sum_{\boldsymbol{x}}\,\langle c_{\boldsymbol{x}}(t_1,t_2)\rangle\,. 
\end{equation}

The \emph{ideal} $\zeta(t_1,t_2)$ is defined from the long-distance decay of $C_{2+2}(\boldsymbol{r}, t,\tw)$:
\begin{equation}
\label{eq:zeta_def}
C_{2+2}(\boldsymbol{r},t_1,t_2) \sim \frac{1}{r^b} \, g( r/ \zeta(t_1,t_2) ) \; ,
\end{equation}
where $g$ is an unknown scaling function. We bypass
our lack of knowledge of $g$ exactly as we solved this problem for $\xi_\mathrm{micro}$:
by using integral estimators, recall Eq.~\eqref{eq:xi_micro_def}.
Note that, by construction, $\zeta(t_1,t_2)$ tends to zero when $t_2$ approaches $t_1$. Conversely, we expect $\zeta(t_1,t_2)$ to grow with the later time $t_2$.

As for the interpretation of the length scale $\zeta$, an analogy with the theory of liquids is of help. We name a \emph{defect} a site where $c_{\boldsymbol{x}}(t_1,t_2)=-1$. Let $n(t_1,t_2)$ be the density of defects [$C(t_1,t_2)=1-2n(t_1,t_2)$] and let $g(\boldsymbol{r})$ be the pair-correlation function for defects: The conditional probability for having a defect at site 
$\boldsymbol{x}+\boldsymbol{r}$, given that a defect is present at site $\boldsymbol{x}$, is $n(t_1,t_2)g(\boldsymbol{r})$ (so that, at long distances, $g(\boldsymbol{r})$ tends to one). Given these definitions, one easily finds that 
\begin{equation}
C_{2+2}(\boldsymbol{r},t_1,t_2) =4\,\overline{n^2(t_1,t_2)\, [g(\boldsymbol{r})-1]}\,.
\end{equation}
In other words, $\zeta$ is the length scale on which defects are correlated. 
Only when $\zeta(t_2,t_1)\approx\xi_{\text{micro}}(t_1)$ does the configuration at time $t_2$ start to differ \emph{structurally}  from the configuration at the earlier time $t_1$.

Finally, let us mention that a length analogous to $\zeta(t_1,t_2)$ can be obtained with the analysis tools of temperature chaos, see Supplementary Note IV.

\subsection{Computation of the chaotic parameter}
\label{sect:chaos}

As we explained in the main text, our goal here is to introduce a correlation parameter that will allow us to compare two different thermal protocols. This comparison should necessarily be local in space. We adapt to that end the procedure introduced in Ref.~\cite{janus:21}.

Specifically, we select $N_{\mathrm{sph}}=8000$ spheres of radius $R$ randomly chosen inside the system and centered at the central points of the elementary cells of the cubic lattice. Now, let us consider two identical systems that are subjected to two different thermal protocols, which we may name protocols $A_1$ and $A_2$. Next one performs a set of independent simulations (\emph{i.e.}, \emph{replicas}) for protocol $A_1$, and another set of independent simulations for protocol $A_2$. Then, the correlation coefficient for protocols $A_1$ and $A_1$ as computed on the $k$-th sphere of radius $R$ is defined as
\begin{equation} 
X^{k,R}_{A_1,A_2} = \dfrac{\langle [q_{A_1,A_2}^{k,R}]^2\rangle_T}{\sqrt{\langle[q_{A_1,A_1}^{k,R}]^2\rangle_T \,\langle[q_{A_2,A_2}^{k,R}]^2\rangle_T}} \>\> . \label{eq:def_chaotic_parameter} 
\end{equation} 
In the above expression, $q_{A_ 1,A_ 2}^{k,R}$ is the overlap between two replicas $\sigma$ and $\tau$ that have undergone thermal protocols $A_ 1$ and $A_ 2$ respectively
\begin{equation}\label{eq:sphere_overlap}
  q_{A_1,A_2}^{k,R} = \dfrac{1}{N_r} \sum_{\mathbf{x}\in B_R^k} s_{\mathbf{x}}^{\sigma,A_ 1} s_{\mathbf{x}}^{\tau,A_2} \>\> ,
\end{equation}
where $N_R$ is the number of spins within the $k$-th sphere $B_R^k$ of radius $R$.

The interpretation of the chaotic parameter is very similar to a correlation coefficient: if $X^{k,R}_{A_ 1,A_ 2}=1$, spin configurations from thermal protocols $A_1$ and $A_2$ are completely indistinguishable inside the sphere $B_ R^k$ (absence of chaos). Instead,  $X^{k,R}_{A_1,A_2}=0$ corresponds to completely different configurations, which is an extremely chaotic situation.

The reader may notice from Eq.~\eqref{eq:def_chaotic_parameter} that the computation of $X^{k,R}_{A_1,A_2}$ involves an exact thermal expectation value (which could be obtained in simulations only if one had simulated an infinite number of replicas).
Unfortunately, we only have $\NR^{\max}=512$ replicas at our disposal. Our choice has been to produce different estimates of $X^{k,R}_{A_1,A_2,\NR}$ by varying $\NR$. Specifically, our procedure has been the following:
\begin{enumerate}
\item For each $\NR < \NR^{\max}$ we randomly order the $\NR^{\max}$ replicas and divide them in $\NR^{\max}/\NR$ groups of $\NR$ replicas.
\item In this way, we get $\NR^{\max}/\NR$ independent estimates of $X^{k,R}_{A_1,A_2,\NR}$.
\item In order to erase the effect of the initial permutation of the $\NR^{\max}$ replicas, we repeat this procedure $10$ times for all $\NR < \NR^{\max}$.
\end{enumerate}

In a nutshell, for every sphere of radius $R$ we obtain $\NTherm(\NR)$ estimates of $X^{k,r}_{A_1,A_2,\NR}$ where
\begin{equation}
\NTherm(\NR<\NR^{\max}) = 10 \times \dfrac{\NR^{\max}}{\NR} \, ,
\end{equation}
or
\begin{equation}
\NTherm(\NR=\NR^{\max}) = 1 \, .
\end{equation}

We average the $\NTherm$ estimates of $X^{k,R}_{A_1,A_2,\NR}$ for every $\NR$ and finally, in a complete analogy with Ref.~\cite{janus:21}, we compute the extrapolation of the chaotic parameter to an infinite number of replicas by means of a simple linear extrapolation
\begin{equation}
X^{k,R}_{A_1,A_2,\NR} = X^{k,R}_{A_1,A_2,\infty} + \dfrac{A^{k,R}_{A_1,A_2}}{\NR} \>\>\> , \label{eq:extrapolacion_lineal}
\end{equation}
where$X^{k,R}_{A_1,A_2,\infty}$ is our best  estimation of $X^{k,R}_{A_1,A_2}$. More complicated extrapolations do not seem to present advantages (see SI in~\cite{janus:21}).

Finally, in order to explore the statistical information carried by the $N_{\mathrm{sph}}=8000$ spheres, we define the distribution function 
\begin{equation}\label{eq:F-def}
  F(\tilde X,A_1,A_2,R)=\text{Probability}[X^{k,R}_{A_1,A_2}<\tilde X]\, .
\end{equation}
Some examples of this distribution function are displayed in Fig.~\ref{fig:null_exp} in the main text.

\begin{acknowledgments}
We acknowledge the precious contributions and ideas of our dear late friend and collaborator Raffaele Tripiccione.
The Janus project would have been impossible without Lele's technical expertise, good sense and kindness,
and we dedicate this work to him.

We thank Prof. R. Orbach for discussions.

This work was partly supported by grants No.~PID2020-112936GB-I00, PID2019-103939RB-I00, 
No.~PGC2018-094684-B-C21 and No.~PGC2018-094684-B-C22  funded by Ministerio de Economía y
Competitividad, Agencia Estatal de Investigaci\'on and Fondo Europeo
de Desarrollo Regional (FEDER) (Spain and European Union), by grants
No.~GR21014 and No.~IB20079 (partially funded by FEDER) funded by
Junta the Extremadura (Spain), and by the Atracción de Talento program
(Ref.~2019-T1/TIC-12776) funded by Comunidad de Madrid and Universidad
Complutense de Madrid (Spain). This project has received funding from  the European Research
Council under the European Union's Horizon 2020 research and innovation programme (grant No. 694925,G. Parisi).  IGAP was supported by MCIU (Spain) through FPU Grant No. FPU18/02665. JMG was supported by the Ministerio de Universidades and the European Union ``NextGeneration EU/PRTR'' through 2021-2023 Margarita Salas grant.
\end{acknowledgments}

\subsection*{Author contributions}
D.I. and  A.T. contributed to the design of the  Janus II project. J.M.G.-N. and D.N. contributed Janus~II/Janus  simulation software.  M.B.-J., E.C., A.C., L.A.F, J.M.G.-N., I.G.-A.P., A.G.-G., D.I., A.M., A.M.-S., I.P., S.P.-G., S.F.S. and  A.T. contributed to Janus~II hardware and software development. L.A.F., E.M., V.M.-M. and  I.P. suggested undertaking this project. L.A.F.,  E.M., V.M.-M., I.-P., F.R.-T. and J.J. R.-L. designed the research. J.M.-G and I.P. analyzed the data. M.B.-J., L.A.F., E.M., V.M.-M., J.M.-G., I.P., G.P., B.S., J.J.R.-L., F.R.-T. and D.Y. discussed the results. L.A.F., E.M.,  V.M.-M., J.M.-G., I.P., J.J.R-L., B.S., F.R.-T. and D.Y. wrote the paper.

\subsection*{Data availability}
The data contained in the figures of this paper, accompanied by the gnuplot script files that generate these figures, are publicly available a  \href{https://github.com/janusII/Rejuvenation_memory.git}{https://github.com/janusII/Rejuvenation\_memory.git}.
The data that support the findings of this study are available from the corresponding author upon reasonable request.
\subsection*{Code availability}
The codes that support the findings of this study are
available from the corresponding author upon reasonable
request
\bibliographystyle{apsrev4-1}

\end{document}


\title{Supplementary Information. Memory and rejuvenation in spin glasses:\\ aging systems are ruled by more than one length-scale}

\author{M.~Baity-Jesi}\affiliation{Eawag, Überlandstrasse 133, CH-8600 Dübendorf, Switzerland}

\author{E.~Calore}\affiliation{Dipartimento di Fisica e Scienze della
  Terra, Universit\`a di Ferrara and INFN, Sezione di Ferrara, I-44122
  Ferrara, Italy}

\author{A.~Cruz}\affiliation{Departamento de F\'\i{}sica Te\'orica,
  Universidad de Zaragoza, 50009 Zaragoza,
  Spain}\affiliation{Instituto de Biocomputaci\'on y F\'{\i}sica de
  Sistemas Complejos (BIFI), 50018 Zaragoza, Spain}

\author{L.A.~Fernandez}\affiliation{Departamento de F\'\i{}sica
  Te\'orica, Universidad Complutense, 28040 Madrid,
  Spain}\affiliation{Instituto de Biocomputaci\'on y F\'{\i}sica de
  Sistemas Complejos (BIFI), 50018 Zaragoza, Spain}

\author{J.M.~Gil-Narvion}\affiliation{Instituto de Biocomputaci\'on y
  F\'{\i}sica de Sistemas Complejos (BIFI), 50018 Zaragoza, Spain}

\author{I.~Gonzalez-Adalid Pemartin}\affiliation{Departamento  de F\'\i{}sica Te\'orica, Universidad Complutense, 28040 Madrid, Spain}

\author{A.~Gordillo-Guerrero}\affiliation{Departamento de
  Ingenier\'{\i}a El\'ectrica, Electr\'onica y Autom\'atica, U. de
  Extremadura, 10003, C\'aceres, Spain}\affiliation{Instituto de
  Computaci\'on Cient\'{\i}fica Avanzada (ICCAEx), Universidad de
  Extremadura, 06006 Badajoz, Spain}\affiliation{Instituto de
  Biocomputaci\'on y F\'{\i}sica de Sistemas Complejos (BIFI), 50018
  Zaragoza, Spain}

\author{D.~I\~niguez}\affiliation{Instituto de Biocomputaci\'on y
  F\'{\i}sica de Sistemas Complejos (BIFI), 50018 Zaragoza,
  Spain}\affiliation{Fundaci\'on ARAID, Diputaci\'on General de
  Arag\'on, 50018 Zaragoza, Spain}

\author{A.~Maiorano}\affiliation{Dipartimento di Biotecnologie, Chimica e
  Farmacia, Universit\`a degli studi di Siena, 53100 Siena,
  Italy and INFN, Sezione di Roma 1, 00185 Rome,
  Italy}\affiliation{Instituto de Biocomputaci\'on y F\'{\i}sica de Sistemas
Complejos (BIFI), 50018 Zaragoza, Spain}

\author{E.~Marinari}\affiliation{Dipartimento di Fisica, Sapienza
  Universit\`a di Roma, and CNR-Nanotec, Rome unit and
  INFN, Sezione di Roma 1, 00185 Rome, Italy}

\author{V.~Martin-Mayor}\affiliation{Departamento de F\'\i{}sica
  Te\'orica, Universidad Complutense, 28040 Madrid,
  Spain}\affiliation{Instituto de Biocomputaci\'on y F\'{\i}sica de
  Sistemas Complejos (BIFI), 50018 Zaragoza, Spain}

\author{J.~Moreno-Gordo}\affiliation{Instituto de Biocomputaci\'on y
  F\'{\i}sica de Sistemas Complejos (BIFI), 50018 Zaragoza,
  Spain}\affiliation{Departamento de F\'\i{}sica Te\'orica,
  Universidad de Zaragoza, 50009 Zaragoza, Spain}\affiliation{Departamento de F\'{\i}sica,
  Universidad de Extremadura, 06006 Badajoz,
  Spain}\affiliation{Instituto de Computaci\'on Cient\'{\i}fica
  Avanzada (ICCAEx), Universidad de Extremadura, 06006 Badajoz,
  Spain}

\author{A.~Mu\~noz~Sudupe}\affiliation{Departamento de F\'\i{}sica
  Te\'orica, Universidad Complutense, 28040 Madrid,
  Spain}\affiliation{Instituto de Biocomputaci\'on y F\'{\i}sica de
  Sistemas Complejos (BIFI), 50018 Zaragoza, Spain}

\author{D.~Navarro}\affiliation{Departamento de Ingenier\'{\i}a,
  Electr\'onica y Comunicaciones and I3A, U. de Zaragoza, 50018
  Zaragoza, Spain}

\author{I.~Paga}\email{ilaria.paga@gmail.com}\affiliation{Dipartimento di Fisica, Sapienza
  Universit\`a di Roma, and CNR-Nanotec, Rome unit, Sezione di Roma 1, 00185 Rome, Italy}

\author{G.~Parisi}\affiliation{Dipartimento di Fisica, Sapienza
  Universit\`a di Roma, and CNR-Nanotec, Rome unit and
  INFN, Sezione di Roma 1, 00185 Rome, Italy}

\author{S.~Perez-Gaviro}\affiliation{Departamento de
  F\'\i{}sica Te\'orica, Universidad de Zaragoza, 50009 Zaragoza, Spain}\affiliation{Instituto de Biocomputaci\'on y F\'{\i}sica de Sistemas
  Complejos (BIFI), 50018 Zaragoza, Spain}

\author{F.~Ricci-Tersenghi}\affiliation{Dipartimento di Fisica, Sapienza
  Universit\`a di Roma, and CNR-Nanotec, Rome unit and
  INFN, Sezione di Roma 1, 00185 Rome, Italy}

\author{J.J.~Ruiz-Lorenzo}\affiliation{Departamento de F\'{\i}sica,
  Universidad de Extremadura, 06006 Badajoz,
  Spain}\affiliation{Instituto de Computaci\'on Cient\'{\i}fica
  Avanzada (ICCAEx), Universidad de Extremadura, 06006 Badajoz,
  Spain}\affiliation{Instituto de Biocomputaci\'on y F\'{\i}sica de
  Sistemas Complejos (BIFI), 50018 Zaragoza, Spain}

\author{S.F.~Schifano}\affiliation{Dipartimento di Scienze Chimiche e Farmaceutiche, Universit\`a di Ferrara e INFN  Sezione di Ferrara, I-44122 Ferrara, Italy}

\author{B.~Seoane}\affiliation{Departamento de F\'\i{}sica
  Te\'orica, Universidad Complutense, 28040 Madrid,
  Spain}\affiliation{Instituto de Biocomputaci\'on y F\'{\i}sica de
  Sistemas Complejos (BIFI), 50018 Zaragoza, Spain}

\author{A.~Tarancon}\affiliation{Departamento de F\'\i{}sica
  Te\'orica, Universidad de Zaragoza, 50009 Zaragoza,
  Spain}\affiliation{Instituto de Biocomputaci\'on y F\'{\i}sica de
  Sistemas Complejos (BIFI), 50018 Zaragoza, Spain}

\author{D.~Yllanes}\affiliation{Chan Zuckerberg Biohub, San Francisco, CA, 94158}
\affiliation{Instituto de Biocomputaci\'on y F\'{\i}sica de
  Sistemas Complejos (BIFI), 50018 Zaragoza, Spain}

\collaboration{Janus Collaboration}

\date{\today}   

\maketitle

\section{The magnetic response is sample independent}

The main quantities used in our \emph{numerical experiments} are the relaxation function, $S_\mathrm{ZFC}(t,\tw;H)$ and the time at which it peaks, the so-called effective time $\teff_H \propto \tw$.
The effective time can be treated as the \emph{bridge} to connect the macroscopic feature of the system to the microscopic one, see \textbf{Methods}. In experiments, all these quantities are computed for a given disorder, and this is the approach we follow as well. However, one may wonder about the variability caused by the particular disorder realization one is looking at. This is the question that we address in this Supplementary Note. 

\begin{figure}[h!]
\centering
\includegraphics[width=\columnwidth]{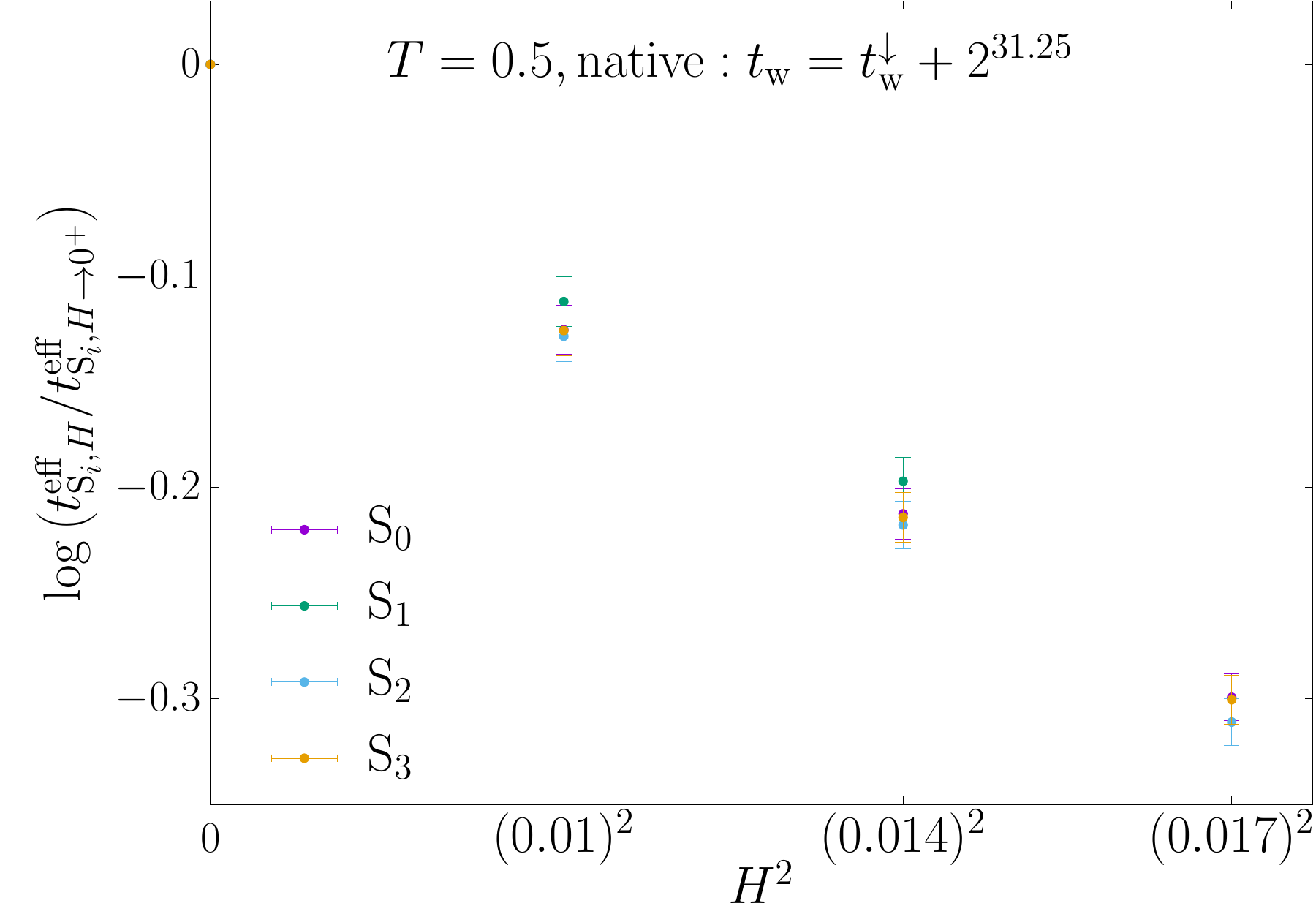}
\caption{Logarithm of the ratio of effective times $\log (\teff_{\Si,H} /\teff_{\Si,H \to 0^+} )$ versus squared magnetic field, as computed for our four independent samples $\mathrm{S}_1, \mathrm{S}_2, \mathrm{S}_3$ and $\mathrm{S}_4$. The system undergoes a Zero-Field Cooled protocol  (i.e. a native protocol) at temperature $T=0.5$ and waiting time $\tw=\tw^\downarrow + 2^{31.25}$ (recall that $\tw\!=\!2^{31.25}$). For each sample, we construct the relaxation function, $S_\mathrm{ZFC}^{\Si}(t,\tw;H)$, and extract from it  the $\teff_{\Si,H}$ values. Errors are computed with a jackknife method applied to the $512$ replicas that we simulate for every sample.}
\label{fig:ratio_log_teff_more_samples}
\end{figure}

As a particularly relevant example, let us consider the position of the peak of the relaxation function for a given magnetic field $H$, namely $\teff_H$. We have simulated four independent samples, i.e. disorder realizations, that we name $\Si$ for $i=1,2,3$ and $4$. For each sample, we extracted the effective time $\teff_{\Si,H}$ following the techniques introduced in Refs.~\citep{zhai-janus:20a,zhai-janus:21}. We specialize to the relevant case of the $H$-dependence of decay of the ratio $\log (\teff_{\Si,H} /\teff_{\Si,H \to 0^+} )$ for $i =1,..,N_\mathrm{S}$  as computed for the native protocols at our coldest temperature ($T=0.5$) and $\tw=\tw^\downarrow + 2^{31.25}$. As the reader can check in Fig.~\ref{fig:ratio_log_teff_more_samples}, the data for the different samples turned out to be statistically compatible.

\section{The zero-field cooling (ZFC) numerical experiment with a jump to $T_2=0.7$}

\begin{figure*}[t]
 \centering
 \includegraphics[width=0.86\textwidth]{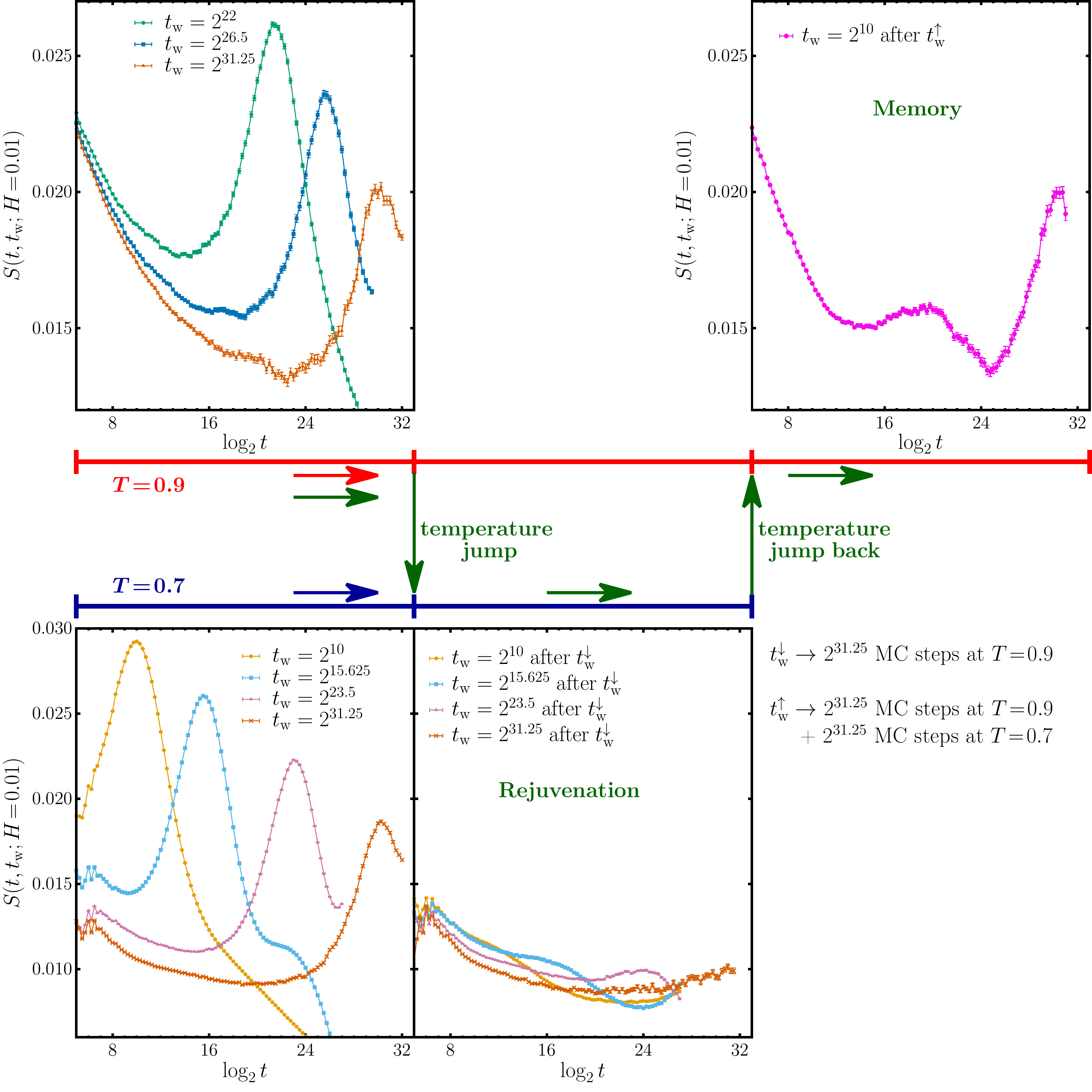}
  \caption{\textbf{The zero-field cooling (ZFC) numerical experiment at the cold temperature $T_2=0.7$.} (This figure is the analog of Figure 1---main text, but the temperature jump is $T_1=0.9\longrightarrow T_2=0.7$). The initial random spin configuration is placed instantaneously at the working temperature and let relax for a time $\tw$ without field.  At time $\tw$, a magnetic field $H=0.01$ is applied and the magnetic density, $M_\mathrm{ZFC}(t,\tw;H)$, is recorded.  \textbf{Left panels} show the relaxation function $S_\mathrm{ZFC}(t,\tw;H)$,  for the \emph{native} runs at the hotter, $T_1=0.9$, and colder, $T_2=0.7$, temperatures (both below the glass temperature $\Tg=1.102(3)$\citep{janus:13}). The physically interesting peak of the
  $S_\mathrm{ZFC}(t,\tw;H)$ is the one at time $\teff_H \simeq \tw$ .
  In our protocol (schematized by the green arrows), after a waiting time $\tw^\downarrow=2^{31.25}$, the temperature is suddenly dropped from the initial temperature $T_1=0.9$ to the colder temperature $T_2=0.7$. Then, the system is let relax at $T_2$ for an additional time, after which the magnetic field is switched on and the function $S_\mathrm{ZFC}(t,\tw;H)$ shown in the \textbf{central-bottom panel} is measured. Waiting times for these \emph{jump} runs are reported in the legend. As we have explained in the \textbf{Main}, the cold temperature $T_2=0.7$ does not satisfy the \textbf{chaos' requirement}, and, so, the \emph{rejuvenation} effect is weak  (see Eq. (2) of the \textbf{Main}).
  In all cases, error bars are one standard deviation.}
  \label{fig:thermal_protocol_07}
\end{figure*}
\twocolumngrid

In Figure  1---main text, we have illustrated the thermal protocol  in our \emph{numerical experiments} for the cold temperature $T_2=0.5$. Now, the reader may rightly wonder about how the relaxation function looks like when the temperature jump does not meet the temperature-chaos requirement expressed by Eq. (2)---main text. The answer to this question can be found in Fig. \ref{fig:thermal_protocol_07}, where the same thermal protocol was considered, but we simulated a more modest temperature jump from $T_1=0.9$ to $T_2=0.7$, which does not meet the chaos' requirement. 

In agreement with the results in the main text,  rejuvenation in the jump to $T_2=0.7$ turns out to be only partial: even for the shortest $\tw$, the aging peak is visible in the form of a shoulder when jumping to $T_2=0.7$, see Fig. \ref{fig:thermal_protocol_07}--central panel. Instead, the aging peak peaks is absent in the native  protocol (not even a shoulder is visible for $\tw=2^{10}$ in Fig. \ref{fig:thermal_protocol_07}--left--bottom).

\section{On the growth of the coherence length, $\zeta(t_1,t_2)$}

As explained in the main text and in \textbf{Methods}, the length scale $\zeta(t_1,t_2)$ allows us to compare quantitatively the spin configurations at the two times $t_1 < t_2$. Indeed, $\zeta(t_1,t_2)$ is the length scale at which significant changes occurred. It is clear, then, that $\zeta$ is very important to rationalize the memory effect found in our \emph{numerical} experiments. However, the time growth of $\zeta$ may be intricate.
In this note we illustrate some features of this growth for the two main protocols considered in this work, namely the native protocol and the
temperature jump protocol.

In fact, for jump protocols only and in shorter length scales,
the growth of $\zeta(t_1\!\!=\!\!\tw,t_2\!\!=\!\!\tw+t)$ was studied in Ref.~\cite{janus:09b}. For $t\ll\tw$, the time growth of $\zeta$ turned out to be power law, with the same exponent found for $\xi_{\text{micro}}$. Yet, when $t$ approached $\tw$ it was found that $\zeta$ saturated at a value approximately equal to $\xi_{\text{micro}}(\tw)$ [given the way we compute it,
we expect that $\zeta(t_1,t_2)$ will be upper-bounded by the smallest of $\xi_{\text{micro}}(t_1)$ and $\xi_{\text{micro}}(t_2)$]. Our aim here will be twofold. First, we shall check for jump protocols whether or not these findings to the larger length scales reached in this work. Our second goal will be presenting, for the first time we believe, results on $\zeta$ as computed for temperature-jump protocols.

\begin{figure}[h]
\centering
\includegraphics[width=\columnwidth]{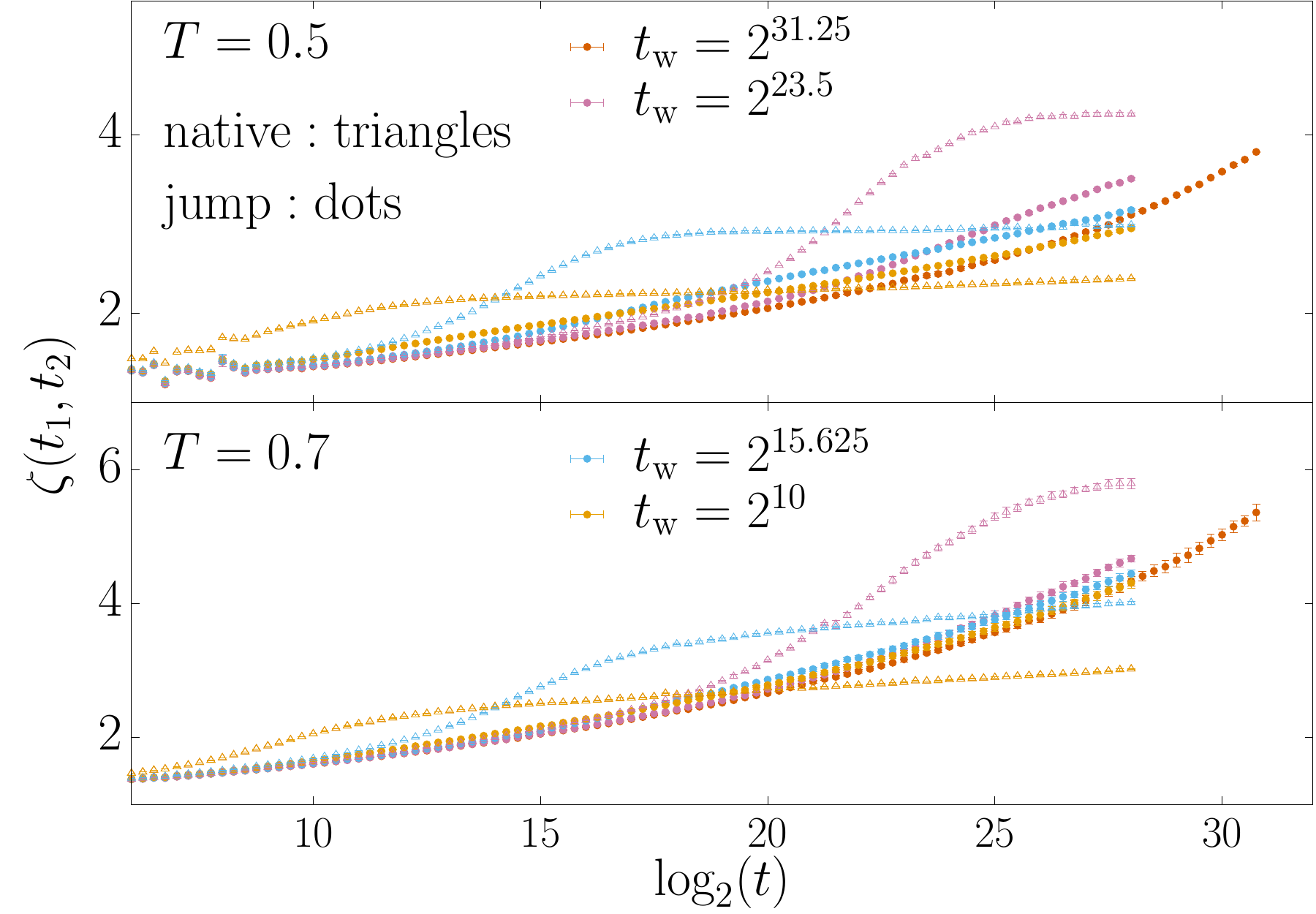}
\caption{Growth of the length scale, $\zeta(t_1,t_2)$, in absence of an external magnetic field, for native protocols (triangles) and for temperature jumps (dots). In both cases, we present results for several waiting times $\tw$. For native protocols, a fully disordered system (i.e. infinite temperature) is placed at the initial time  working temperature $T$ and let to evolve there ($t_1=\tw$ and $t_2=t+\tw$ for this protocols). Instead, in temperature jumps a fully disordered system is suddenly placed at $T=0.9$ where it is let to relax for a time $\tw^\downarrow$. Next,
the temperature is instantaneously lowered to $T$. For jump protocols, $t_1=\tw^\downarrow+\tw$ and $t_2=t_1+t$.
\textbf{On the top}, data obtained for the measuring temperature $T=0.5$; \textbf{on the bottom}, data for  at the hot temperature $T=0.7$.  Errors are obtained through Jack-knife propagation of the sample-to-sample fluctuations.}
\label{fig:growth_zeta}
\end{figure}

Indeed, the two thermal protocols are compared on figure~\ref{fig:growth_zeta}. We find rather different time-behaviours for both kinds of thermal histories:
\begin{itemize}
\item For native protocols, the curves for the different
$\tw$ start to significantly differ for time $t\sim\tw$, when
a saturation value is approached.
\item For our jump protocols instead, the curves corresponding
to the different waiting time are almost statistically compatible.
\end{itemize}

\begin{figure}[h!]
\centering
\includegraphics[width=\columnwidth]{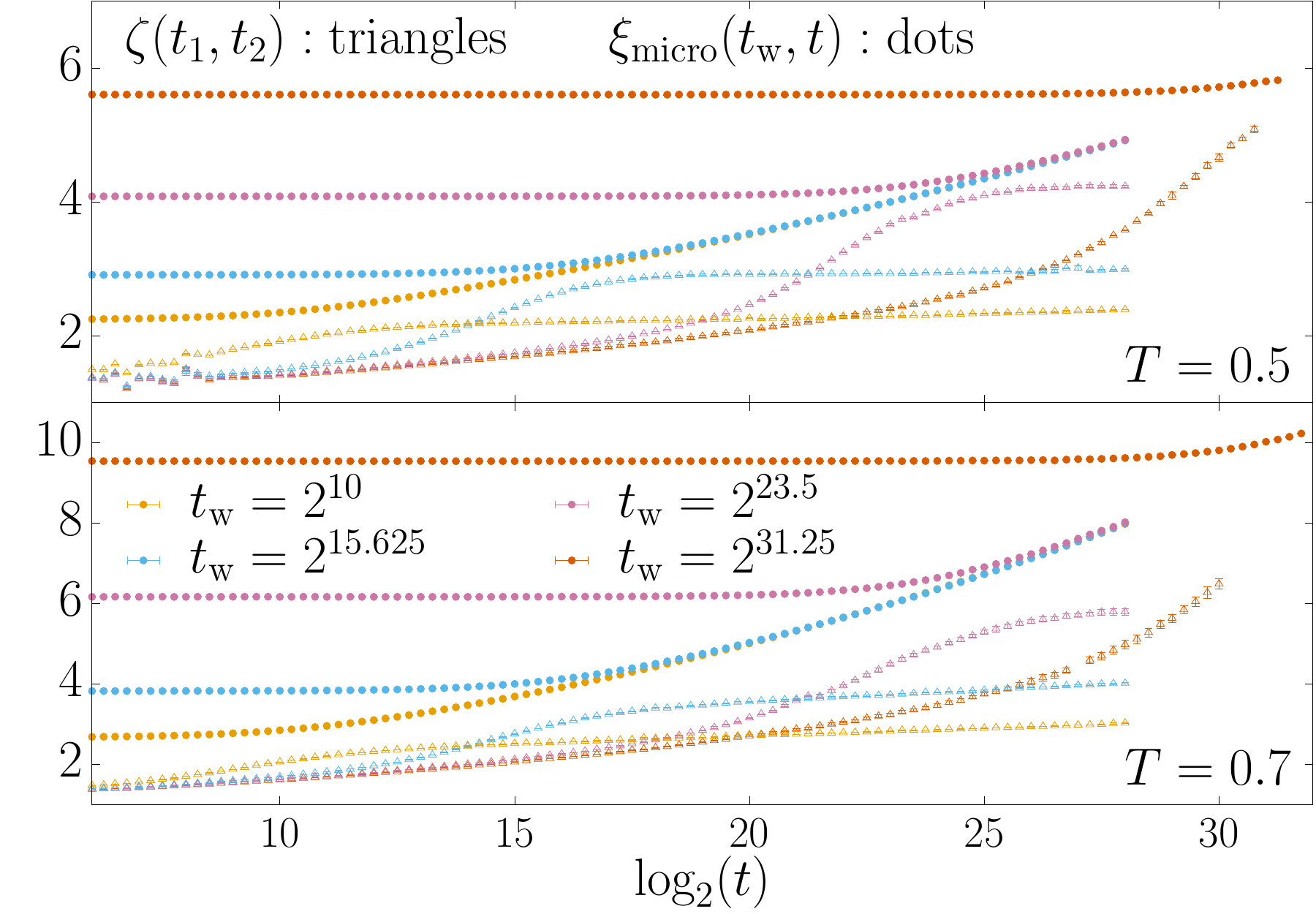}
\caption{The data for $\zeta(t_1=\tw,t_2=t+\tw)$ (triangles) from
Fig.~\ref{fig:growth_zeta}, \emph{native} protocols only, are
compared with the lengths
$\xi_{\text{micro}}(t_2=t+\tw)$ computed for the same simulations.}
\label{fig:comparison_xi_micro_zeta}
\end{figure}

In other words, native protocols behave in agreement with our expectations from older simulations~\cite{janus:09b}. Indeed,
Fig.~\ref{fig:comparison_xi_micro_zeta} confirms that $\zeta(t_1=\tw,t_2=t_1+t)$ saturates to the
$\xi_\mathrm{micro} (\tw)$ value. Instead,
for the jump-protocols, $\zeta(t_1,t_2)$ is significantly smaller than $\xi_\text{micro}$ (and, accordingly, data show no sign of saturation).

\section{Re-obtaining $\mathbf{\zeta}$ with the tools of Temperature Chaos}

Our goal here will be identifying the length scale $\zeta(t_1,t_2)$
discussed in the main text, in {\bf Methods} and in the Supplementary Note II, using the analysis tools that we employ to discuss temperature chaos. 

The main difference will be in that in the
computation of $\zeta(t_1,t_2)$ the spin configurations of the $\emph{single}$ replica are compared at times $t_1$ and $t_2$. Instead, here we shall be considering two independent replicas, one taken at time $t_1$, the other at time $t_2$. The key point is that
the distribution function for the chaotic correlation parameter $X$ turns out to be strongly dependent on the radius of the sphere (remember that $X$ is computed by restricting our attention to the interior points of randomly chosen spheres of radius $R$, see {\bf Methods}). Now, it will be found that the distribution function extending to smaller values of $X$ (i.e. the strongest chaotic signal) is found, precisely, when $R\approx \zeta(t_1,t_2)$.

Specifically, we shall compare configurations for different times $\tw$ and for different sphere radius $R$. In order to lighten the notation we give a different label in Table~\ref{tab:configurations_chaotic_comparison} to all the protocols used to compute the chaotic correlation parameter $X$ (remember that $X$ is always computed for a \emph{pair} of protocols).  

\begin{figure*}[h!]
\centering
\includegraphics[scale=0.65]{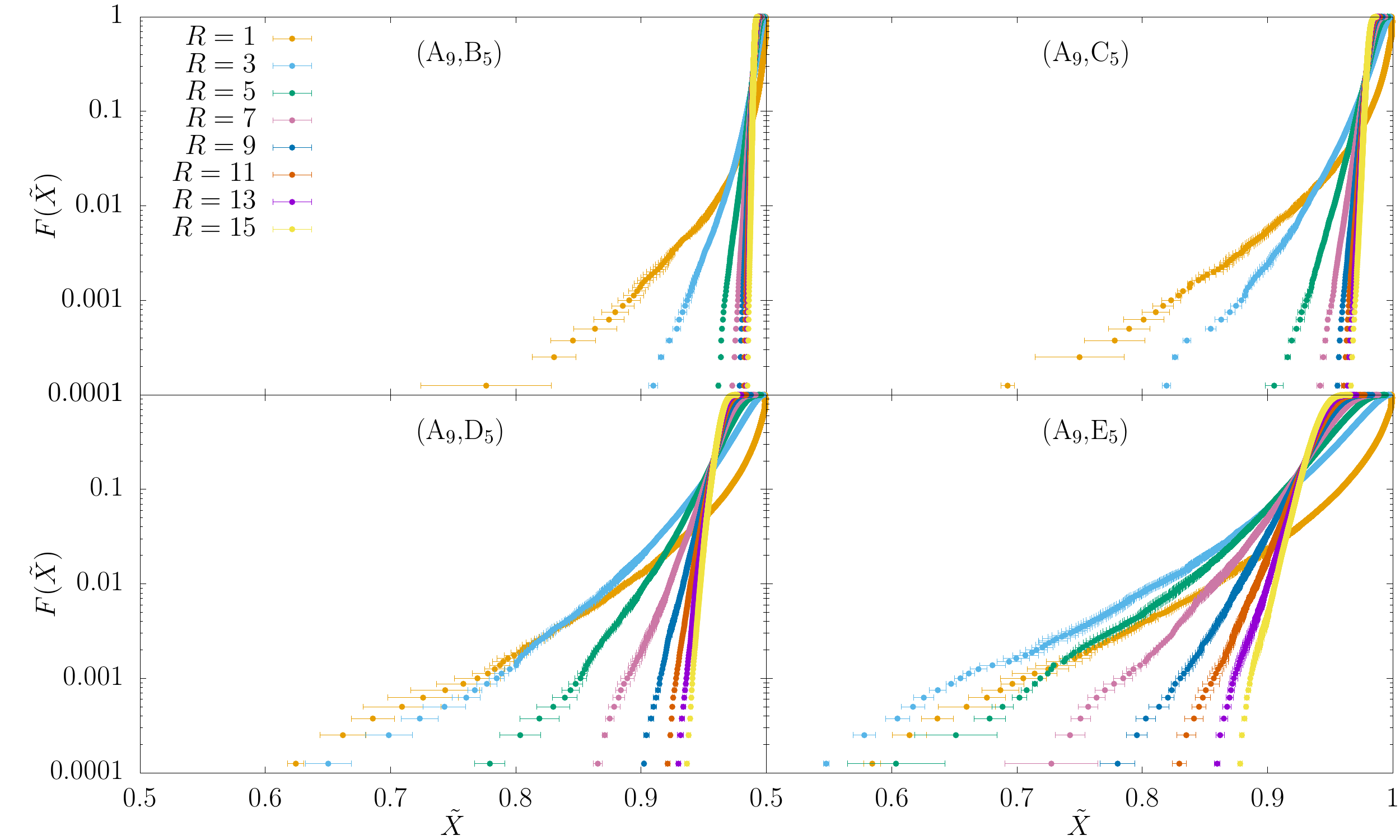}
\caption{Fraction of the spheres with a chaotic parameter $X<\tilde X$, as obtained for different pairs of protocols and different radius of the studied spherical regions. The compared configurations belong to the temperature jump from $T=0.9$ to $T=0.5$. Every panel shows in the title the used pair of configurations according to the notation introduced in Table~\ref{tab:configurations_chaotic_comparison}.}
\label{fig:rmax_jump_to_T5}
\end{figure*}

\begin{figure*}[h!]
\centering
\includegraphics[scale=0.65]{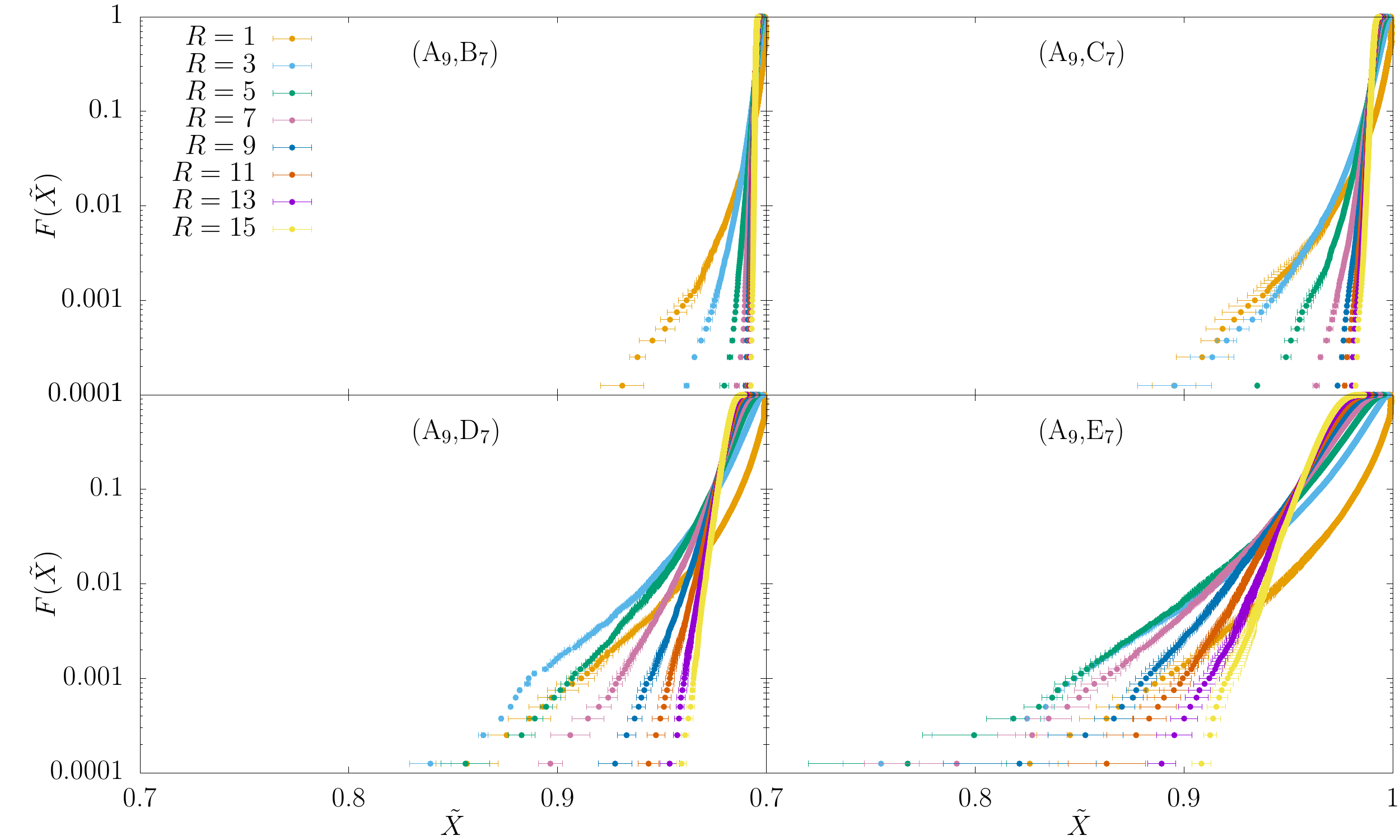}
\caption{Fraction of the spheres with a chaotic parameter $X<\tilde X$, as obtained for different pairs of protocols and different radius of the studied spherical regions. The compared configurations belong to the temperature jump from $T=0.9$ to $T=0.7$. Every panel shows in the title the used pair of configurations according to the notation introduced in Table~\ref{tab:configurations_chaotic_comparison}.}
\label{fig:rmax_jump_to_T7}
\end{figure*}

In Fig.~\ref{fig:rmax_jump_to_T5} we show the comparison between configurations at $T=0.9$ just before the temperature jump against configurations at $T=0.5$  for several times $\tw$ after the temperature jump. The analogous data for the jump to $T=0.7$ can be seen in Fig.~\ref{fig:rmax_jump_to_T7}. For the shortest $\tw$
(when $\zeta$ is very small), i.e. protocols B$_5$ and B$_7$, most
of the spheres have $X\approx 1$. We can only find a tail of smaller correlation parameters for spheres of radius $R=1$.
However, at later times $\tw$ after the temperature jump, corresponding to configurations E$_5$ and E$_7$, the relevant scale to observe the chaotic signal is $R=3$ for the temperature jump to $T=0.5$ and $R \in [3,5]$ for the temperature jump to $T=0.7$.

In conclusion, we find that the optimal $R$ for comparing the configurations right before the jump with configurations at time
$\tw$ after the jump goes from $R\approx 1$ for very short
$\tw$ to $R=3$ (for our largest $\tw$ in the jump to $0.9$) or
$R \in [3,5]$ for the temperature jump to $T=0.7$. The comparison
of these results with the values of $\zeta$ in Fig.~\ref{fig:growth_zeta} suggests that, indeed, the two approaches are featuring the same length scale.

\begin{table}[b] \centering

\begin{tabular}{c c c c c c c c c}
\hline\hline
 & $\,\,\,$ & System &  $\,\,\,$ & $T$ &  $\,\,\,$ & type &  $\,\,\,$ & $\tw$ \\
\hline \\
A$_5$ & & SG & & $0.5$ & & \textsc{native} & & $\tw^{\downarrow}+2^{31.25}$ \\
B$_5$ & & SG & & $0.5$ & & \textsc{jump} & & $\tw^{\downarrow}+2^{10}$ \\
C$_5$ & & SG & & $0.5$ & & \textsc{jump} & & $\tw^{\downarrow}+2^{15.625}$ \\
D$_5$ & & SG & & $0.5$ & & \textsc{jump} & & $\tw^{\downarrow}+2^{23.5}$ \\
E$_5$ & & SG & & $0.5$ & & \textsc{jump} & & $\tw^{\downarrow}+2^{31.25}$ \\
\hline\\
A$_7$ & & SG & & $0.7$ & & \textsc{native} & & $\tw^{\downarrow}+2^{31.25}$ \\
B$_7$ & & SG & & $0.7$ & & \textsc{jump} & &  $\tw^{\downarrow}+2^{10}$ \\
C$_7$ & & SG & & $0.7$ & & \textsc{jump} & & $\tw^{\downarrow}+2^{15.625}$ \\
D$_7$ & & SG & & $0.7$ & & \textsc{jump} & & $\tw^{\downarrow}+2^{23.5}$ \\
E$_7$ & & SG & & $0.7$ & & \textsc{jump} & & $\tw^{\downarrow}+2^{31.25}$ \\
\hline\\
A$_9$ & & SG & & $0.9$ & & \textsc{native} & & $\tw^{\downarrow}$ \\
B$_9$ & & SG & & $0.9$ & & \textsc{jump/back} & & $\tw^{\downarrow}+\tw^{\uparrow}+2^{10}$ \\
C$_9$ & & SG & & $0.9$ & & \textsc{jump/back} & & $\tw^{\downarrow}+\tw^{\uparrow}+2^{15.625}$ \\
D$_9$ & & SG & & $0.9$ & & \textsc{jump/back} & & $\tw^{\downarrow}+\tw^{\uparrow}+2^{23.5}$ \\
E$_9$ & & SG & & $0.9$ & & \textsc{jump/back} & & $\tw^{\downarrow}+\tw^{\uparrow}+2^{31}$ \\
\hline\\ [-1.5ex]
A'$_5$ & & DIM & & $0.5$ & & \textsc{native} & & $76$ \\
B'$_5$ & & DIM & & $0.5$ & & \textsc{jump} & & $430+69$ \\
A'$_7$ & & DIM & & $0.7$ & & \textsc{native} & & $197$ \\
B'$_7$ & & DIM & & $0.7$ & & \textsc{jump} & & $430+165$ \\
A'$_9$ & & DIM & & $0.9$ & & \textsc{native} & & $430$ \\
\hline\hline
\end{tabular}
\caption{ Main parameters for each of the numerical simulations. SG stands for \emph{Spin Glass}, while DIM means \emph{Diluted Ising Model}. $\tw^{\downarrow}$ corresponds to $2^{31.25}$ steps at $T=0.9$, $\tw^{\uparrow}$ corresponds to $2^{31.25}$ steps at $T=0.5$ as appears in the main text. For details about the thermal protocol see the main text, in particular Fig. 1, where we explicit the difference between the fixed-temperature protocol (native), the jump protocol, and the jump-back protocol.}
\label{tab:configurations_chaotic_comparison}
\end{table}

\section{The temperature-chaos viewpoint on memory}
In this supplementary note, we show a long numerical simulation that allows us to observe memory effects from the Temperature Chaos point of view.

The studied simulation starts at temperature $T=0.9$ where it ages for $\tw^{\downarrow}=2^{31.25}$, then, the temperature of the simulation jumps to $T=0.5$ where it ages for a time $\tw^{\uparrow}=2^{31.25}$. Finally, the temperature of the thermal bath in the simulation raises up again to $T=0.9$ where it ages for a time $\tw=2^{31}$. 

We select spheres of radius $R=3$ and we compute the chaotic parameter restricted to those spheres, just as explained in the main text.

For the computation of the chaotic parameter, we select as a reference the configurations of the system at temperature $T=0.9$ just before the temperature jump to $T=0.5$ which correspond with configuration A$_9$ in Table~\ref{tab:configurations_chaotic_comparison}, and we compare them with configurations spaced along the above described simulations, namely B$_5$, C$_5$, D$_5$, E$_5$, B$_9$, C$_9$, D$_9$, and E$_9$ in Table~\ref{tab:configurations_chaotic_comparison}.

\begin{figure*}[h!]
\centering
\includegraphics[scale=0.68]{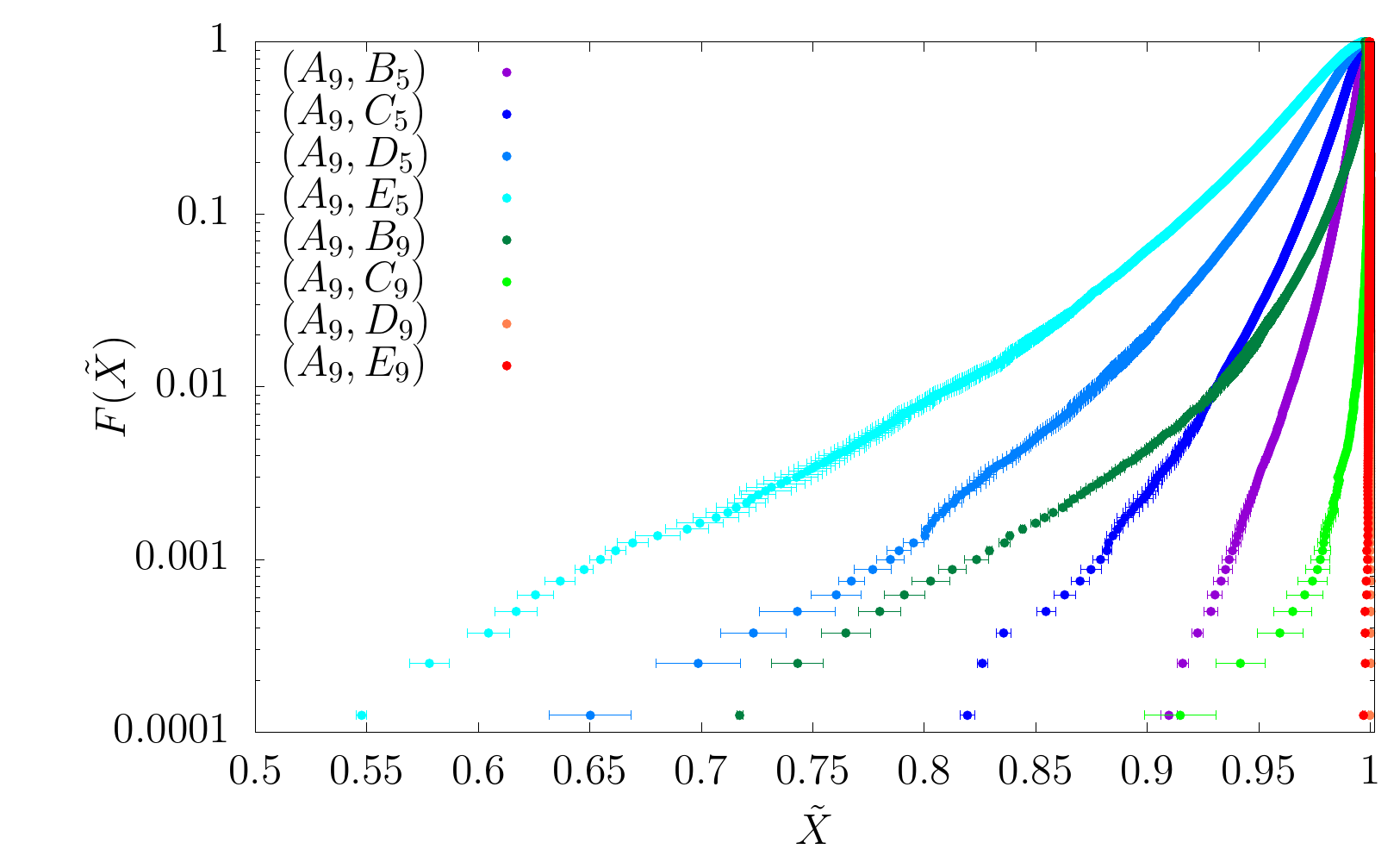}
\caption{Distribution function of the chaotic parameter computed in spheres of radius $R=3$. We simulate a very long simulation experimenting two temperature jumps and we compare the last configurations before the temperature jump from $T=0.9$ to $T=0.5$ (i.e. configurations A$_9$ in \ref{tab:configurations_chaotic_comparison}) with several configurations along the simulation (i.e. B$_9$,C$_9$,D$_9$, and E$_9$ in Table~\ref{tab:configurations_chaotic_comparison}).}
\label{fig:jump_back}
\end{figure*}

The results of these computations are shown in Fig.~\ref{fig:jump_back}. The initial situation, corresponding to the purple points, represents a non-chaotic situation in which configurations at $T=0.9$ and $T=0.5$ are similar (i.e. $X\approx 1$ for most spheres). When the color becomes lighter, the simulation at $T=0.5$ ages, and the configurations differ more and more from the reference configuration at $T=0.9$. Indeed, the distribution function shows a small tail at smaller values of $X$ (however, even in the most extreme situation $x<0.8$ only for
1\% of the spheres).

When the temperature jumps back to $T=0.9$, the distribution function evolves very quickly to a non-chaotic situation. The final situation, corresponding to the distribution function displayed with red points, is a perfect example of memory effects in Spin Glasses. The system at this point explores the same configurations that used to explore before the temperature jump, just as its excursion to $T=0.5$ never happened.

\section{The optimal sphere size when $\xi_{\text{micro}}$ is truly different}

In the main text (specifically, in the analysis leading to Fig. 4 in the main text)  we needed to face a problem never encountered previously in numerical simulations. Specifically, we wanted to compare spin configurations from a thermal protocol producing 
$\xi_{\text{micro}}\approx 16.6\,a_0$ with configurations with a protocol reaching $\xi_{\text{micro}}\approx 5.8\,a_0$.

\begin{figure*}[h!]
\centering
\includegraphics[scale=0.59]{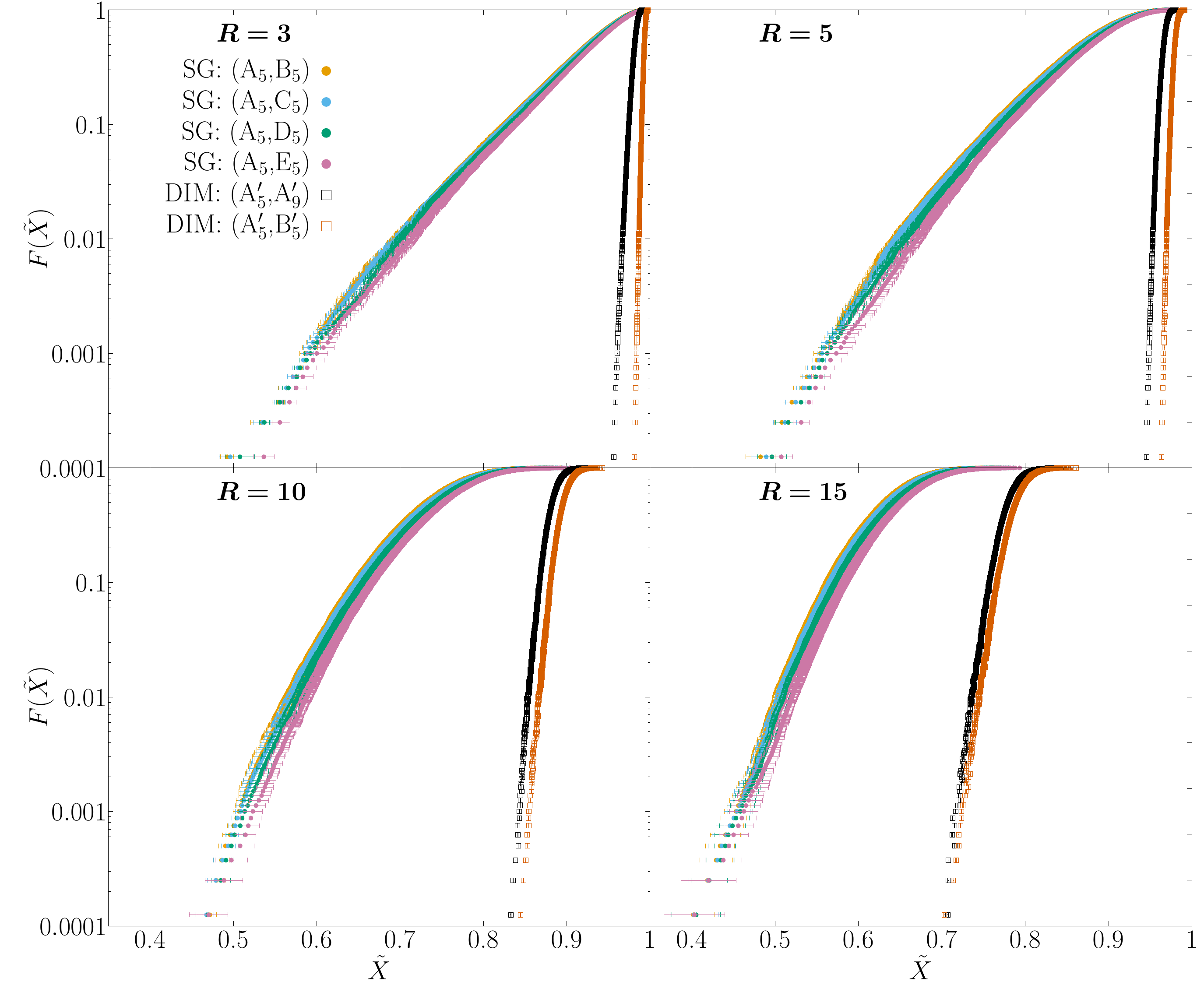}
\caption{Fraction of the spheres with a chaotic parameter $X<\tilde X$, as obtained for different pairs of protocols and different radius of the studied spherical regions. The compared configurations are exactly the same that appear in the top part of Fig. 4 in the main text (for the reader's convenience we give as well in Table~\ref{tab:configurations_chaotic_comparison} the protocols labelling). We include in the plots a comparison with the
Diluted Ising Model (DIM) where no temperature chaos is expected.}
\label{fig:best_r_with_chaos}
\end{figure*}

To address this problem, we computed the distribution function $F(\tilde X)$ for various sphere radius for two models. One of the models is a spin glass (the Edwards-Anderson model considered in this work), the other is the Diluted Ising Model (DIM; this is a ferromagnetic model where temperature chaos is absent). As explained in the main text and in {\bf Methods}, the conditions are carefully matched in both computations. Hence, we consider that a  given $R$ is safe to use when the DIM distribution function concentrates at $X\approx 1$. As Fig.~\ref{fig:best_r_with_chaos} shows, this is the case for $R=3\,a_0$ and $5\,a_0$. Yet, when
$R$ becomes significantly larger than the smallest coherence length $\xi_{\text{micro}}\approx 5.8\,a_0$, we start finding smaller values of the correlation parameter $X$ even for the chaos-free DIM. This suggests that including in the analysis spheres larger than $\xi_{\text{micro}}$ may produce misleading results (although in all cases the spin-glass  distributions have a much larger weight at small $X$). The good news are that, for safe sphere radius $R\leq\xi_{\text{micro}}$, the distribution function show a very weak dependence on $R$ (see Fig.~\ref{fig:best_r_with_chaos}). 

\bibliographystyle{apsrev4-1}
\bibliography{biblio}